\documentclass[10pt,oneside,twocolumn,aps,pra,preprintnumbers,bibnotes10pt,superscriptaddress]{revtex4-2}

\usepackage[dvipsnames]{xcolor}
\usepackage{subfig}

\usepackage{graphicx}
\usepackage{tikz}
\usepackage[absolute,overlay]{textpos}
\usepackage[T1]{fontenc}
\usepackage[utf8]{inputenc}
\usepackage{url}
\usepackage{lmodern}

\usepackage{enumitem}
\usepackage{braket}
\usepackage{booktabs}
\usepackage{amssymb}
\usepackage{amsmath}
\usepackage{amsthm}
\usepackage{float}
\usepackage{lineno}
\usepackage{wrapfig}
\usepackage{comment}
\usepackage[colorlinks=true, allcolors=blue]{hyperref}

\usepackage[font=footnotesize]{caption}
\usepackage{textcomp}

\begin{document}

\title{Energetics of Fault-Tolerant Spin-Qubit Quantum Computation}

\author{João Santos}
\affiliation{PQI -- Portuguese Quantum Institute, Portugal}

\author{Jayshankar Nath}
\affiliation{Quobly, Grenoble, France}

\author{Marco Pezzutto}
\affiliation{PQI -- Portuguese Quantum Institute, Portugal}
\affiliation{Physics of Information and Quantum Technologies Group, Centro de Física e Engenharia de Materiais Avançados (CeFEMA), Portugal}
\affiliation{LaPMET -- Laboratory of Physics for Materials and Emerging Technologies, Portugal}

\author{Valentin Savin}
\affiliation{Quobly, Grenoble, France}

\author{Tristan Meunier}
\affiliation{Quobly, Grenoble, France}

\author{Yasser Omar}
\affiliation{Instituto Superior Técnico, Universidade de Lisboa, Portugal}
\affiliation{PQI -- Portuguese Quantum Institute, Portugal}
\affiliation{Physics of Information and Quantum Technologies Group, Centro de Física e Engenharia de Materiais Avançados (CeFEMA), Portugal}
\affiliation{LaPMET -- Laboratory of Physics for Materials and Emerging Technologies, Portugal}
\affiliation{Quantum Green Computing, Ltd., Portugal}

\date{\today}

\begin{abstract}
Semiconductor spin qubits offer a promising route to compact and scalable quantum computers, combining long coherence times in isotopically purified silicon with compatibility with semiconductor manufacturing. Beyond computational speedup, their energy efficiency is an important consideration for future large-scale systems. We evaluate the power and energy requirements of a Loss-DiVincenzo spin-qubit quantum computer by analyzing the Quantum Fourier Transform (QFT) and benchmarking it against the classical Fast Fourier Transform (FFT). We develop a detailed energy budget, identifying cryogenic cooling as the dominant baseline contribution before quantum error correction (QEC). The comparison between the QFT and the classical FFT reveals that, for any quantum workload operating under a polynomially bounded sampling regime, the crossover in energy consumption occurs at smaller problem sizes than the crossover in execution time. Thus, quantum computing can become energetically competitive before achieving a speedup in execution time. We then incorporate surface-code QEC and analyze the energy costs of QEC cycles, gate synthesis, and decoding. At scale, instrumentation and decoding become dominant contributions, while cryogenic cooling imposes a constraint on scalability. We show that optimizing the control-signal demultiplexing ratio can reduce the overall energy cost by approximately two orders of magnitude. For a system with about one million physical qubits, we estimate a power consumption of approximately 100 kW, at least one order of magnitude below that of today’s highest-performance classical computing systems.
\end{abstract}

\maketitle

\section{Introduction}

In past decades, classical computers have been steadily optimized for energy efficiency, with modern architectures approaching the physical limits of what can be achieved with classical transistors. Yet, the rapid growth of High Performance Computing and Artificial Intelligence has shown that, even with continued optimization, classical computing will remain unable to offset the rising energetic burden of new technologies \cite{Prieto2024,Zhao2023,Wang2025,Dauner2025,Chen2025}. This becomes an increasingly relevant issue as the global energy demand of information and communication technologies increases. In this context, quantum computers emerged as a potential way to address this challenge, in addition to the computational speed-up that has been pursued ever since their proposal in the 1980s. For example, the energy requirements of the Sycamore quantum processor by Google, for solving the problem presented in ~\cite{Arute2019} in 2019, were approximately five orders of magnitude smaller than those of a classical computer solving a similar problem, hinting at a possible energetic advantage.

From a fundamental perspective, this possibility is supported by the fact that gate-based quantum computing is a reversible model of computation~\cite{FredkinToffoli}, due to the inherent reversibility of the Schr\"odinger dynamics underlying quantum unitary gates. This is in contrast with classical computing, which makes extensive use of logically irreversible gates, such as AND and OR gates which require a certain minimum heat dissipation per irreversible single-bit operation~\cite{Landauer1961}. However, in practice, although quantum gates are logically reversible, their physical implementation involves irreversible processes. Furthermore, other stages of quantum computation, such as qubit initialization and measurement, are also physically irreversible. 


From a more concrete perspective, in classical computers the energy dissipation comes primarily from processes not connected with the computation itself, such as from dissipation through Joule effect during information transfer \cite{Gammaitoni2015,Theis2017}. For similar reasons, it is unrealistic to expect quantum computers to dissipate no energy at all, even in the distant future, because both the process of sending and retrieving information from the quantum computer, and the classical control of the system require a classical, energy-dissipative process \cite{Banacloche2002,carrascocodina2026}. 

In fact, as shown in this work, taking one quantum gate and one Floating Point Operation (FLOP) as the fundamental units of quantum and classical computation, an energetic comparison between the two currently favors classical computers. Nonetheless, if a quantum computer demonstrates an algorithmic advantage, it may still achieve a lower total energy cost, simply by requiring fewer operations. However, an algorithmic advantage alone does not guarantee an energetic benefit or that the algorithm is energetically feasible in practice, as it is possible that the reduction in total cost is only realized at very large problem sizes, either beyond the range of relevant applications, or at a point where the energetic cost is already prohibitively high. Furthermore, whether such an algorithmic advantage exists is itself dependent on the specific computational problem under study. 

These considerations show that an energetic advantage cannot be inferred from asymptotic scaling alone, but depends on the relative cost of the fundamental elements involved in the computation. Quantifying the energetic costs associated with these elements is therefore one of the main motivations for this in-depth study of energetics. More broadly, this shifts the immediate question from whether quantum computing can be inherently more energy efficient to whether its energy requirements can be made compatible with the execution of useful computation at relevant scales, or if energy consumption may constitute a limiting factor. Establishing this energetic feasibility is a necessary first step before assessing the possibility for the energy consumption of quantum computers to be competitive with classical computers, which remains to be explored.

Until now, the efforts have been focused primarily on achieving reliable quantum computers capable of performing useful computations with increasing input size and computational performance, and minimal attention has been paid to evaluating and optimizing the energetic resources required for quantum computing. Only recently did the energetic performance of quantum processors become a topic of discussion \cite{Moutinho2023,Gois2024,ramos2026,EERA_WP2026}. A complete study of the energetics of quantum computation should start by identifying all the components that contribute to energy expenditure: the initialization and readout of the qubits, the execution of the quantum gates, the baseline costs of running the experimental setup (e.g. magnetic field, cryogenics, electronics, etc.), the classical control of the experiment, and the additional classical computing supporting the quantum computation at all stages \cite{Gois2024,Oscar2024,ramos2026}, which are the focus of this work. Future work should extend this research agenda to include the costs of generating non-trivial initial states and the repetitions required for extracting the necessary information of the final state.

One must consider that the energetic costs will depend on the chosen platform, requiring dedicated studies \cite{Gois2024,Oscar2024,ramos2026}. Within this context, this work aims at evaluating the energetic requirements of a quantum computer with a single QPU based on semiconductor spin qubits \cite{Burkard2023,Vandersypen2019,Meunier2025,Maurand2016}, assessing the feasibility of an algorithm in a relatively large system. 

As a means of comparison between the platform and classical computers, many of the algorithms that have already been proposed have classical counterparts, giving us a reference point that relate quantum and classical computers performing an analogous task. A notable example is the Quantum Fourier Transform (QFT)~\cite{Coppersmith1994}, which is a common sub-routine of key algorithms such as Shor's algorithm and Quantum Phase Estimation (QPE), and offers an exponential speedup. For such a benchmark, in this work we use the classical Fast Fourier Transform (FFT).

To evaluate the energy cost of our quantum system, we divide the energy estimates into four operational stages:\begin{itemize}
    \item \textbf{Initialization:} operations required to initialize the qubits and prepare them for performing operations, not including the generation of non-trivial initial states.
    \item \textbf{Computation:} execution of the gates that comprise the algorithm.
    \item \textbf{Readout:} measurement and signal acquisition of the qubits and retrieval of the classical bitstrings.
    \item \textbf{Baseline:} static power consumed by cryogenics, control instrumentation, magnetic field, and other auxiliary equipment to maintain the quantum computer in an active or idle state.
\end{itemize}

Using this framework, we estimate the energy cost to run QFT on the spin qubit platform, and benchmark it against classical FFT implementations, considering the energy consumption of state-of-the-art, high-performance classical computers. Our primary contributions are summarized as follows:
\begin{itemize}
    \item \textbf{Detailed energy evaluation of the spin qubit platform:} We estimate a detailed energy budget and show that the baseline costs associated with the cryogenic cooling is the dominant contributor when QEC is not included.
    \item \textbf{Early energy crossover:} We show that energy required to execute a QFT circuit drops below that of a classical FFT at practically attainable system sizes, well before reaching a crossover in execution time. This holds generally for any quantum workload with a complexity advantage, provided its state preparation and shot sampling complexities remain polynomially bounded.
    \item \textbf{Physical connectivity and noise overheads:} We evaluate the energetic penalties arising from the connectivity constraints of the system. We also estimate the overhead from error suppression techniques to mitigate decoherence arising from noise.
    \item \textbf{Large scale fault-tolerant systems:} Incorporating surface code QEC, we evaluate the energy budget of transpiled QEC cycles, rotation gate synthesis, and MWPM decoding. We show that the rotation gate synthesis gives rise to a transition from a quadratic to a linear scaling of algorithmic complexity with the number of logical qubits, mirroring the system-level impact of angular rotation cutoffs. We determine the baseline energy from instrumentation to be the largest contributor, increasing with larger code distances, which can be optimized with faster gate speeds.
    \item \textbf{Scaling with cryo-electronics:} We show that scaling Fault-Tolerant Quantum Computing (FTQC) architectures quickly hits the thermal budget set by the cryostat, making cryo-electronic control and readout mandatory. We determine that the baseline energy contributions from both decoding and instrumentation play a dominant role in these systems. The energy can be optimized using the control demultiplexing ratio, balancing contributions from cooling, instrumentation, and decoding, resulting in a reduction of around two orders of magnitude.
    \item \textbf{System-level resource estimation:} For a benchmark system of approximately 1 million physical qubits, with code distances of 13 and 25, which fall within the set error thresholds, and a demultiplexing ratio of 1:500, within the cryostat thermal budget, we estimate the total system power draw to be around 100 kW, at least an order of magnitude lower than that of the current highest performing classical computing systems.
\end{itemize}

The rest of this article is structured as follows. In section II we review the QFT, and establish the criteria under which the performance of a system implementing the QFT is evaluated, along the classical FFT as a benchmark. Section III provides an overview of the semiconductor spin qubit platform, and discusses the control methods considered in this work. Section IV evaluates in detail the energy requirements associated with each of the four different contributions highlighted above. Section V presents the overall energy requirements of executing the algorithm and a first comparison with classical machines. Section VI investigates the impact of quantum error correction on the energetics, including the effects arising from using a transpiled QEC circuit with native gate set, the synthesis of the rotation gates, and decoding. Finally, section VII focuses on the role of cryo-electronics in the efficient scaling of the platform and the energy and power advantages associated with their use.

\section{The Quantum Fourier Transform}

\begin{figure}[ht]
    \centering
\includegraphics[width=0.9\linewidth]{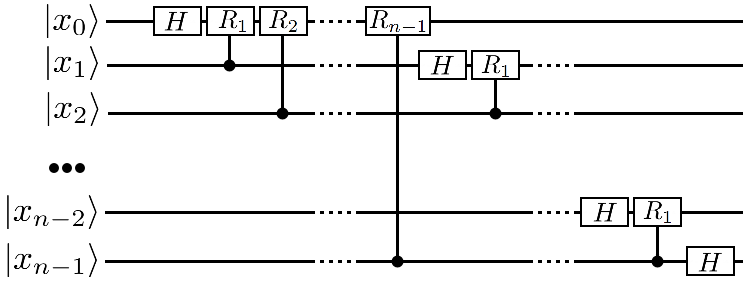}
    \caption{QFT circuit}
    \label{qft_circuit}
\end{figure}

The Quantum Fourier Transform (QFT) sub-routine \cite{Coppersmith1994} maps a state $\ket{x}$ to the superposition $\frac{1}{\sqrt{N}}\sum_{k=0}^{N-1}e^{2\pi i\frac{xk}{N}}\ket{k}$, with $\{\ket{0},...,\ket{N-1}\}$ a basis of $N$ elements, which can be represented in binary bitstrings and encoded in $q=\lceil \log_2{N} \rceil$ qubits. If the values of a function $f(x)$ are encoded in a quantum state given by $\sum_{x=0}^{N-1}f(x)\ket{x}$, the result of the QFT will be 
\begin{equation}
    \sum_{x=0}^{N-1}f(x)\ket{x}\to\sum_{k=0}^{N-1}\frac{1}{\sqrt{N}}\sum_{x=0}^{N-1}e^{2\pi i\frac{kx}{N}}f(x)\ket{k}.
\end{equation}
Thus, the algorithm maps the amplitudes $f(x)$ to the amplitudes of its discrete Fourier transform $F(k)$. This operation is unitary and can be represented by an $N\times N$ matrix. Although $N$ grows exponentially with the number of qubits, the circuit to implement the algorithm (Fig.~\ref{qft_circuit}) only requires $q$ Hadamard gates and $\frac{q(q-1)}{2}$ controlled-phase gates, so the number of gates scales with $O(q^2)=O(\log_2(N)^2)$. In terms of circuit complexity, this represents an exponential reduction in the number of elementary operations compared with the $O(N\log_2(N))$ arithmetic operations required for the full FFT \cite{Cooley1965}.

As a classical counterpart of the QFT, the FFT provides a natural reference benchmark for a particular problem of size $N$, against which the performance of the QFT can be assessed. The time and energy costs of executing the QFT can therefore be evaluated and compared against the costs of executing the FFT on state-of-the-art classical supercomputers.

Importantly, in these estimates, we do not account for the costs associated with preparing non-trivial input states. Furthermore, we assume a fixed number of 1024 repetition shots, which is reasonably more than sufficient for a polynomial scaling of shots within the system sizes considered in this work. We also do not consider tomographic reconstruction of the final states, therefore a full reconstruction of the spectra is precluded. The analysis instead focuses on the resources required to execute the QFT, which serves as a subroutine of several important algorithms, including Shor's algorithm and Quantum Phase Estimation (QPE). We therefore interpret the QFT primarily in this context, rather than as a standalone algorithm, as this would require different assumptions specific to the problem, input state, and use of the resulting final state. The assumptions used in this work allow the analysis to focus on the intrinsic costs of the quantum algorithm and, in particular, on the energetic and power requirements of the quantum system.

Given these considerations, although the FFT is the classical counterpart of the QFT, the comparison between the two should not be interpreted as a comparison between equivalent computational tasks. The full FFT would compute all Fourier coefficients explicitly, whereas, under the assumptions considered in this work, the QFT could only be used to estimate them up to a certain accuracy. The FFT is therefore used solely as a classical benchmark. Consequently, any crossover between the estimated QFT performance and the classical reference should be interpreted with respect to this benchmark, rather than as evidence of a definitive quantum advantage, as this requires a broader study accounting for specific use cases.

\section{Semiconductor Spin qubits}

The spin of a charge was early identified as a natural two-level system that could be used to make a qubit. By taking advantage of the existing industry of semiconductors and their fabrication processes, semiconductor spin qubits can be built and controlled within nanometer scale, allowing for very high density of qubits, long coherence times \cite{Burkard2023,Paz2024} and seamless connection to classical control in integrated circuits \cite{Toubeix2024}.

They were first proposed by Daniel Loss and David DiVicenzo in 1997 \cite{Loss1998}. The manipulation of the spin itself borrows ideas from nuclear magnetic resonance (NMR) spectroscopy. The first step to obtain a Loss-DiVicenzo (LD) qubit is to trap an electron in a confined region called quantum dot, and apply a magnetic field $B_0$ in the $z$ axis. Then, the states $\ket{0}$ and $\ket{1}$ are mapped to the spin down/up respectively and become separated by an energy of
\begin{equation}
    \Delta E=g\mu_B B_0 \approx 10^{-4}\,\text{eV} \quad \text{for $B_0=1$ T.}
\end{equation}
This energy contribution is negligible compared to that of other components, as the energy necessary to execute the transition between the levels is dominated by the control mechanism. Under the Zeeman Hamiltonian, the spins also precess around the $z$ axis with the Larmor frequency 
\begin{equation}
    w=\frac{g\mu_B}{\hslash} B_0=1.76\times10^{11}\,\text{rad/s} ,\quad\text{for $B_0=1$ T}.
\end{equation}
Individual spins can then be controlled with oscillating magnetic or electric fields perpendicular to the static magnetic field, allowing us to perform one-qubit gates. Other than the Zeeman Hamiltonian, the spin dynamics of adjacent dots are modeled by a Hubbard Hamiltonian \cite{Hubbard1963,Hensgens2017} which, in the limit of tight electronic confinement (1 electron per dot), can be simplified to the Ising exchange Hamiltonian 
\begin{equation}
    H_E=J_{ij}S_z^iS_z^j.
\end{equation}
The constants $J_{ij}$ are the coupling constants and they can be manipulated using gate voltages applied to the barrier gates between the dots. This forms the basis for two-qubit gates in this platform.



\begin{figure}[ht]
  \centering
\includegraphics[width=0.45\textwidth]{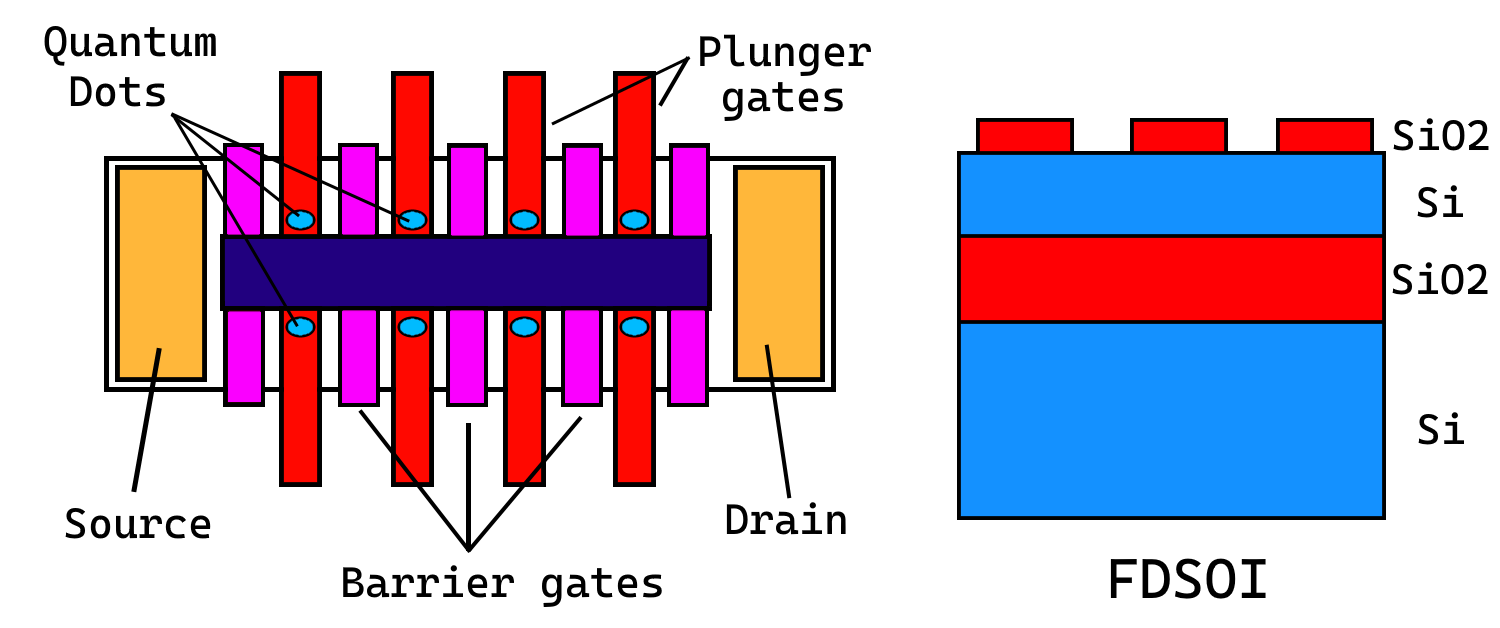}   
  \hfill
\caption{On the left, a diagram of a quantum dot and readout dot array. In a 1D architecture, each qubit has 1 plunger plus one half-barrier on each side; on the right Cross-sectional view of the Fully Depleted Silicon-on-Insulator (FDSOI) device structure.}  
\label{dot_diagram} 
\end{figure}

In general, the electron system is formed as a 2D electron gas in a thin layer, or at the interface between two layers of semiconducting material. In the case of Si-MOS devices (Fig.~\ref{dot_diagram}), the dot is initialized in a thin Silicon nanowire. The confinement in the $z$ direction is defined by the thickness of the nanowire as well as back-gate potential. Confinement in the $xy$ plane is defined by the width of the nanowire, and by potential barriers. The potential in the quantum dot is controlled by plunger gates, while the potential between dots is controlled by barrier gates.

In this work, we consider two different device architectures: a 1D architecture and a dense 2D architecture.

The 1D architecture consists of a linear array of quantum dots, with one single-electron transistor (SET) for each qubit. Each dot has two neighboring dots and is controlled by one plunger and two barrier gates, one on each side. For our estimates, we attribute half the plunger to each neighboring dot. Thus, each individual dot is controlled by two gate electrodes, each requiring a DC and an RF output. Scaling this architecture to multiple source and drain contacts or nanowire channels is not considered, as the required implementation depends on the specific large-scale architecture employed and is beyond the scope of this study.

The 2D architecture consists of a dense array of quantum dots, with the readout SET positioned directly below each physical qubit \cite{Vinet_2018}. Each dot is controlled by one plunger gate and four barrier gates, one on each side. By attributing half of each barrier gate to neighboring dots, each individual physical qubit thus requires three gate electrodes. This architecture has nearest neighbor connectivity between physical qubits. For the analysis with quantum error correction, we consider that connectivity between logical patches is established through patch deformation \cite{Litinski2019} and the QPU layout is based on the fast data block and the 15-to-1 distillation protocol introduced in \cite{Litinski2019}.

The two architectures are used at different stages of the analysis. The simpler 1D architecture is considered for the energy estimates without quantum error correction, where we first assume all-to-all connectivity between qubits at the algorithm level. We then consider a more realistic nearest neighbor connectivity, in which distant qubits are brought together using SWAP gates. The dense 2D architecture is used for the analysis including quantum error correction, where the physical layout and its importance to the energetics are explicitly taken into account.

\section{Energetic considerations} 

To quantify the energy necessary to run a quantum algorithm, we will consider the four steps identified in the introduction: initialization; computation; readout and baseline. 

To make these calculations, we will consider that the source, drain, plunger, and barrier gates are connected to the control electronics via unattenuated fast DC cables (up to around 100 MHz bandwidth). The plungers and the barrier gates, i.e. the gates that require fast pulsing are also connected via high frequency (RF) cables. The RF and DC cables are then connected via a bias-tee before connecting onto the device. The RF line connecting to the ESR line is also unattenuated. Furthermore, here we consider a global ESR control. Finally, given the small footprint of silicon spin qubits, we can scale up to a large number of physical qubits in a single cryostat, as long as the energy dissipated by the system is within the cooling budget of the cryostat \cite{pauka2019}

\subsection{Initialization}

Initialization is the process by which the array is filled with electrons. In case of spin qubits, this process needs to be performed once at the beginning. To move the electron into the dot, the dot can be thought of as a transistor, where the electrons move from the source to the drain one by one. By tuning the plunger gate voltage, individual electrons can be trapped under the gate. To ensure this degree of precision the electrons must be cooled down to a temperature of $0.1\text{ K}-1.5\text{ K}$. Additionally, by applying large positive voltages on the barrier gates adjacent to the source and drain, the electron system can be isolated from rest of the system. This prevents electrons from tunneling in and out of the system. Thus, once loaded, the electrons stay trapped for a long period and need not be periodically refreshed during execution or between repetitions of the algorithm.

There are two steps that must be considered when calculating the initialization energy. First, the dots within an array are loaded with electrons, by lowering the potential in each dot by $V$, usually of the order of some mV (i.e. $100$ mV). The voltage is applied using RT instrumentation connected to the gates with a line of capacitance $C$ (around $100$ pF). This results in a line charging energy of
\begin{equation}
    E_{\text{charge}}=\frac{1}{2}CV^2=3.125\,\text{MeV}.
\end{equation}
This applied voltage also results in an effective dot potential that is determined by the lever-arm ratio $\alpha\approx0.15\,\text{eV/V}$, but it is much smaller than the line charging energy itself. The device geometry influences the lever-arm ratio via the gate capacitance, and hence the values can vary depending on the implementation.

Once the electrons are in the dots, the potentials on the two ends of the array are lowered to $\approx -1$ V, effectively raising two potential barriers for the electrons, so that the they become isolated from the source and drain. As these dots are fabricated with CMOS technology \cite{Hamonic2024}, the system can be engineered to minimize leakage current and there is no expenditure to maintain the potential barrier, other than the energy necessary to generate the voltage swing: 
\begin{equation}
    E_{\text{isol}}=2\cdot\frac{1}{2}CV^2=625\,\text{MeV}.
\end{equation}
In order to systematically load electrons into the array, the dots may be loaded from one end of the array to the other, sequentially filling two dots at a time. Each time two dots are filled, there is an addition of $E_{\text{isol}}=625\,\text{MeV}$ to isolate them from the unfilled part of the array and prevent the electrons from tunneling out of it. The total initialization energy per qubit becomes
\begin{equation}
E_{\text{init}}=E_{\text{charge}}+E_{\text{isol}}=628.125\,\text{MeV}.
\end{equation}

\subsection{Computation}

Computation refers to the energy necessary to execute the one qubit gates and two-qubit gates, which forms the gates sequence comprising the quantum circuit. There are many physical mechanisms that can be used to drive Rabi oscillations for the one qubit $x$ and $y$ gates. Among them, Electron Spin Resonance (ESR) is a method similar to the one used in nuclear magnetic resonance to implement single qubit rotation gates. In this method, an oscillating magnetic field $B=B_1(\cos(wt+\phi)\hat x + \sin(wt+\phi)\hat y)$ is applied, where $w$ is the Larmor frequency. In the rotating frame, the Hamiltonian is $H_{\text{ESR}}=\frac{1}{2}g\mu_B B_1(\cos(\phi)\sigma_{x}+\sin(\phi)\sigma_{y})$ which results in a Rabi frequency
\begin{equation}
\Omega=2\pi f=\frac{g\mu_B B_1}{\hslash}.
\end{equation}
In order for the complete transition $\ket0\to\ket1$ to be executed in $\approx1\,\mu s$, corresponding to $f=\frac{\Omega}{2\pi}\approx 500\,\text{KHz}$, we require a field of $B_1\approx18\,\mu \text{T}$. This microwave (MW) field is generated using a stripline antenna with an RF current and is linearly polarized, $B_1=B_{\text{ESR}}\frac{e^{iwt}+e^{-iwt}}{2}$. Under the rotating wave approximation, only the first term contributes to the rotation and thus we must generate a linear field twice the magnitude of $B_1$: $B_{\text{ESR}}\approx36\,\mu \text{T}$. If the distance between the line and the dot is $r=1\,\mu\text{m}$ \cite{Vinet_2018}, then $I=\frac{2\pi r B_{\text{ESR}}}{\mu_0}\approx 180\,\mu\text{A}$. This current is oscillating, so we can take the root-mean-square current $I_{\text{rms}}=\frac{I}{\sqrt{2}}\approx 127.3\,\mu\text{A}$. Assuming an impedance $R$ of around 100 Ohms, including losses in practice, the energy for the ESR field during the transition is:
\begin{equation}
    E_{\text{ESR}}=\Delta t\cdot R\left(\frac{2\pi r B_{\text{ESR}}}{\sqrt{2}\cdot\mu_0}\right)^2=10\,\text{MeV}.
\end{equation}
In a global ESR control scheme, the field is shared among the full qubit array. Individual qubits are addressed by bringing the target dots into or out of resonance with the global ESR field. This means that, if $m$ $R_{x/y}$ gates are performed at the same time in $m$ different qubits, the total cost of the gate is reduced to the energy of bringing the $m$ qubits into resonance plus the energy of a single ESR field application. This allows for the parallelization of the RF pulse, decreasing the cost of applying simultaneous $R_{x/y}$ gates.

Rotations in $z$ can be applied with $g$-factor modulation \cite{Chan_2018,Tanttu_2019}, in which the effective $g-$factor of the electron in the dot is temporarily modified by $\delta g$, resulting in a shift of the Larmor frequency by $\delta w=\frac{\delta g\mu_B}{\hslash} B_0$. This induces a phase shift in the target qubit. Alternatively, depending on the algorithm, $R_z(\theta)$ rotations can be implemented virtually. Here, we consider the energy necessary to induce the variation of the $g-$factor. Electrons in silicon dots have a tunability of few 10s of MHz/V \cite{Chan_2018,Tanttu_2019} so, to achieve a comparable frequency to ESR, a voltage of $\approx 100\,\text{mV}$ is applied in the target dot, resulting in a line charging energy of 
\begin{equation}
    E_Z=\frac{1}{2}CV^2=3.125\,\text{MeV}.
\end{equation}
This is the same effect utilized in ESR to bring the dots in and out of the magnetic field resonance, such that the energy of bringing the target dot into resonance is also $3.125$ MeV.

A two-qubit gate can be realized by controlling the coupling between two adjacent dots \cite{Burkard1999,Meunier2025,Bartee2025}, by modulating the voltage on the barrier gate. This corresponds to an evolution operator
\begin{equation}
    U(\theta)=e^{-\frac{i \, t }{4\hbar} J\sigma_z^1\sigma_z^2}=e^{-\frac{i}{2}  \theta \, \sigma_z^1\sigma_z^2},\quad \theta=\frac{Jt}{2\hslash}.
\end{equation}
The energy dissipated to perform a two-qubit gate thus depends on the energy associated with the modification of the exchange interaction $J$ between the dots. As $J$ is modeled by the potential in the barrier between the two dots, and considering again the same voltage swing of $100\,\text{mV}$, we have that $E_J=3.125\,\text{MeV}$.

With $x,y,z$ rotations and the $U(\theta)$ gate, any other gate can be decomposed and executed (decompositions in Appendix~\ref{sec:app_gates}). In fact, just the set \{H,S,CNOT\} plus a gate outside the Clifford group (usually the T gate) can be used to approximate any gate up to arbitrary accuracy \cite{Dawson2005,Kitaev1997,NielsenChuang2010}. For simplicity, we assume that the time required to perform any of these three gates is fixed at $1\,\mu \text{s}$. Although different rotation angles can be implemented by varying the amplitude of the pulse, in the circuit-level calculations we assign each gate the energy that was previously calculated, regardless of the rotation angle.

\subsection{Readout}

The readout is performed via spin-to-charge conversion of the electrons in the dots. This conversion is performed using an auxiliary dot to which the target qubit can tunnel into \cite{Hamonic2024}. If the initial state is the (1,1) singlet, the final state after the readout has occupancies (2,0) or (0,2). On the other hand, because the energy of the triplet is higher by an amount corresponding to the exchange coupling constant $J$, the initial (1,1) triplet state stays blocked, up to a finite decay time. The charge configuration can be probed with an adjacent electrometer. This electrometer is a Single Electron Transistor (SET) operating in the transport regime, i.e. with a current flowing through the SET from the source to the drain of the nanowire. The SET is capacitively coupled to two isolated dots, thus, the current flowing through the SET depends on their state \cite{Elzerman2004,Urdampilleta2018,Hamonic2024}. An example of a possible configuration for placing the electrometer face-to-face to a quantum dot is presented in Fig.~\ref{dot_diagram}. The current flowing through the SET can be approximated to $I\approx1$ nA, with a source-drain bias $V\approx1\,\text{mV}$. For a readout time $t_{\text{readout}}=1\,\mu$s, the energy dissipated in the SET is 
\begin{equation}
E_{\text{current}}=VIt_{\text{readout}}=6.25\,\text{eV}
\end{equation}
The energy needed to apply the gate voltage to the SET is 
\begin{equation}
E_{\text{SET}}=\frac{1}{2}CV^2=3.125\,\text{MeV}.
\end{equation}
Additionally we assume the use of frequency-multiplexed cryogenic CMOS transimpedence amplifiers in our readout chain \cite{Schmidt2024}, which consume $7.4\,\mu$W for a time of $1\,\mu$s, corresponding to $46.26\,\text{MeV}$ per qubit, bringing the total readout cost to
\begin{equation}
E_{\text{readout}}=49.385\,\text{MeV}.
\end{equation}

\subsection{Baseline Costs}

\subsubsection{Establishing a magnetic field}

The global magnetic field is initially generated with a current in a superconducting solenoidal magnet. The total energy stored will be $E_{B_0}=\frac{1}{2}LI^2$ where $L\approx10\,\mu \text{H}$ is the inductance and $I\approx8\,\text{A}$ is the current necessary to generate a magnetic field $B_0=1\,\text{T}$. The final expression becomes 
\begin{equation}
    E_{B_0}=\frac{1}{2}LI^2\approx 2\times10^{15}\,\text{eV}.
\end{equation}
Because this depends heavily on the characteristics of the coil, we can take a higher bound of $E_{B_0}=1\,\text{mJ}=6.25\times10^{15}\,\text{eV}$. Once the magnetic field is established, the coil is cooled down until it reaches a superconducting state. No energy other than the cooling cost of the entire system is necessary to maintain the magnetic field further. Similarly to the initialization, this process is done only once at the preparation stage.

\subsubsection{Cooling power}

The cooling energy is often one of the dominant contributors to the energy expenditure. We can expect the cryostat to require between 5 and 10 kW of input power. Because we can scale up to a large number of physical qubits in a single cryostat, this power is constant and the energy is dependent on running time.

\subsubsection{Instrumentation}

All the equipments used to control the QPU are also considered part of the baseline costs, including the microwave source, the DC source and the arbitrary waveform generator (AWG)/RFSoC circuit. The last two components (DC and RF) will be referred to as "instrumentation", and depend on the number of DC and RF outputs required to control the entire dot array. As representative of typical DC sources, we will consider that each DC source has 50 outputs and consumes 20 W. Similarly, we consider an RF source with 8 outputs which consumes 40 W. This instrument includes the Analog-to-Digital Converters (ADC) needed for signal acquisition. We can conservatively account for 220 W for other control instrumentation, which is an additional fixed cost, independent of the number of qubits. Although there are other components that are necessary to keep the circuit running, these form the essential components of the platform.

\section{Cost of the QFT on spin qubits}

\begin{table*}[t]
    \centering
    \footnotesize
    \begin{tabular}{cccc}
        \toprule
        Operation & executions & Time & Cost (eV or eV/s) \\
        \midrule
        Loading dot & 1 per qubit & $100\, \mu s$ & $3.125\times10^6$ eV \\
        Isolation & 1 per qubit & negligible & $625\times10^6$ eV \\
        ESR MW field & 1 per $R_{x/y}$ layer & $1\,\mu s$ & $10\times10^6$ eV \\
        ESR Stark shift & 1 per $R_{x/y}$ & $1\,\mu s$ & $3.125\times10^6$ eV \\
        Z rotation & 1 per $R_z$ & $1\,\mu s$ & $3.125\times10^6$ eV \\
        Two qubit gate & 1 per $U(\pi/2)$ & $1\,\mu s$ & $3.125\times10^6$ eV \\
        Readout & 1 per qubit per shot & $1\,\mu s$ & $49.39\times 10^{6}$ eV \\
        Magnetic field & 1 & -- & $6.25\times 10^{15}$ eV \\
        Cooling power & -- & -- & $3-6\quad( \times10^{22})$ eV/s\\  
        Instrumentation & -- & -- & $3.375\times10^{19}$ eV/s per qubit \\
        Other baseline & -- & -- & $1.375\times 10^{21}$ eV/s\\ 
        \bottomrule
    \end{tabular}
    \caption{Energy or power values for individual operations. The initialization of the qubits as well as the preparation of the magnetic field are performed only once. The readout of the qubits is performed at every shot. The ESR field is global, needing to be sustained only once for a layer of parallel $R_{x/y}$ gates. The baseline components (magnetic field, cooling cost, instrumentation and other baseline) do not impact the running time, but need to be sustained during the entire duration of the algorithm.}
    \label{table1}
\end{table*}

The energy budget per operation on the spin qubit platform is tabulated in table~\ref{table1}. Knowing that, for $q$ qubits, the QFT includes $q$ Hadamard gates and $\frac{q(q-1)}{2}$ controlled phase gates, we can use the decompositions in Appendix~\ref{sec:app_gates} to determine the total number of native gates needed to run the QFT: $4q^2-2q$ $R_x$ gates; $\frac{11}{2}q^2-\frac{9}{2}q$ $R_z$ gates and $q^2-q$ $U(\pi/2)$ gates. The total execution time and energy are then obtained by summing the contribution of loading, isolating and performing the readout of $q$ qubits, the contributions of each native gate, and the contributions of the different components of baseline. The resulting total is then multiplied by the number of shots required to obtain the output, which in this work is taken as 1024.

In this section, we consider a NISQ application without error correction, as the energetic costs associated with error correction are addressed in a later section. We assume that qubit initialization and establishment of the magnetic field are performed only once, whereas the readout of the $q$ qubits is executed at the end of each shot. We consider the 1D architecture introduced previously and initially assume all-to-all connectivity. We then account for additional swap gates required for nearest-neighbor connectivity in Section~\ref{sec:connectivity}.

\subsection{Benchmark using classical machines}

The total time and energy of the QFT can be benchmarked using the fastest and most efficient supercomputers ranked in the TOP500~\cite{top500} as of March 2026, performing the FFT. El Capitan, the fastest supercomputer ($1742$ PFLOP/s), will be used to benchmark the performance in terms of time, while the KAIROS, the most energy efficient ($46\,\text{kW}, \,3.1\,\text{PFLOP/s}$), will be used to benchmark the performance in terms of energy, so that the best possible parameters for state-of-the-art classical machines are used. These numbers are estimated from the performance and power data published by the TOP500 project. Because the actual time and energy consumption per FLOP depend on factors such as clock frequency, queuing delays, utilization and the benchmark application, they should be regarded as indicative values.

To make this benchmark, we note that $q$ qubits can encode a problem of size $N=2^q$. In this way, we can relate the two quantities: $\log_2(N)\equiv\textbf{"Number of Qubits"}$. Thus, we quantify the performance of the QFT to be $O(q^2)$ and that of the classical $FFT$ to be $O(q\cdot 2^q)$. It must be noted that we would need more bits than qubits to encode the same input data because, in the quantum algorithm, the values of the input are encoded in the superposition of the total state, while, for the classical version, we would need dedicated bits for each entry of the input. However, by fixing the value of $\log_2(N)$ for both the classical and quantum Fourier transform, we are able to estimate the performance of both algorithms working to compute the same function and use the classical FFT as a reference value for each $N$.

\subsection{Time and energy performance of the QFT}

\begin{figure*}[t]
  \centering
  \subfloat[]{
\includegraphics[width=0.32\textwidth]{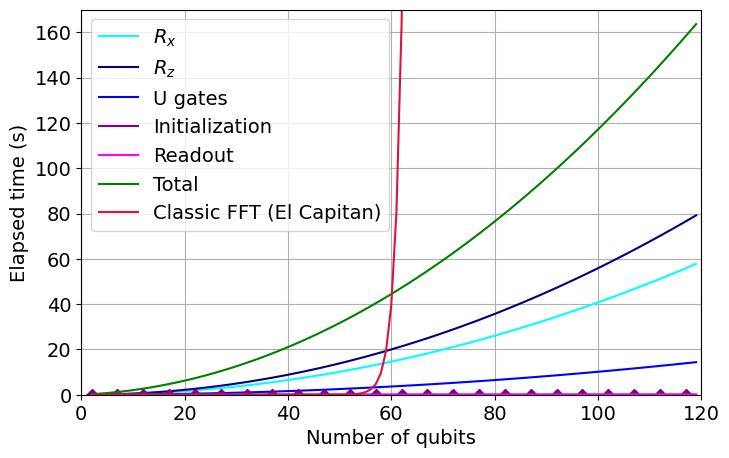}
    \label{no_ec_time}    
    }
  \hfill
    \subfloat[]{
\includegraphics[width=0.32\textwidth]{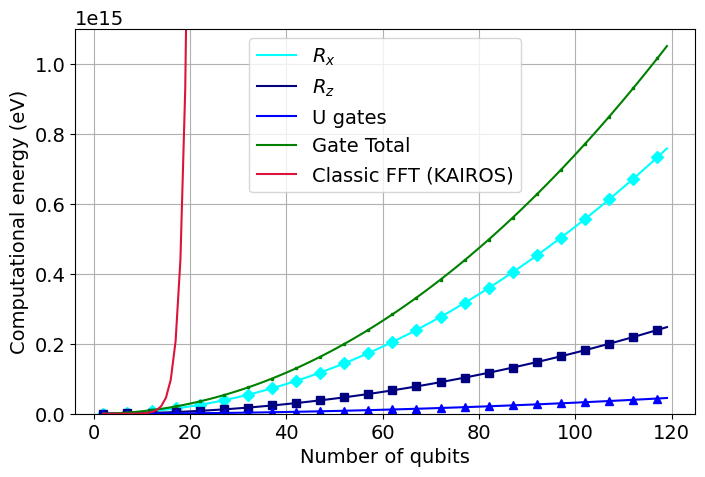}
    \label{no_ec_comp}
    }
   \hfill
    \subfloat[]{
\includegraphics[width=0.32\textwidth]{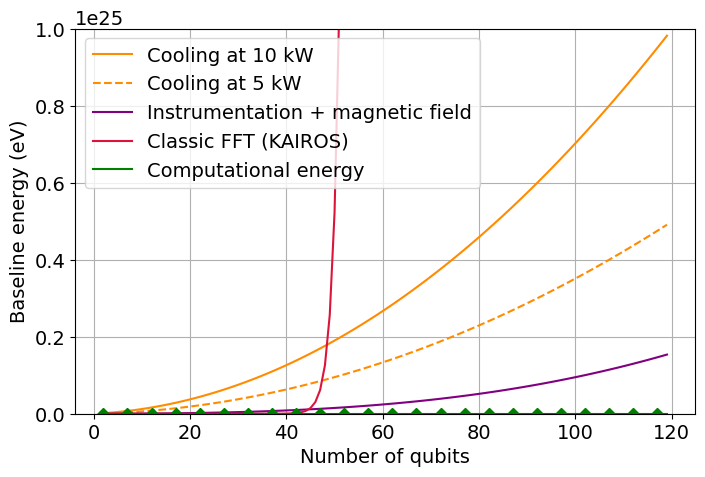}
    \label{no_ec_total}
    }
\caption{QFT (1024 shots) vs FFT: a) Algorithm running time: The complexity of the QFT is given by $q^2$ while the complexity of the FFT is $q2^q$, by assuming that both the QFT and FFT work with the same input data with $N=2^q$ entries; b) Computational energy: the cost of on-chip execution of all the gates of the algorithm. We can see that the $R_x$ gate with ESR is the most expensive gate; c) System-level energy budget: The system-level energy includes the baseline energy necessary to establish the magnetic field; the cooling cost, for which the power consumption is constant; and the instrumentation cost, for which the power consumption is proportional to the number of qubits. The dominant component of baseline could either be the cooling, which scales only with the running time (for lower number of qubits) or the instrumentation energy (for larger number of qubits). The baseline energy represents by far the largest part of the energy consumption and scales with time, so the energy becomes $E_{\text{total}}\approx P_{\text{baseline}}t_{\text{total}}$. Here we consider $P_{\text{cooling}}=10$ kW. The apparent difference between the energy of the classical FFT in b and c is due to the different scales of the $y$ axis.
}
\end{figure*}

Figs.~\ref{no_ec_time} and \ref{no_ec_comp} show, respectively, the running time and computational energy of performing the QFT and FFT, separated in different contributions. The computational energy accounts exclusively for the energetic cost of the on-chip quantum gate operations. ESR dominates the computational cost, accounting for $\approx70\%$ of the computational energy. This is partly due to the fact that, in the QFT circuit considered in Fig.~\ref{qft_circuit}, the rotation $R_{x/y}$ gates are to be applied sequentially. This limits parallel gate operations and the related energy gain from using a global ESR. 

In order to obtain a system-level energy budget, we need to also consider the baseline energy costs associated with the external magnetic field, the cryogenic cooling and the instrumentation. The system-level energy costs are plotted in Fig.~\ref{no_ec_total}. The baseline energy is much larger than the energy associated with the quantum gates and, in fact, accounts for approximately the entire energetic cost. From the different baseline contributions, the cooling cost represents the biggest fraction of the total energy cost.

Benchmarking this system-level cost to run the QFT against the classical FFT, we see that the energy requirements of the QFT fall below those of the full classical FFT beyond a certain problem size. This crossover is consistent with the algorithmic advantage of the quantum algorithm when compared to the classical analogue, and reflects the crossing point at which the lower scaling of the QFT results in lower total energy requirements than the used benchmark. 

While this crossover with respect to the classical full FFT should be taken as indicative rather than a definitive energy advantage, the problem size at which it happens is an important metric. This is because it is not \textit{a priori} guaranteed that the crossing point is achievable for a small input size; in principle it could require a number of qubits so large as to render the algorithm energetically unfeasible. Here, instead, our results indicate that it can be achieved for a relatively small number of qubits.

\begin{table}[t]
    \centering
    \begin{tabular}{ccccc}
        \toprule
        \textbf{Component} & Cost (eV) \\
        \midrule
        $R_x$ gates & $5.35\times10^{14}$ \\
        $R_z$ gates & $1.75\times10^{14}$ \\
        $U$ gates   & $3.17\times10^{13}$ \\
        Initialization & $6.31\times10^{10}$ \\
        Readout & $5.06\times10^{12}$ \\
        \midrule
        \textbf{Total algorithm cost} & $7.46\times10^{14}$ \\
        \midrule
        Static Magnetic field & $6.25\times10^{15}$ \\
        Cooling & $7.02\times10^{24}$ \\
        Instrumentation & $9.49\times10^{23}$ \\
        \midrule
        \textbf{Total QFT cost} & $7.97\times10^{24}$ \\
        \midrule
        \textbf{Classic FFT} & $1.17\times10^{40}$\\
        \bottomrule
    \end{tabular}
    \caption{QFT (1024 shots, 100 qubits) vs. FFT with equivalent problem size $N=2^{100}$}
    \label{table2}
\end{table}

At 100 qubits (see Table \ref{table2}), the cost of the full FFT is many orders of magnitude above the QFT. An important component of the baseline is the contribution of instrumentation. At $\approx 100$ qubits, this contribution is small compared to the cooling cost. However, because the number of DC and RF sources increases with the number of qubits, so does the power consumption associated with these instruments, $P_{\text{inst}}\propto q$. Furthermore, as the complexity of the algorithm is $O(q^2)$, the running time scales quadratically with the number of qubits. Hence, the instrumentation energy $E_{\text{inst}}=P_{\text{inst}}t$ would scale as $O(q^3)$. In contrast, the power needed for cryogenics was assumed to be constant, so the cooling energy scales only with the running time, as $O(q^2)$. Due to this difference, if we extrapolate the energy up to a large number of qubits, within the cooling power of commercial dilution refrigerators, the instrumentation energy would surpass the cooling energy in this regime.

The rest of the baseline includes the energy necessary to establish the static magnetic field, which is a fixed cost, independent of the number of qubits and insignificant compared to the other contributions, especially when we are considering a large number of qubits and repetitions. In contrast with classical computers, many components of the quantum computer are not yet fully optimized, so we can expect that the energy performance will be improved further as the platform develops.

\subsection{Energy and running time crossover points}

\begin{figure}[t]
    \centering
 \includegraphics[width=0.45\textwidth]{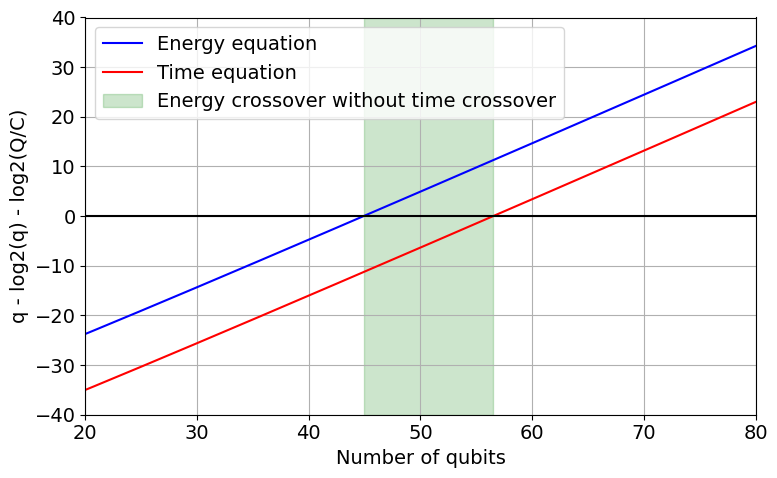}
    \caption{Lines representing the function $q-\log_2(q)-\log_2(Q/C)$. In the "Energy equation" line, in blue, $Q=1024\times 0.01=10\,\text{J/Gate}$ and $C=1.37\times10^{-11}\,\text{J/FLOP}$. In the "Time equation" line, in red, $Q=1024\times 10^{-6}=10^{-3}\,\text{s/Gate}$ and $C=5.74\times 10^{-19}\,\text{s/FLOP}$. The value of $q$ for which each line crosses 0 (45 for the blue line and 57 for the red line) represents the point for which the QFT is below the FFT benchmark, in terms of energy (blue line) and time (red line). The QFT is energetically below the reference line in energy before the same happens for running time, because the ratio in cost of 1 individual gate and 1 individual FLOP in terms of energy is much smaller than the same ratio in terms of running time, which guarantees that there is a window in which the energy of the QFT is below the energy of the FFT benchmark before the time crossing point has been reached. This window is represented in green.}
    \label{crossingpoints}
\end{figure}

Interestingly, while the crossover point in running time comes at around 60 qubits (Fig.~\ref{no_ec_time}), the crossover in total energy happens earlier, at fewer than 50 qubits (Fig.~\ref{no_ec_total}). This indicates that there is a regime where the  energy requirements of the quantum algorithm fall below those of the classical FFT benchmark, despite the QFT not yet outperforming the FFT in terms of running time. The reason behind this is that, if we compare the cost of 1 FLOP in the KAIROS supercomputer ($1.37\times10^{-11}\,\text{J/FLOP}$) with the cost of a single gate including baseline at $\approx10$ kW ($\frac{10000\,\text{J/s}}{10^6\,\text{Gates/s}}=10^{-2}\,\text{J/Gate}$), we see that 1 gate is approximately $10^9$ times more costly than 1 FLOP. If we do the same comparison for the time requirements, we see that 1 quantum gate ($10^{-6}\,\text{s}$) takes $10^{12}$ times more time than 1 FLOP ($\frac{1}{1742\,\text{PFLOP/s}}=5.74\times 10^{-19}\,\text{s}$ per FLOP in the El Capitan). 

As the scalings in time and energy are approximately the same, the relative positions of the two crossover points are determined by the gate-to-FLOP ratios. Since the ratio in time between 1 gate and 1 FLOP is much larger than the ratio in energy, there is an interval in number of qubits or input size where the QFT is energetically below the reference line associated with the full FFT despite not being favorable yet in terms of time. As long as this relation between the ratios remains true, a similar effect would be observed for any other application that exhibits a quantum-classical crossover. In this particular example, as the FFT benchmark scales exponentially with qubit count, an early energy crossover remains guaranteed, provided that the total workload of the QFT, including additional costs such as initial state preparation and variable number of shots, remains polynomially bounded. Such applications are precisely those in which the algorithmic advantages of quantum algorithms are expected. In these specific cases, this phenomenon illustrates that an energy gain relative to a classical reference can emerge even before any advantage in terms of running time has been achieved.

We can estimate analytically the interval between the energy and time crossover points by considering the complexity of the QFT in terms of the number of gates, $O(q^2)$ and the complexity of the FFT in terms of the number of floating point operations, $O(q2^q)$. The crossing point, for which the operations have the same cost, is approximately given by the solution of the equation $Qq^2=Cq2^q$, where $Q$ is a factor related to the QFT cost and $C$ is a factor related with the classical FFT cost. If we are comparing the energetic cost of the QFT for 1024 shots and the FFT, we have $Q=1024\text{ shots}\times 0.01\text{ J/gate}=10\,\text{J/Gate}$ and $C=1.37\times10^{-11}\,\text{J/FLOP}$. If we are comparing the running time, $Q=1024\text{ shots}\times 10^{-6}\text{ s/gate}=10^{-3}\,\text{s/Gate}$ and $C=5.74\times 10^{-19}\,\text{s/FLOP}$. The solution of the equation is then the value of $q$ for which 
\begin{equation}
    q-\log_2(q)-\log_2(Q/C)=0.
\end{equation}
This is shown in Fig.~\ref{crossingpoints}. This solution only exists because the quantum algorithm demonstrates an advantage relative to the classical benchmark as, otherwise, the lines in Fig.~\ref{crossingpoints} would not cross 0. The crossing point is estimated at 45 qubits in terms of energy, and 57 qubits in terms of running time, which, considering that this equation is a simplification, is consistent with the explicit calculations in Figs.~\ref{no_ec_time} and \ref{no_ec_total}. Although this gap appears to be numerically small, it represents a substantial reduction in the amount of exponentially scaling classical computation required before the quantum energetic resources fall below the reference line associated with the full FFT. If the crossing point in energy was only realized after the crossing point in time, at 57 qubits, that would imply that up to $\approx 1600$ quantum gates or $\approx 8\times10^{18}$ classical operations are needed before the quantum computer falls below the benchmark. By comparison, the fact that the crossover point in energy is achieved at 45 qubits means that only $\approx 2\times10^{15}$ classical operations or $\approx 990$ quantum gates are needed before the quantum computer is below the reference line, which represents a decrease by a factor of $10^3$ in classical resources.

\subsection{Importance of connectivity}
\label{sec:connectivity}

\begin{figure*}[ht]
  \centering
  \subfloat[]{
\includegraphics[width=0.32\textwidth]{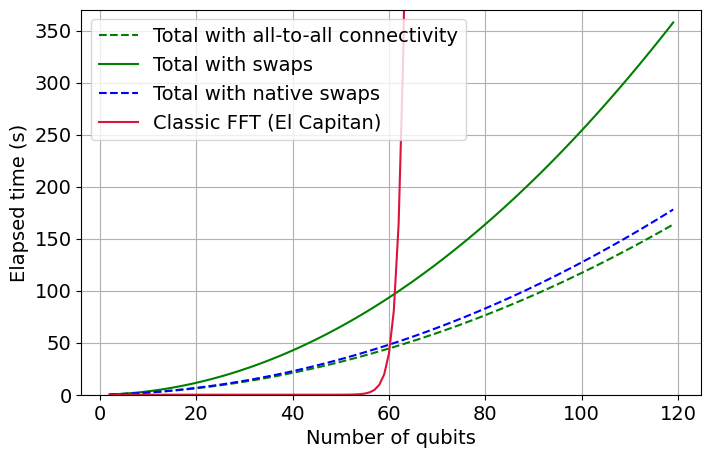}
    \label{swap_time}    
    }
  \hfill
    \subfloat[]{
\includegraphics[width=0.32\textwidth]{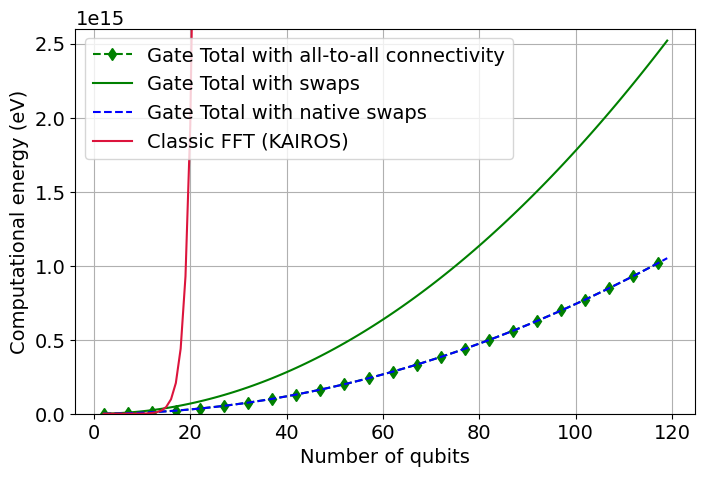}
    \label{swap_comp}
    }
   \hfill
    \subfloat[]{
\includegraphics[width=0.32\textwidth]{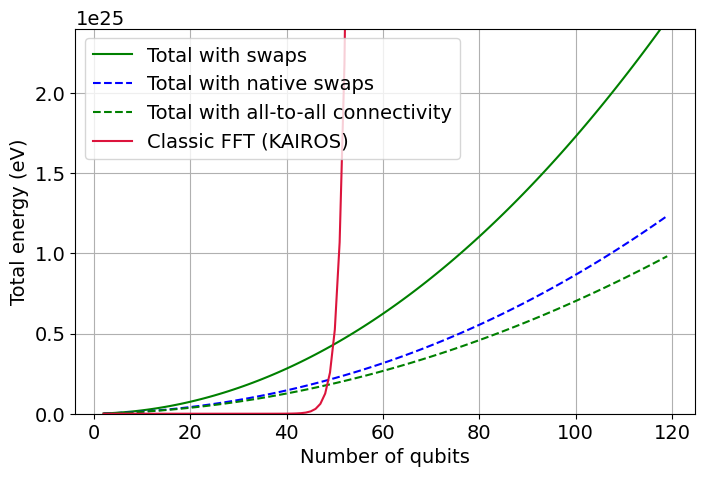}
    \label{swap_total}
    }
\caption{QFT with and without SWAP gates vs FFT: (a) running time: Native SWAP gates increase the running time by $10\%$, while non-native (decomposed) SWAP gates result in an increase by a factor of $\approx5/2$, compared to the case with 1D all-to-all connectivity. This is because the controlled phase gate involves 2 CNOT gates, while the SWAP gate involves 3 CNOT gates. 1 SWAP is added after each phase gate, increasing the number of CNOT gates per controlled phase gate from 2 to 5 (b) Computational energy: The cost with native SWAPS is approximately the same as without SWAP gates, while non-native (decomposed) SWAP gates result in an increase by a factor of $\approx5/2$; (c) total energy: The increase with Native SWAP gates and non-native SWAP gates comes from the increase in running time, because baseline is the most significant cost of the algorithm.
}
\end{figure*}

Previously, we assumed that qubits have all-to-all connectivity, so that they can interact even if they are not direct neighbors. However, this is not the case for the spin qubit architectures under consideration (Fig.~\ref{dot_diagram}) and we must establish connectivity between distant qubits with SWAP gates. The SWAP can be decomposed into 3 CNOT gates, following Appendix~\ref{sec:app_gates}, from which the cost of each SWAP can be estimated to be $204\times10^6\,\text{eV}$. For the QFT, the SWAP gates can be arranged such that one SWAP gate is applied after each controlled phase gate, ensuring that the qubits involved in every controlled phase gate are adjacent. Since the controlled phase gate involves 2 CNOT gates, while the SWAP gate involves 3 CNOT gates, we can estimate that the decomposed SWAP gates increases the cost by at most a factor of $\approx 5/2$, compared to an ideal case with all-to-all connectivity in which no SWAP gates are needed. This is confirmed in Fig.~\ref{swap_comp} which shows the computational energy of the QFT, for both 1D cases, with all-to-all connectivity versus requiring SWAP gates. 

However, the SWAP gate can be implemented through the exchange interaction \cite{Ni_2025}, reducing its energy to the order of the two-qubit gate energy. In this situation, adding the native SWAP gates would correspond to an increase of just 10\% in the total energy (Fig.~\ref{swap_total}), with the increase coming exclusively from the increase in time (Fig.~\ref{swap_time}), as the computational cost of the native SWAP ($3.125\,$MeV) is much smaller than the cost of the controlled phase gate ($145.625\,$MeV).

\subsection{Scaling with error suppression}

\begin{figure}[t]
  \centering
  \subfloat[]{
\includegraphics[width=0.45\textwidth]{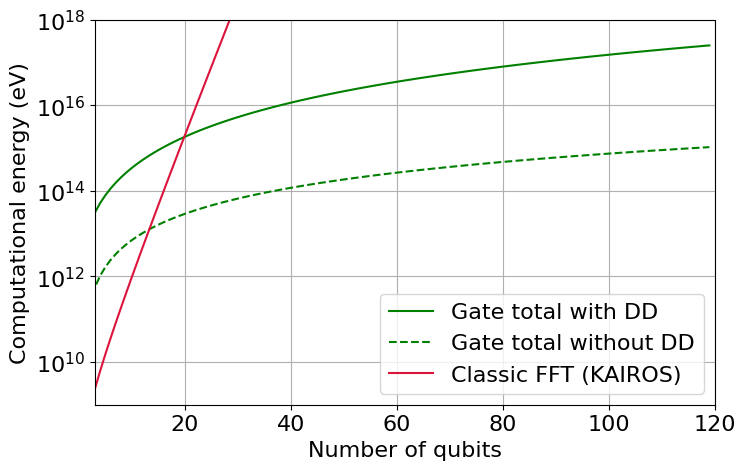}
    \label{dd_comp}    
    }
  \hfill
    \subfloat[]{
\includegraphics[width=0.45\textwidth]{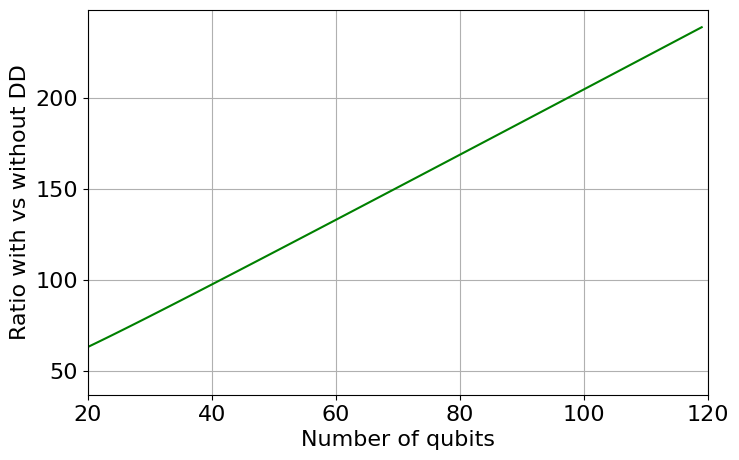}
    \label{dd_ratio}
    }
\caption{QFT (1024 shots) with vs without error suppression: a) computational energy with and without DD. DD increases the computational cost, but if the DD sequence is applied to idle qubits so that the running time is not affected, the total energy cost would be unaffected, as baseline scales only with time and is the most significant fraction of energy expenditure; b) ratio between computational energy with DD and without DD. This ratio increases linearly with the number of qubits.}  
\end{figure}

The biggest challenge in quantum computing is that each qubit and each operation is subject to a small error probability, and over the course of execution of an entire algorithm, it accumulates to reduce the accuracy of the final result. In spin qubits, charge and magnetic noise can lead to decoherence, causing such errors. However, error suppression using spin-echo techniques has been shown to extend coherence times by up to two orders of magnitude \cite{Bluhm2010}, at the added cost of more operations.

In this work, we consider the dynamical decoupling (DD) sequence XY-4: $\tau-X-\tau-Y-\tau-X-\tau-Y$, applied to idle qubits during QFT execution. This sequence can be executed for all the qubits in parallel, so its cost is reduced to that of 1 ESR field, which is used for $4\,\mu$s and thus costs 40 MeV, plus the cost of bringing the qubits into resonance for each rotation (3.125 MeV per rotation per qubit). This gives $E_{DD}=40+12.5q$ (MeV) per round of DD, where $q$ is the number of qubits. 

Without DD, the coherence time is $T_2^*\approx10\,\mu s$ \cite{Maurand2016} so, as a lower bound, we can assume that we apply 1 round of DD per physical gate, to keep the error accumulation from dephasing small. Because the QFT has a complexity of $O(q^2)$ in terms of number of gates, this means that $q\cdot O(q^2)=O(q^3)$ DD sequences in individual qubits are performed in total. To simplify the comparison, we assume all-to-all connectivity. The computational energy, accounting for the required gates of the algorithm and without considering running time, includes the contribution of these sequences and is plotted in Fig.~\ref{dd_comp}.

The large number of gates that are added to execute this sequence for every qubit increases the computational energy by a factor of up to 2 orders of magnitude, at 120 qubits. The increase in computational energy, shown in Fig.~\ref{dd_ratio}, is approximately linear with the number of qubits. This comes from the fact that, because the sequence is applied to the qubits in parallel, the energy per round of DD sequence, $E_{DD}=40+12.5q$, is also linear with the number of qubits. As such, each added qubit represents an additional amount of gates that must be executed during the sequence. This can also be understood from the number of operations involved in the algorithm, which scales with $O(q^2)$, while the number of individual DD sequences scales as $O(q^3)$. 

However, as long as the sequence is applied to idle qubits, the total running time of the algorithm is not largely affected. In that case, the total energy would be practically the same, because the baseline power is still the biggest fraction of the energy cost and it scales only with time. The change in total cost coming from the increase of the computational cost, which is shown in Fig.~\ref{dd_comp} is, in this case, negligible.

\section{Scaling with Error Correction}

Using dynamical decoupling, the dephasing time in Silicon 28 has been extended to few tens of ms \cite{Veldhorst2014}. This however, remains short compared to computation times, resulting in each qubit being subjected to small idling errors. Additionally, each gate operation suffers from control errors and noise, yielding a characteristic error rate of $p$ per operation. These two factors require the use of quantum error correction (QEC) codes to ensure that the accuracy of the algorithm stays above a minimum threshold \cite{Devitt2013}. Hence, in order for the results to be representative of the real energetic requirements of a quantum computing platform, the QEC costs must be considered \cite{stevens2026}. 

Early resource estimates for fault-tolerant quantum computing suggested that upwards of millions of qubits may be needed for practical applications, with a significant fraction of them being for error correction \cite{Preskill2018,Preskill2023}. Although recent work has focused on reducing these estimates \cite{cain2026}, a system with one million physical qubits is still recognized as a benchmark for large-scale systems, and will be considered as such in this work. The error correction code considered is the surface code \cite{Bombin2013,Fowler2012,Litinski2019} on a dense 2D architecture. We first analyze how the results are modified by the additional contributions that arise from the inclusion of error correction. We then compare it with the results obtained with no QEC and explore possible optimization routes that could reduce the final energy cost.

\subsection{Energetic considerations for the surface code}

Executing the QFT in a fault-tolerant manner requires the inclusion of several additional contributions to the total energy budget. We detail them individually in the following subsections. 

We begin with the QEC cycle, for which we adopt the ZXXZ surface-code cycle of Ref.~\cite{google2024}. We then consider the synthesis algorithm necessary to approximate gates outside the logical gate set, namely arbitrary $R_z(\theta)$ rotations. This approximation is done up to some error $\epsilon$, which in this work is taken as $\epsilon=10^{-8}$. Next, we describe the QPU architecture introduced previously, the fast data block, and the 15-to-1 magic-state distillation protocol \cite{Litinski2019}, and its importance to the energy budget. We then evaluate the resulting error budget, including the approximation error, distillation error and logical error, assuming a physical error probability $p=10^{-4}$, from which we determine the code distance. Finally, we account for the classical decoding cost, using the MWPM decoder model of Ref.~\cite{Camps2024}, before discussing the limitations of brute-force scaling and possible optimization routes.

\subsubsection{Cost of error correction cycles}

The QEC cycle is a sequence of physical operations performed within a logical qubit, at the end of which syndrome data is extracted. If the operations are done in parallel, we can estimate the energy and time of 1 QEC cycle for 1 logical qubit using the gate sequence shown in Fig. S12 of Ref.~\cite{google2024}, which displays a ZXXZ surface-code cycle composed of alternating layers of parallel H and CZ gates. The time taken by the cycle is the time of executing these layers and a series of measurements at the end. The decomposition of the entire QEC cycle is based on Ref.~\cite{gravier2025} and is presented in Appendix~\ref{sec:app_cycle}. 

Considering this decomposition, the cycle is composed of 9 layers. In 4 of these layers, $R_{x/y}$ are applied to the physical qubits, while in the remaining 4 layers, two-qubit $U(-\pi/2)\equiv U$ gates are applied between data and ancilla qubits. To mitigate phase noise accumulated during the two-qubit gate, a spin refocusing sequence can be incorporated into the implementation of the $U$ gate \cite{gravier2025}. Additional phase corrections can be characterized and handled in software during compile time and incorporated into subsequent single-qubit gates. The refocused operation is 
\begin{equation}
    U_{\text{refocused}}=(X_1\otimes X_2)\sqrt{U}(X_1\otimes X_2)\sqrt{U}
\end{equation}
in which the $\sqrt{U}$ gate can be implemented by halving the application time of the $U$ gate. This enables the cycle to not only be used to correct logical errors, but also to extend the dephasing time of the qubits, effectively serving the role of a DD sequence. 

\begin{figure}[h]
    \centering
    \includegraphics[width=0.8\linewidth]{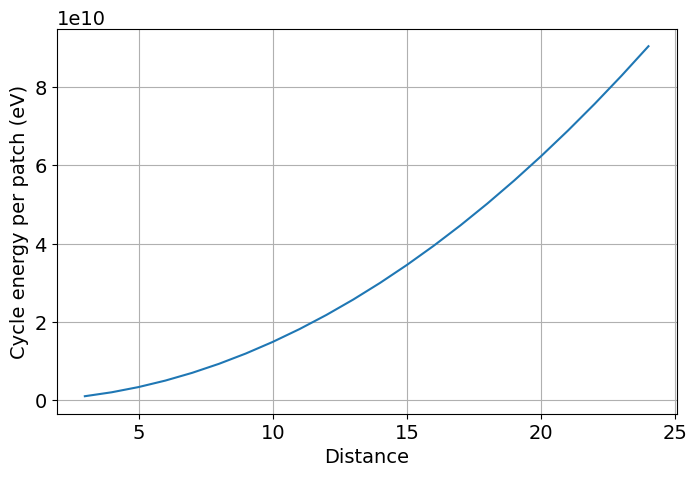}
    \caption{Energy of the QEC cycle per logical qubit as a function of distance, accounting for the operations in Table~\ref{table_cycle_3} in App.~\ref{sec:app_cycle} and the measurements of $d^2-1$ ancilla qubits at the end. The global ESR field contributes with an additional power of $1.6\,\mu\text{W}$, which is sustained during $10\,\mu$s per cycle. This contribution is not accounted for in this figure, because the ESR field is shared among all physical qubits and its energy cost therefore does not scale with the number of logical qubits, assuming that the cycles are executed in parallel.}
    \label{cycle_energy}
\end{figure}

The last layer is the measurement of all the ancilla qubits. The complete cycle takes $t_{\text{cycle}}=15\,\mu$s. The energy involved in the operations of the cycle per patch scales with the square of the code distance $d$ and is presented in Fig.~\ref{cycle_energy}. The layers of the cycle that require the application of parallel $R_{x/y}$ gates, as well as the refocusing sequence in $U_{\text{refocused}}$, result in an additional power contribution required to sustain the global ESR field during the cycle, given by $1.6\,\mu$W, which is sustained during $10\,\mu$s in total, per cycle. 

Each gate requires a fixed number of cycles \cite{Litinski2019}: $O(d)$ cycles for logical Z and X gates, $O(2d)$ for H and S gates, $O(5d)$ cycles for CNOT gates using lattice surgery and $O(d)$ cycles for patch deformation. As such, the cost of each gate can be approximated as the cost of the cycle multiplied by the number of cycles for that gate.

\subsubsection{Synthesis of rotation gates}

\begin{figure}[ht]
  \centering
  \subfloat[]{
\includegraphics[width=0.39\textwidth]{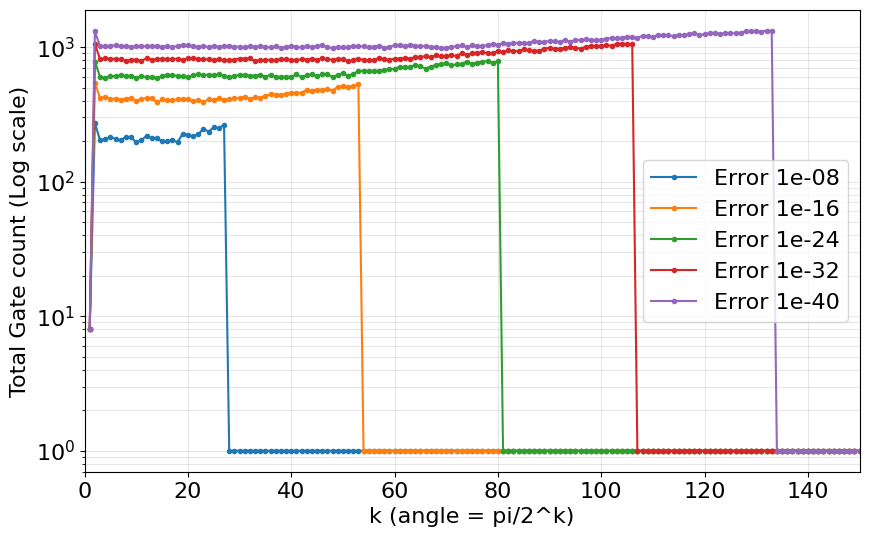}
    \label{approx_gates}    
    }
    \\
  \subfloat[]{
\includegraphics[width=0.39\textwidth]{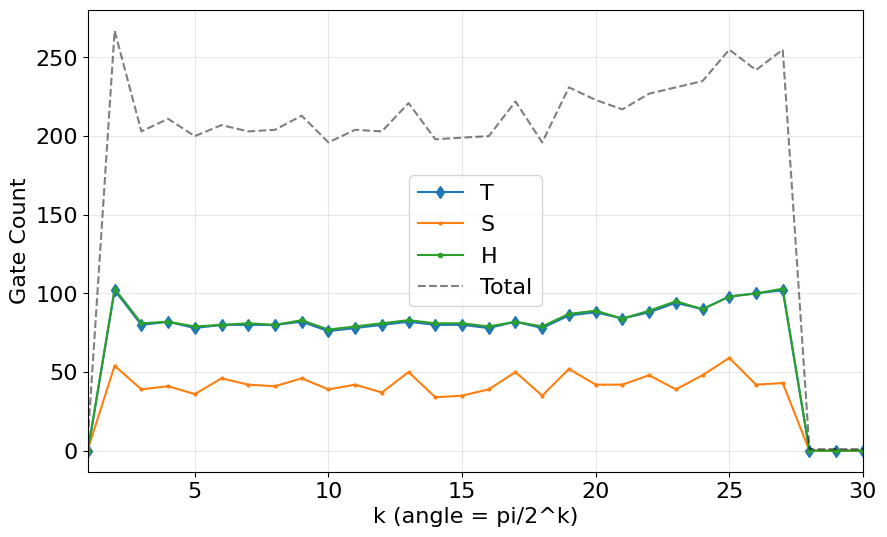}
    \label{approx_gates_1e8}    
    }
    \\
  \subfloat[]{
\includegraphics[width=0.39\textwidth]{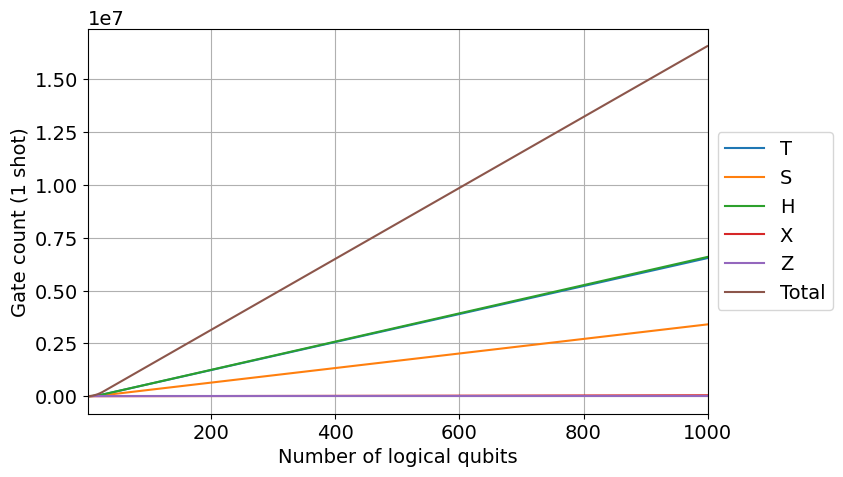}
    \label{total_gate_count}    
    }
\caption{a) Total gate count for different synthesis errors and values of $k$, corresponding to rotation angles $\theta=\frac{\pi}{2^k}$. 
The smaller the error $\epsilon$, the more gates are required to achieve the synthesis. 
When the rotation angle is smaller than the error, $\theta<\epsilon$, the best synthesis is simply the identity, 
requiring no operation on the qubit; 
b) Total gate count for different values of $k$ and $\epsilon=10^{-8}$ separated in the contributions of T, H and S gates. This error requires 200-250 gates, of which 40\% are H gates, 40\% are T gates and 20\% are S gates; c) Total gate count, as the number of qubits $q$ increases. The transition from quadratic to linear happens for 27 qubits. 
The T gate count scales as $T_{\text{count}}\approx6500\cdot q$, while the total gate count scales as $G\approx 16500\cdot q$.}
\end{figure}

The use of an error correction code restricts the possible operations to a small set, which in this case is assumed to be \{Z,X,H,S,CNOT\}. To achieve a universal gate set, the T gate is generated up to a small error with magic state distillation and injection \cite{Litinski2019}. Logical gates must then be done with some synthesis algorithm which decomposes the target gate into the available gate set. In the case of the QFT, the only gate that must be synthesized is the $R_z(\theta)$ gate, using the algorithm described in Refs.~\cite{ross2016,Selinger2016}. The synthesis is done up to some error $\epsilon$ (see Appendix~\ref{sec:app_synthesis}). 

Fig.~\ref{approx_gates} shows the total number of gates (vertical axis) required to synthesize a certain angle given by $\theta=\pi/2^k$ where $k$ will be interpreted as a parameter that determines the angle. The larger the value of $k$, the smaller the angle resolution. Furthermore, the smaller the synthesis error is fixed at, the more gates are required. In particular, the number of gates scales as $\approx O(\log(1/\epsilon))$. Furthermore, one can also note that the synthesis error introduces a cutoff for small angles, in the sense that, if the angle $\theta$ is smaller than $\epsilon$, the best synthesis for $R_z(\theta)$ is simply the identity. The angles in the QFT are of the form $\theta=\frac{\pi}{2^{\Delta+1}}$, considering the phase gate decomposition in Appendix~\ref{sec:app_gates}, so this cutoff happens when $\Delta$, the distance between the target and the control qubit, is $\Delta_{\text{max}}=\log_2(\pi/\epsilon)-1$.

\begin{figure}[th]
    \centering
\includegraphics[width=0.45\textwidth]{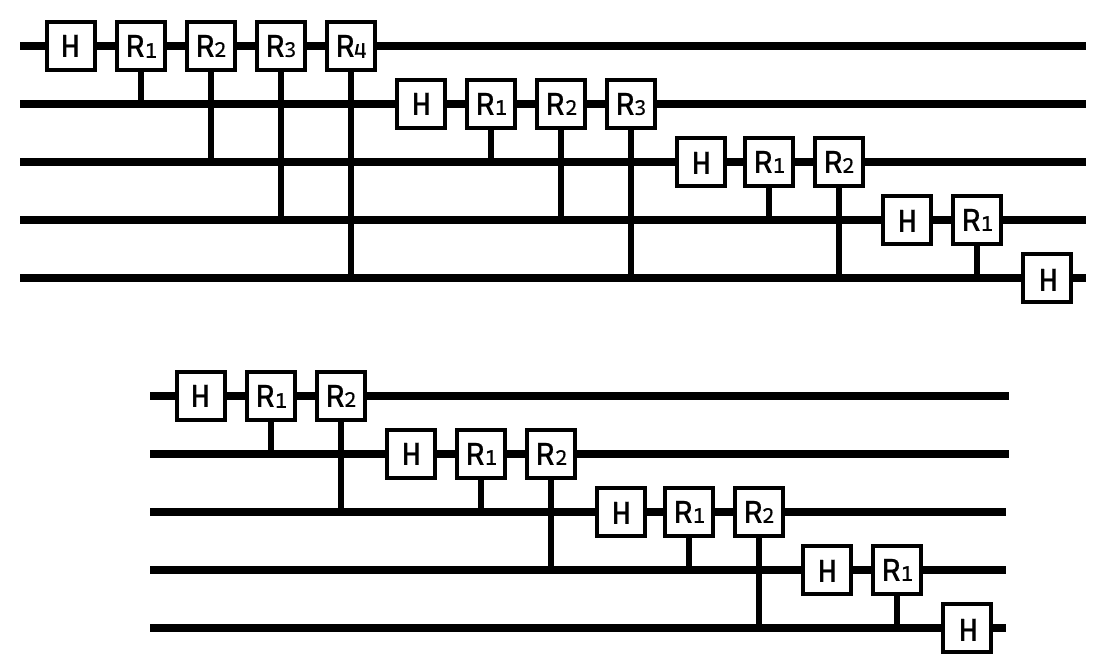}
\caption{Example of the effect of the synthesis algorithm on the QFT: The top panel corresponds to the exact QFT for 5 qubits. The bottom panel corresponds to the QFT after an approximation in which angles lower than $\frac{\pi}{8}$ are ignored. This approximation brings the complexity of the QFT from quadratic to linear in the number of qubits and is a consequence of the synthesis.}
\label{approx_qft}
\end{figure}

The resulting cutoff in the rotation angles has an important effect on the QFT, because this cutoff for small angles corresponds to a change of complexity, from quadratic to linear, once the number of qubits is larger than $\Delta_{\text{max}}$. This approximation is illustrated in Fig.~\ref{approx_qft} and can be reasonably applied even without error correction \cite{Nam2020}, as there is usually a limit to how small the rotation angle can be, depending on the precision of the rotation gates implemented in practice, but here it appears naturally as a consequence of the synthesis algorithm.

One could consider the value of $\epsilon$ to depend on the value of the synthesized angle $\theta$, e.g., requiring increased resolution for smaller angles. However, in this work, we assumed that each rotation was synthesized up to a constant error $\epsilon=10^{-8}$ with a sequence of 200-250 gates as in Fig.~\ref{approx_gates_1e8}, of which approximately 40\% are T gates. The quadratic-to-linear transition happens after 27 logical qubits. While the majority of the rotations will be approximated as the identity and therefore ignored, the operations actually performed will represent the biggest fraction of the total operational time, therefore impacting the energetic cost. After the quadratic-to-linear transition, the gate count is approximately $G\approx16500\cdot q$, as seen in Fig.~\ref{total_gate_count}.

\subsubsection{Effect of the QPU layout on instrumentation}

At the qubit level, each logical qubit is encoded in $d^2$ data qubits with an additional $d^2-1$ ancilla qubits used for syndrome extraction, for a total of $2d^2-1$ physical qubits per logical qubit. We denote as $q$ the number of logical qubits, such that the number of physical qubits is $q(2d^2-1)$.

The fact that each logical qubit requires a large number of physical qubits results in a significant instrumentation overhead that increases quadratically with distance. In contrast, because we assume that all the qubits can be controlled using one single cryostat, the cooling energy scales only with time. Because of this discrepancy, contrarily to what was verified in the case without QEC, the cost of instrumentation becomes much more relevant than the cooling cost.

At the QPU level, the architecture considered consists of two blocks: a data block, where the qubits that are used for computation are situated, and a distillation block, where magic states will be distilled. Both blocks are described in Ref.~\cite{Litinski2019}. Here, we will use the fast data block and the 15-to-1 protocol for distillation, with an example configuration displayed in Fig.~23 of Ref.~\cite{Litinski2019}. 

This choice determines how fast the magic states can be distilled and consumed, as well as the error associated with the distillation of the magic states, and how many qubits will be needed for distillation and storage, or used as logical data or ancilla qubits. Here, we denote by $n$ the total number of surface-code patches, including those used for logical data and ancilla qubits, distillation and storage.

The fast data block uses $n_{\text{data}}=2q+\sqrt{8q}+1$ patches, and is able to consume one magic state for every $d$ cycles. The 15-to-1 protocol uses 11 patches and produces a magic state with an error of $35p^3$, where $p=10^{-4}$ is the physical error probability. On average, one magic state is produced for every $11d$ cycles. Although we do not perform any T gate in parallel, to ensure that a magic state is always available and that the magic state production is not a bottleneck, we assume that we have access to 11 distillation blocks and additional storage patches, as in Ref.~\cite{Camps2024}, for a total number of patches:
\begin{equation}
    n=2q+\sqrt{8q}+133.
\end{equation}
Because this layout requires 133 additional patches, this imposes a minimum overhead of 133 logical qubits, even for small $q$, each of which also contributes to the overall instrumentation requirements.

\subsubsection{Errors and determination of distance}

Before considering decoding, three sources of error contribute to the algorithm-level error: the approximation error, associated with the approximation of gates; the distillation error, associated with the distillation of magic states; and the logical error, associated with the execution of gates in general.

First, we consider the approximation error. Because the synthesis error $\epsilon=10^{-8}$ corresponds to the error associated with one rotation, we can estimate the final error of the algorithm to be $\epsilon\cdot R_{\text{count}}$, in which $R_{\text{count}}\approx q^2$ is the number of rotations, including the synthesized and the ignored rotations. This gives $10^{-4}$ for 100 logical qubits or $10^{-2}$ for 1000 logical qubits.

For the distillation error, knowing that the distillation protocol produces a magic state with an error of $35p^3$, the final error of the algorithm due to faulty magic states can then be approximated as $P_T=T_{\text{count}}\cdot 35p^3$, in which $T_{\text{count}}$ is the total number of $T$ gates. In our case, using $\epsilon=10^{-8}$, the number of T gates grows approximately linearly with the number of qubits: $T_{\text{count}}\approx 6500\cdot q$, as seen in Fig.~\ref{total_gate_count}. Taking $p= 10^{-4}$, we can estimate that the error of a single magic state is $35p^3=3.5\cdot10^{-11}$. Therefore, at 1000 logical qubits, the probability of error of the algorithm due to faulty T gates is approximately $35p^3\cdot 6500q\approx 0.022\%$. This shows that the 15-to-1 protocol is appropriate to our study.

Finally, the algorithm might also fail due to faulty gates in general. The logical error probability per cycle is approximately $p_L=0.1\cdot(\frac{p}{p_{\text{threshold}}})^{\frac{d+1}{2}}$ \cite{Litinski2019}, where $p_{\text{threshold}}\approx 0.01$ \cite{Fowler2012} is the error threshold for which increasing the distance results in an overall improvement of the accuracy. In our case, since the gates are applied sequentially and the number of gates in the algorithm grows approximately linearly with the number of qubits, we can take $G=16500\cdot q$, as seen in Fig.~\ref{total_gate_count}. The logical algorithm-level error can then be upper-bounded by the total number of gates $P_L\approx n\cdot G\cdot d\cdot p_L$ \cite{Litinski2019}.

Since this is the only error contribution that depends on the code distance, the code distance can be determined from the target logical error probability. One can either choose the distance such that the algorithm-level error due to logical error is on the same order as the approximation and distillation errors, or accept a lower distance, which would require less resources, but with the logical errors becoming the limiting factor of the final accuracy.

The choice of decoder also affects the resources required, including the distance and number of qubits. In particular, the decoder introduces an additional error contribution coming from the decoding interruption time, which may become the limiting factor, in which case a smaller distance could achieve the same algorithm-level accuracy. In other words, even though we can choose a distance such that the logical error is below the target error, the decoding error may be the limiting factor to the algorithm accuracy. However, in this work, we determine the distance in the first place, independently of the decoding requirements, which are evaluated only after the distance has been fixed.

Table \ref{error_probs} in Appendix~\ref{sec:app_errors} shows the algorithm error due to faulty T gates ($P_T$) and the error due to logical errors ($P_L$) for different number of logical qubits $q$, values of $p$ and values of distance $d$, corresponding to a fixed number of 1 million physical qubits and assuming $\epsilon=10^{-8}$.

\subsubsection{Decoding energy}

We have to consider the cost of decoding so that we can apply the necessary corrections. For reference, at each QEC cycle, $n\cdot (d^2-1)$ syndrome bits are generated. At $10^6$ physical qubits (corresponding to, for example, $d=25$ and $n=800$) this corresponds to $\frac{n\cdot (d^2-1)}{t_{\text{cycle}}}\approx33$ Gbits/s. Thus, interpreting this data and identifying the errors requires a powerful classical decoder, whose energy expenditure cannot be ignored.

To estimate the decoding power necessary for decoding, we adopt the MWPM decoder model of Ref.~\cite{Camps2024}. The syndrome data is divided in time windows of size $3d$, which require on average $34d^{4.3}$ FLOPs for each patch to decode. The time to decode a window depends on the CPU that is used. Here, we assume an AMD EPYC 7763 processor with a peak performance of $3.584$ TFLOPs/s distributed over 64 cores and a physical error $p=10^{-4}$. The decoder is allowed to run for at most $M$ QEC cycles before being interrupted. This interruption time influences the probability of failure of the decoder and translates into an algorithm-level error probability. In this case, we choose $M$ such that this error is limited to a target value of 0.01, if $P_L<0.01$, or limited to $P_L$, if $P_L>0.01$. The corresponding value of M is then used to determine the computational power required for decoding. The complete derivation is provided in Appendix~\ref{sec:app_decoding}.

\begin{figure}[t]
  \centering
  \subfloat[]{
\includegraphics[width=0.4\textwidth]{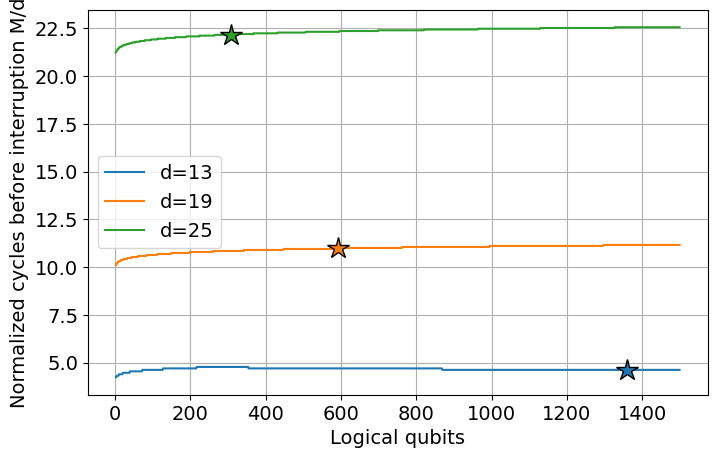}
    \label{decoding_M}    
    }
    \hfill
    \subfloat[]{
\includegraphics[width=0.45\textwidth]{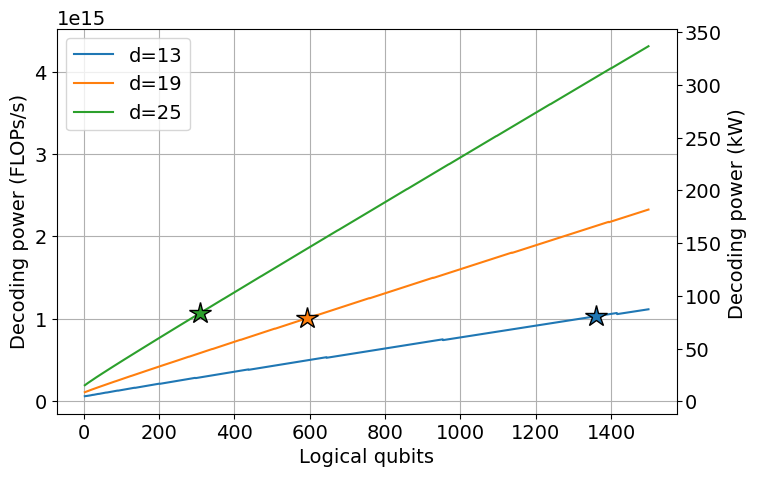}
    \label{decoding_power}
    }
\caption{Decoding parameters as a function of the number of logical qubits and distance $d$, assuming that an AMD EPYC 7763 processor is used, $p=10^{-4}$, $t_{\text{cycle}}=15\,\mu$s and target decoding error $P_L^{\text{dec}}=0.01$. The points marked by a star represent the configuration (in terms of distance and number of logical qubits) corresponding to 1 million physical qubits: a) Interruption time $M$ before the decoding fails divided by the distance; b) Decoding power in FLOPs/s and in W.}
\end{figure}

The resulting $M$ divided by the distance and decoding power in FLOPs/s are presented in Figs.~\ref{decoding_M} and \ref{decoding_power} respectively. Then, by dividing the result by the FLOPs/s of the CPU and multiplying by its power (280 W for the AMD EPYC 7763), we can determine how many CPUs are needed and the resulting power in W, which is displayed in Fig.~\ref{decoding_power}. We see that, for a configuration with 1 million physical qubits, the required computational power is of the order of $100$ kW and that this value increases with distance and number of logical qubits. This estimate depends on the choice of decoder and decoding parameters, such as target error, and would be different if, for example, neural-network-based decoders are used \cite{caldwell2025}.

From this, we confirm that the energy of decoding becomes a relevant fraction of the total cost, similar to the baseline energy.

\subsubsection{Limitations of brute-force scaling}

In practice, the number of qubits that can be integrated in a single cryostat can be limited by several factors, including maximum cooling power of the cryostat, the dense cabling requirements etc. The wiring to each physical qubit brings-in an additional amount of heat that must be extracted \cite{pauka2019}. Assuming that the system is operated at $\approx1$ K, we can consider the maximum cooling capacity to be of approximately $700$ mW, consistent with commercially available systems such as the Bluefors XLDHehp \cite{bluefors_xldhehp}. As the dissipation comes primarily from the heat transfer through the cables, the thermal budget can be approximated by considering the static load of each cable that enters the system. The dynamical contribution would be negligible and therefore not considered \cite{Raicu2025}. Each qubit requires a minimal configuration of 3 DC and 3 RF cables, with an estimated static load of $200$ nW for 1 DC cable and $15$ nW for 1 RF cable at the mixing chamber \cite{Raicu2025}. Then, at most 20000-30000 physical qubits could be accommodated in the cryostat. This value depends on the characteristics of the cables, cryostat, and architecture of the system, so it should be interpreted as an order of magnitude upper bound of $\approx10^4-10^5$ rather than a strict limit. A QFT algorithm on $q=20$ logical qubits, requiring a total of $n=186$ patches with distance $d=9$, would already require almost 30000 qubits, which is at the estimated limit. Such a brute-forced system would also have to physically accommodate all these cabling inside the cryostat. More importantly, at these large qubit numbers, the power requirement of the instrumentation needed to control all these qubits would be significant, approaching $\approx15$ MW for a million physical qubits. This provides a strong impetus to optimize the system along these design axes. Operating the system at higher temperatures of $\approx 1$ K may have an impact on qubit coherence times and gate fidelities, and is the subject of ongoing research \cite{steinacker20264}. However, doing so provides opportunities to scale efficiently with respect to power and at the same time maintain the required fidelities at an added cost. For these reasons, we will make the use of cryoelectronics and multiplexing techniques to scale to larger systems.

With these considerations in mind we first present the results corresponding to no more than 100 logical qubits and distance $d=9$, under the naive assumption that the system can be operated within a single cryostat, noting that this assumption becomes more reasonable once multiplexing is included. Then, in Sec.~\ref{sec:multiplexing}, we extend the analysis by including cryoelectronics and multiplexing, which enable the consideration of much larger fault-tolerant systems.

\subsection{Total scaling with and without QEC}

\begin{figure}[t]
  \centering
  \subfloat[]{
\includegraphics[width=0.45\textwidth]{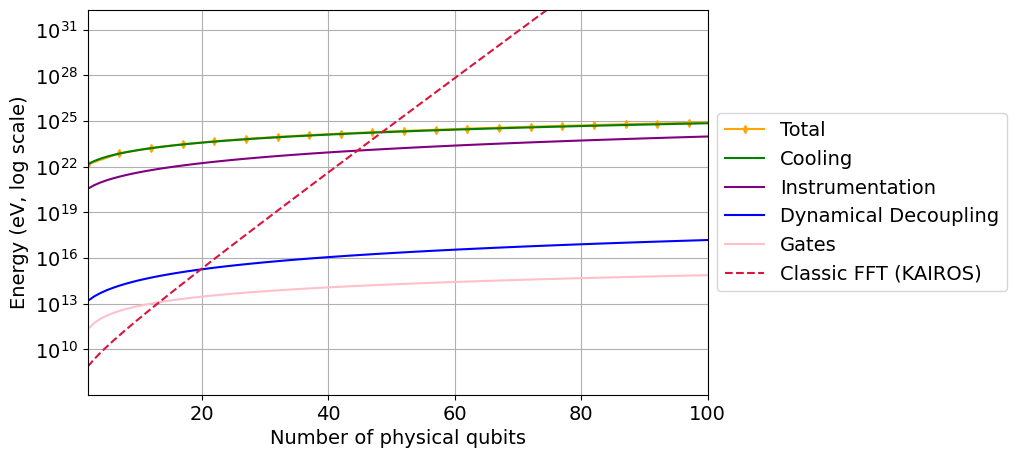}
    \label{ec_total_no_ec}    
    }
    \hfill
    \subfloat[]{
\includegraphics[width=0.45\textwidth]{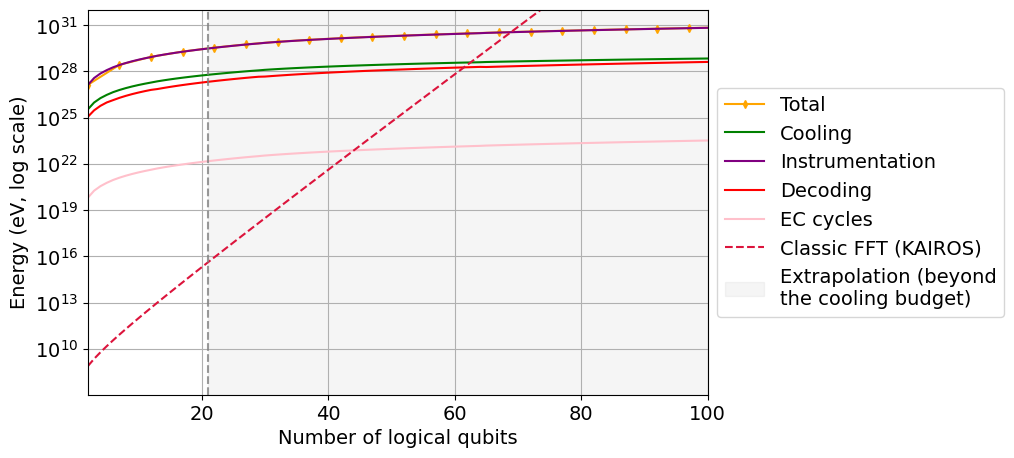}
    \label{ec_total_d9}
    }
\caption{QFT (1024 shots) without vs with error correction at $d=9$: a) total energy without QEC separated in different components; b) total energy with QEC separated in different contributions, and in log scale. The baseline is still the most significant fraction of energy cost. Accounting for the increase in time by a factor of $\approx 10000$ (at 100 logical qubits) and the added instrumentation, which represents the dominant contribution, the baseline energy is increased by a factor of around $8.5\times10^5$ when QEC is included. With error correction, there are two additional components that must be considered. The first is the cost of QEC cycles, which represents approximately $4.4\times10^8$ times the cost of the computational cost without QEC, at 100 logical qubits. The second is the decoding cost, which is not as expensive as baseline, but more expensive than the energy necessary for QEC cycles, and larger than the total cost without QEC. Baseline energy is the single dominant factor of the energy requirement. The shaded region in gray represents an extrapolation beyond the estimates of the cooling budget.}  
\end{figure}

\begin{figure}[h]
  \centering
\includegraphics[width=0.45\textwidth]{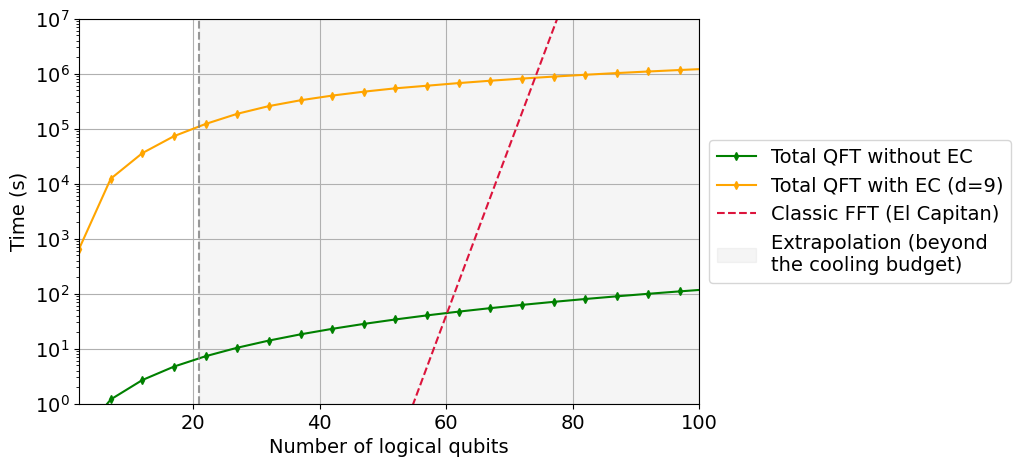}
\caption{Running time (1024 shots) with vs without error correction at $d=9$: error correction leads to an increase in running time by 4 orders of magnitude. Moreover, even with error correction, the crossover point in running time appears after the crossover point in energy, but only beyond the estimated limit of qubits within a single cryostat (shaded region starting at $\approx 20$ qubits).}
\label{ec_time_d9}
\end{figure}

Fig.~\ref{ec_total_no_ec} and \ref{ec_total_d9} show, respectively the scaling of energy without error correction and the same scaling when error correction is included for $d=9$. As predicted, QEC comes with a significant increase in energetic cost, with the dominant contribution coming mainly from the instrumentation, which is considerably larger than in the case without QEC. This is further amplified by the increase in running time by 4 orders of magnitude, as seen in Fig.~\ref{ec_time_d9}. The QEC cycles increase the computational energy by roughly 8 orders (a factor of $4.6\cdot10^8$ at 20 logical qubits and $4.4\cdot 10^8$ at 100 qubits). We can also see that the decoding energy is significantly larger than both the QEC cycles cost and the cost of DD and close to the baseline energy. In total, when comparing the algorithm without QEC and the version with QEC for this distance, we get an increase by a factor of $E_{\text{with QEC}}/ E_{\text{without QEC}}\approx 7.2\times10^5$ at 20 qubits or $8.6\cdot 10^5$ at 100 qubits. The increase from 20 qubits to 100 qubits is smaller than initially expected. The reason for this is the fact that, even at a small number of qubits, we are assuming an overhead of 133 patches for distillation. Thus, the increase from 20 to 100 qubits should be interpreted as an increase from 186 to 361 total patches. 

Furthermore, Fig.~\ref{ec_time_d9} shows that the crossing point in time (at 75 qubits) still happens after the crossing point in energy (at 69 qubits), but the gap between the two crossing points is smaller, when compared to the case without QEC, as the energy is much larger than the case without QEC. Furthermore, this crossing point is only reached for a system much larger than what could be reasonably operated inside one single cryostat without the use of cryoelectronics. Thus, a concrete analysis of both crossing points is only possible once further optimization is considered, by the use of cryogenic demultiplexing. In that case, the discrepancy becomes more obvious once again, as an optimization in energy does not represent an optimization in terms of time.

\begin{figure}[h]
  \centering
  \subfloat[]{
\includegraphics[width=0.45\textwidth]{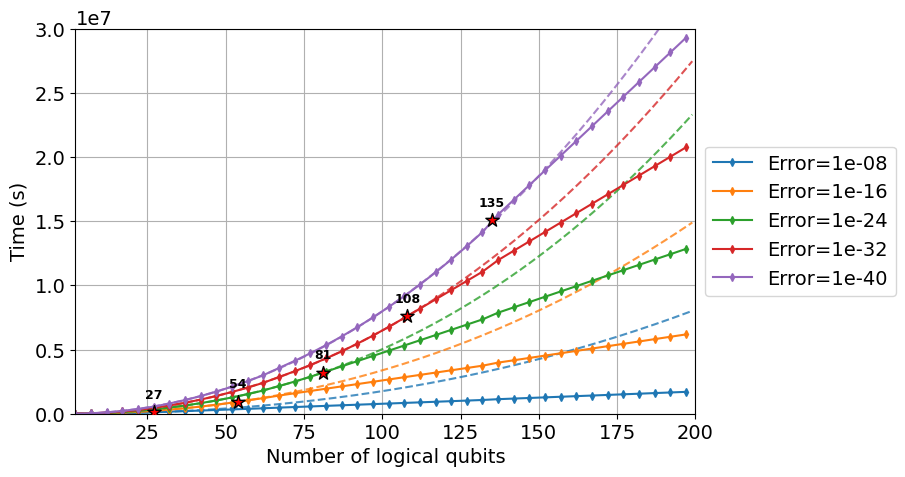}
    \label{ec_total_d9_error_time}    
    }
    \hfill
    \subfloat[]{
\includegraphics[width=0.45\textwidth]{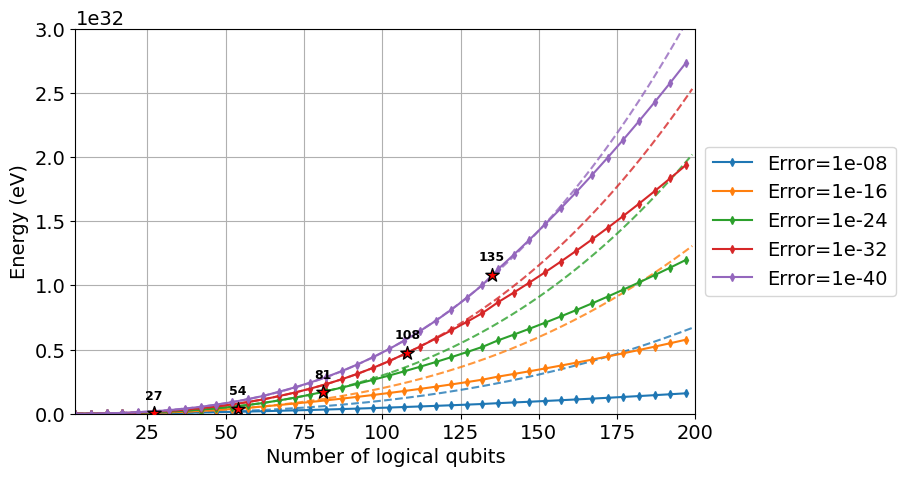}
    \label{ec_total_d9_error_energy}
    }
\caption{QFT (1024 shots) with error correction at $d=9$: Total time (a) and energy (b), respectively, for different values of the error $\epsilon$: As $\epsilon$ decreases, more gates are required to achieve the synthesis, which increases the total energy. For each value of $\epsilon$, there is a cutoff for small angles that are not synthesized, which results in a quadratic-to-linear transition in terms of the complexity of the algorithm past a certain number of qubits (marked by a star for each $\epsilon$). The dashed line represents an estimate of the total running time or energy, if the cutoff did not happen. In terms of energy, this translates into a cubic-to-quadratic transition in the total energy scaling.}  
\label{ec_total_d9_error}
\end{figure}

Finally, in the synthesis algorithm, we used $\epsilon=10^{-8}$, which results in a quadratic-to-linear transition in complexity after 27 qubits. We can use a smaller approximation error $\epsilon$ e.g. $10^{-16}$, to further improve the algorithm accuracy. Figs.~\ref{ec_total_d9_error_time} and \ref{ec_total_d9_error_energy} shows what the scaling in time and energy would look like for different values of $\epsilon$, identified by color. As expected, a smaller error would require more gates, increase the overall cost and increase the qubit count for which the transition happens. One must also consider that the increased number of $T$ gates would likely require a better distillation algorithm which requires more resources. 

The stars mark the number of qubits for which the complexity of the algorithm transitions from quadratic to linear. These are, essentially, the results that would be obtained if the cutoff was done simply as an algorithmic approximation \cite{Nam2020}, but here it appears naturally as a consequence of the synthesis. This shows the system level impact of synthesis of rotation gates and has a considerable effect on energetics. As the complexity in time transitions from quadratic to linear, the dominant term of the energy scaling, which is the instrumentation, goes from cubic to quadratic.

The dashed lines represent an extrapolation of the scaling if the transition did not happen or, equivalently, if the synthesis did not result in the cutoff for small angles and, in particular, we can see that the dashed line and the solid line drift further from each other as the number of qubits increases. In the case of this work, where we took $\epsilon=10^{-8}$, at 100 logical qubits the approximation results in a reduction in energy of 50\% compared to the dashed line.

\subsection{Energy optimization with gate frequency}
\begin{figure}[h]
  \centering
  \subfloat[]{
\includegraphics[width=0.45\textwidth]{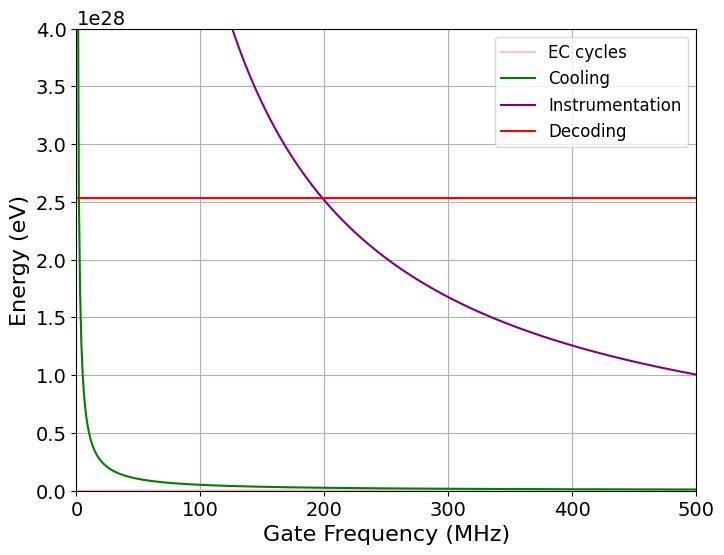}
    \label{freq1}    
    }
    \hfill
    \subfloat[]{
\includegraphics[width=0.45\textwidth]{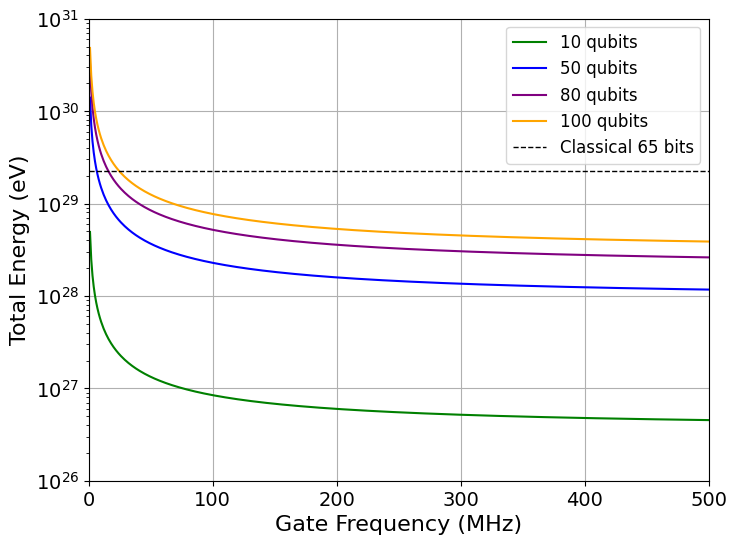}
    \label{freq2}
    }
\caption{QFT energy with gate frequency (1024 shots) at $d=9$: (a) Energy for 100 logical qubits discriminated by different contributions. As the frequency increases, the gates become more expensive, which affects the QEC cycles directly. On the other hand, the time decreases, which affects the total cost of baseline components; Decoding depends on the cost of operations and on the running time. Assuming that the decoding interruption time in cycles is the same, it remains constant with frequency. (b) Total energy for varying number of qubits: by balancing the reduction in cost of baseline with the increase in cost of the QEC cycles, we can reduce the total cost of the algorithm by 2 orders of magnitude by increasing the gate frequency.}  
\end{figure}

Presently, there are engineering challenges that must be solved before semiconductor systems can be scaled \cite{Gonzalez_Zalba2021,Vinet_2018} and controlled at higher frequencies. However, we could predict that, considering that baseline requires a great fraction of energy, decreasing the running time would make a noticeable difference. The most direct way to do this is to increase the gate frequency.

However, the most expensive gates, the $R_x$ gates, would become more expensive by the same factor, as the cost of ESR is proportional to the Rabi frequency. At first, this seems to be irrelevant because the control cost is a small fraction of the total cost. But if we include error correction, the increase in computational cost coming from QEC cycles could become relevant and, at higher Rabi frequencies, they could even become comparable to the baseline energy. Although, this also depends on how the remaining components related to QEC scale with frequency, namely decoding and number of correction cycles per gate.

As a first approximation, if the cycles are being completed faster, we need a proportionally more powerful decoder. However, the algorithm time decreases by the same factor, which means that, under the assumption that we have access to an arbitrarily fast decoder, and assuming that the number of decoding cycles before interruption is not largely affected when the frequency is increased, the total decoding energy is approximately the same, and it is the limiting factor in terms of energy.

The main changes come, firstly, from the QEC cycle whose cost increases with the frequency because the gates involved become more expensive by the same factor. And secondly, from the baseline energy, which decreases with frequency because the instrumentation is utilized for less time. Under these assumptions, one could imagine that there is a point at which the cost of the different contributions becomes balanced, as seen in Fig.~\ref{freq1}, thus minimizing the total cost. Fig.~\ref{freq2} shows the total energy, with varying frequency for different numbers of qubits. By increasing the gate frequency up to a certain point, the cost can be reduced by a factor of 2 orders of magnitude.

\section{Practical route to energy efficient scaling using cryoelectronics}
\label{sec:multiplexing}

In the 2D architecture discussed above, each gate electrode requires one RF output. However, this requirement can be relaxed by introducing cryoelectronics, specifically control demultiplexers, which allow the gate electrodes to be controlled using fewer RF cables. This enables the integration of more logical qubits with increased distance without the need of increasing the cooling budget of the cryostat \cite{pauka2019}. This significantly strengthens the assumption that the system can be operated with a single cryostat and allows the crossover between the quantum algorithm and the classical FFT benchmark to be reached in practical scenarios. From this perspective, demultiplexing is not only an optimization route in terms of energy, but a necessary step towards scalability.

In a similar manner, we can use readout multiplexers to reduce the number of readout cables and associated instrumentation. These cryoelectronic circuitries can possibly be monolithically co-integrated directly onto the qubit die, leveraging QSOI technology \cite{Meunier2026}.


\subsection{Control demultiplexing}

On the control side, the control demultiplexers take an input from the room temperature instrumentation and fan out to multiple qubit electrodes. The ratio between the number of inputs to the outputs of the demultiplexer is known as the demultiplexing ratio $m$. Because each demultiplexer is connected to one RF instrumentation output, multiple qubit-gate electrodes can therefore share the same RF output. Thus, we can also refer to the demultiplexing ratio $m$ as the ratio between the number of RF outputs required with control demultiplexing and the number of RF outputs required without demultiplexing, where the latter assumes that there is one dedicated RF output for every gate electrode, allowing all qubits to be controlled at the same time. Because the power consumption of the demultiplexer itself is very low, the demultiplexing ratio is also equivalent to the ratio between the instrumentation power with and without demultiplexing. 

\subsubsection{QEC cycle time overhead}

\begin{figure}[h]
  \centering
\includegraphics[width=0.45\textwidth]{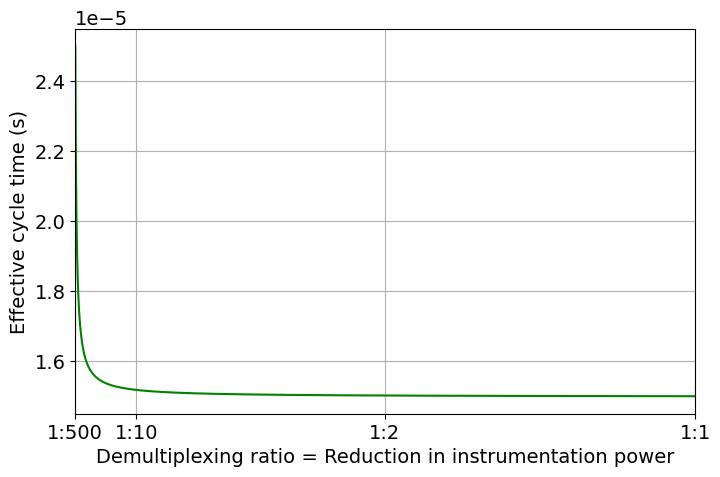}   
\caption{Effective QEC cycle time as a function of the demultiplexing ratio $m$, in the worst case scenario for which $1/m < d^2$: As demultiplexing is increased, less instrumentation is needed, but less qubits can be controlled simultaneously, so the cycle is extended in time. However, for the same demultiplexing ratio, the increase in cycle time is much smaller than the decrease in instrumentation power. When the demultiplexing ratio is 1, this is equivalent to no demultiplexing being used, so the effective cycle time tends to $t_{\text{cycle}}=15\,\mu$s.}
\label{mltplx_ratio} 
\end{figure}

A cryogenic time-division demultiplexer running at a frequency $F$ can divide a time interval of $1/f$ in $F/f$ time slots, where $f$ is the gate frequency. The shortest gate we need to perform for the QEC cycle is that of the $\pi/2$ rotations, which takes 500 ns, resulting in $f=2$ MHz. In this work, we consider a demultiplexer consisting of a counter, which allows a voltage pulse to be routed to a different output gate electrode, cyclically, during each time slot \cite{Jadot2023}. Each output includes a holding capacitor that maintains this electrode voltage until the output is addressed again  \cite{pauka2019}. This allows us to apply a voltage on one gate electrode during one time slot, and maintain the voltage while the demultiplexer cycles through the other gate electrodes, for rest of the time slots. The multiplexer then returns to the same gate electrode at the end of the pulse duration, to turn off the voltage. Thus, for a pulse duration of $1/f$, one demultiplexer can execute $F/f$ pulses before having to return back to the same output. The first pulse starts at the first time slot, while the pulse sent at the last utilized window only starts at the $1/m$-th time slot, exactly $\frac{1}{F}(\frac{1}{m}-1)$ seconds after the first pulse. Thus, instead of applying a layer of $1/m$ operations in parallel, they need to be applied sequentially, extending the time per layer by an additional $\frac{1}{F}(\frac{1}{m}-1)$ seconds. In our case, the cycle for 1 logical qubit involves a total of 20 pulses, so the time of the cycle becomes
\begin{equation}
    t_{\text{cycle}}^{\text{demux}}=t_{\text{cycle}}+20\frac{1}{F}(\frac{1}{m}-1)
\end{equation}
We consider a control demultiplexer clocked at $F=1$ GHz, giving $F/f=500$ time slots for each demultiplexer cycle. Although higher demultiplexer frequencies $F$ can give more time slots, resulting in an increased demultiplexing ratio $m$, it also gives rise to a larger dissipated energy, adding to the thermal budget of the system \cite{pauka2019}. The effective cycle time when demultiplexing is used is illustrated in Fig.~\ref{mltplx_ratio}. At $t_{\text{cycle}}=15\,\mu$s, which corresponds to all the windows of the demultiplexer being utilized, this gives $t_{\text{cycle}}^{\text{demux}}=25\,\mu$s, which is an extension in time by a factor of 5/3, but the instrumentation power is reduced by a factor of 500, equal to the demultiplexing ratio.

\subsection{Cryoelectronic readout multiplexing}

Similar considerations apply to the readout. Unlike the control multiplexer, a time-division multiplexer has to stay connected to the same readout channel during the entire readout. The output of this multiplexer is then passed to a frequency multiplexer, which is capable of reading out many such signals in parallel, as discussed previously. During the readout, the readout signal is continuously integrated for $1\,\mu$s to ensure a good signal-to-noise ratio. Thus, it would be impractical to sequentially readout each qubit within a patch using the time-division multiplexer. The strongest requirements arise at the end of the algorithm, during which both the data and ancilla qubits of the logical qubit need to be measured in parallel, for a total of $n(d^2-1)+qd^2$ readout lines. However, because the readout is executed only once at the end of each cycle, these multiplexers can be reused for $t^{\text{demux}}_{\text{cycle}}/1\,\mu$s QEC patches. In other words, the number of QEC patches that can be readout using a single multiplexer depends on the effective cycle time. Thus for a cycle time $t_{\text{cycle}}^{\text{demux}}=25\,\mu$s, a readout multiplexer can sequentially readout 25 different qubits from different patches. The total number of multiplexers for readout is thus 
\begin{equation}
N_{RO}=(n(d^2-1)+qd^2)\frac{1\,\mu\text{s}}{t^{\text{demux}}_{\text{cycle}}} 
\label{N_RO}
\end{equation}
We also note that, as this number depends on the effective cycle time, it depends on the demultiplexing ratio as well. Furthermore, since the multiplexers are reused sequentially, the QEC cycles of groups of patches are offset in time by $1\,\mu$s rather than synchronized. Synchronization between two patches are only required during certain operations such as lattice surgery.

\subsection{Optimizing energy with demultiplexing}

\begin{figure}[h]
  \centering
  \subfloat[]{
\includegraphics[width=0.45\textwidth]{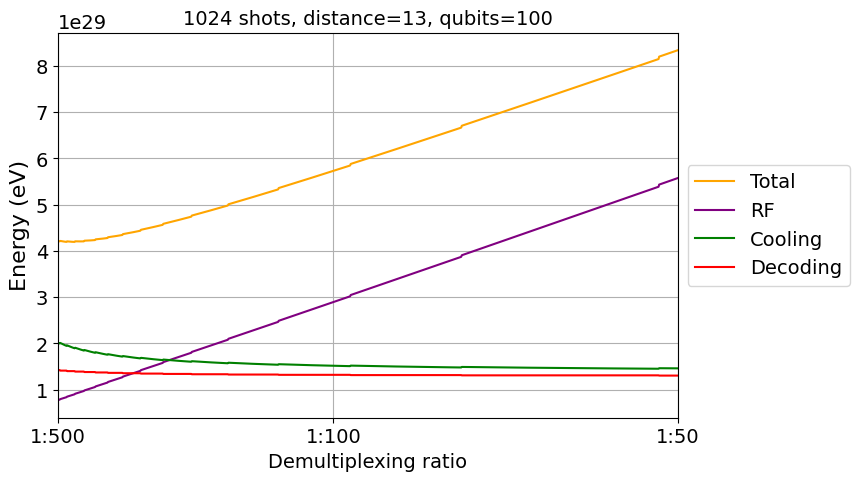}
    \label{mltplx_d13q100}    
    }
    \hfill
    \subfloat[]{
\includegraphics[width=0.45\textwidth]{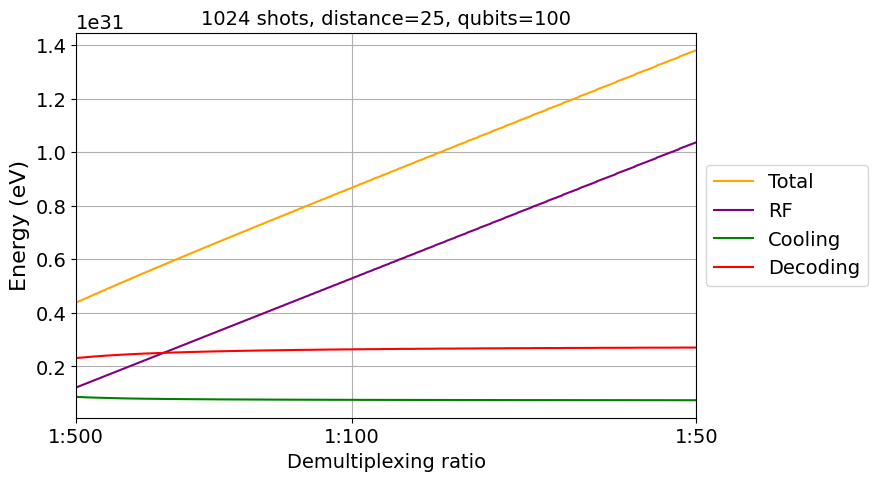}
    \label{mltplx_d25q100} 
    }
\caption{Comparison between decoding, cooling, instrumentation (RF) and the total energy as a function of the demultiplexing ratio for 100 logical qubits, 1024 shots: a) $d=13$ and b) $d=25$. For larger distances, the instrumentation energy (RF), which is the main component that is minimized with demultiplexing, is much higher than the remaining contributions. Thus, increasing the demultiplexing ratio from 1:1 to 1:500 is more impactful. This directly affects the ideal demultiplexing ratio.}  
\end{figure}

When demultiplexing is utilized, the three components affected are the instrumentation energy, the cooling energy and the decoding energy, as seen in Figs.~\ref{mltplx_d13q100} and \ref{mltplx_d25q100}, which are precisely the three dominant contributors to the energetic cost. In particular, while the instrumentation power is reduced by a factor equal to the demultiplexing ratio, the cooling energy, which is associated with a constant power, increases due to the longer effective cycle time. However, for a given demultiplexing ratio, the reduction in instrumentation power is considerably larger than the extension of the cycle time. Furthermore, because the instrumentation is the dominant contributor to the total energy consumption of the system, reducing its power becomes increasingly impactful, especially as the distance increases. As such, increasing the demultiplexing ratio is energetically beneficial, at least up to a certain point.

\begin{figure*}[ht]
  \centering
  \subfloat[]{
\includegraphics[width=0.3\textwidth]{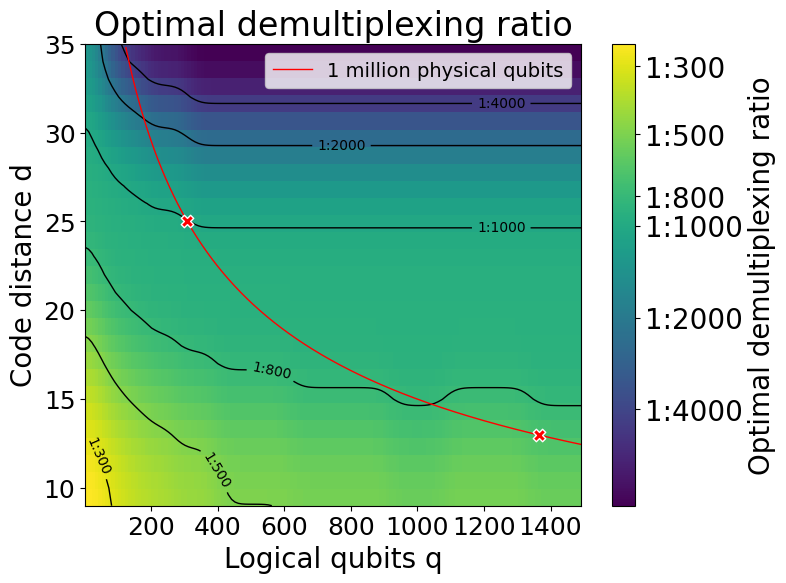}
    \label{mltplx_optimal_ratio}    
    }
  \subfloat[]{
\includegraphics[width=0.3\textwidth]{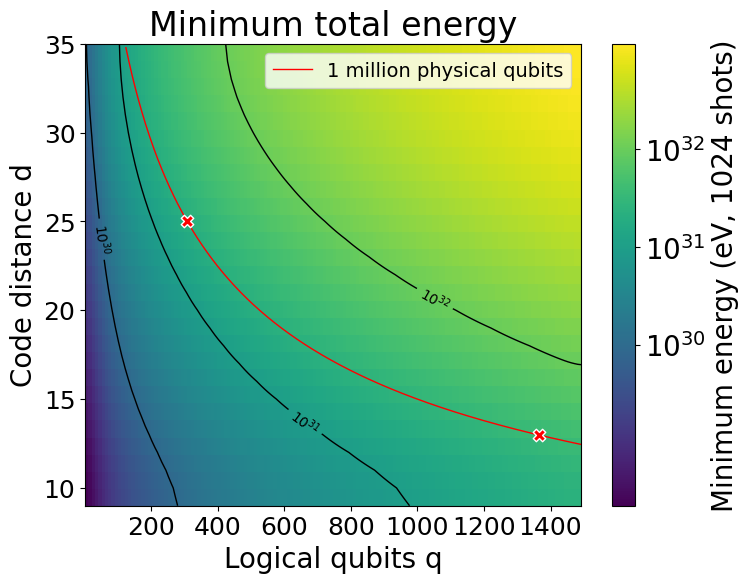}
    \label{mltplx_optimal_energy}    
    }
  \subfloat[]{
\includegraphics[width=0.3\textwidth]{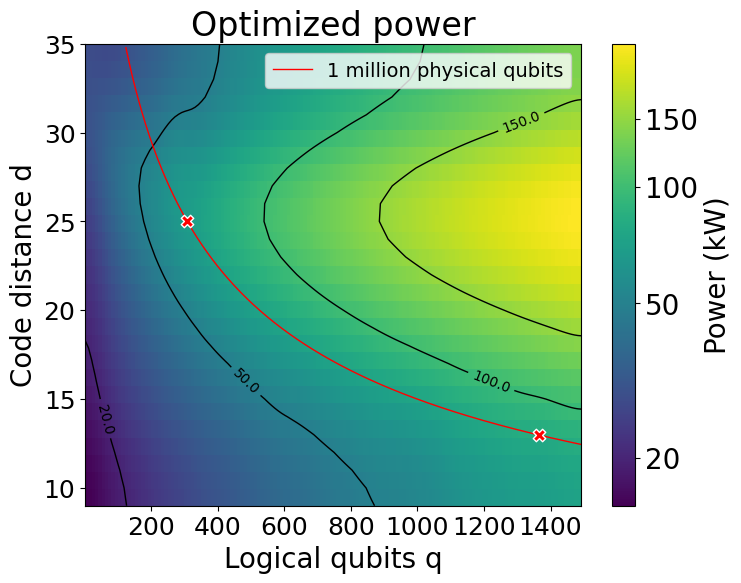}
    \label{mltplx_optimal_power}    
    }
\caption{a) Optimal demultiplexing ratio for varying number of logical qubits and distance. The optimal demultiplexing ratio can range from $1:200$ for small distance and small number of logical qubits, to as low as $1:4000$, for large distances and large number of logical qubits. The red line corresponds to the configuration with approximately $10^6$ physical qubits, with the highlighted points indicating the $d=13, q=1365$ and $d=25,q=308$ configurations; b) total energy (1024 shots) when optimal demultiplexing is applied; c) Total power when the optimal demultiplexing ratio is applied: even a large scale system can operate with a total power consumption well below 1 MW, substantially lower than the power requirements of the highest-performing supercomputers listed in the TOP500 ranking \cite{top500}.}
\end{figure*}

\begin{figure*}[ht]
  \centering
  \subfloat[]{
\includegraphics[width=0.25\textwidth]{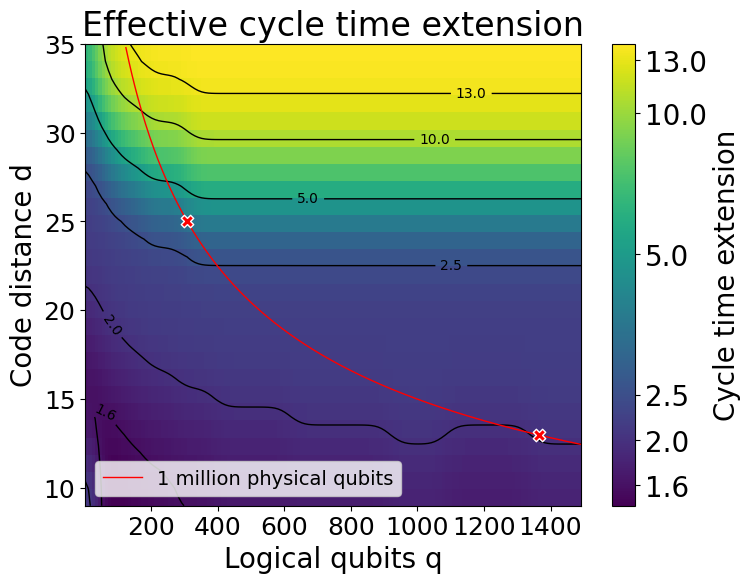}
    \label{mltplx_optimal_cycle}    
    }
  \subfloat[]{
\includegraphics[width=0.25\textwidth]{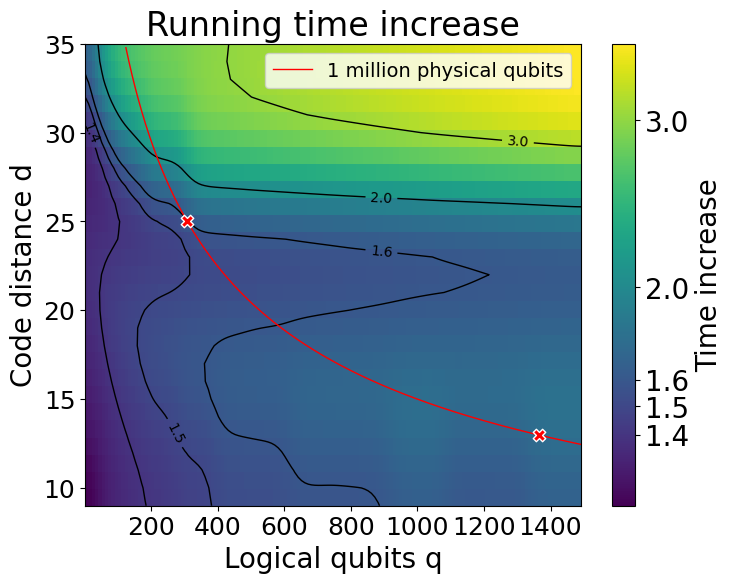}
    \label{mltplx_optimal_time}    
    }
  \subfloat[]{
\includegraphics[width=0.25\textwidth]{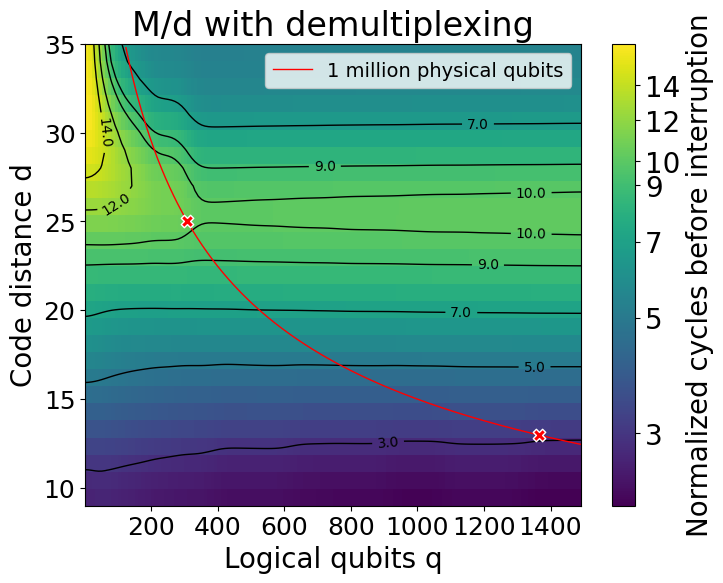}
    \label{mltplx_optimal_M}    
    }
  \subfloat[]{
\includegraphics[width=0.25\textwidth]{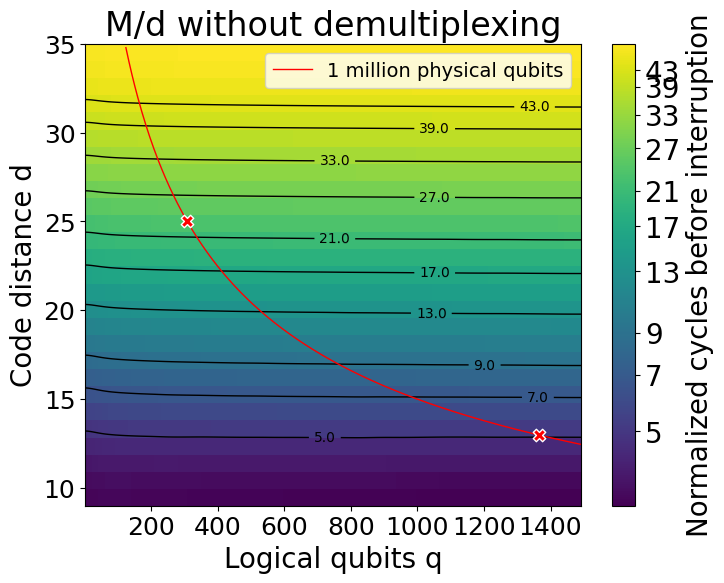}
    \label{original_M}    
    }
\caption{a) Increase in effective cycle time when the optimal demultiplexing ratio is applied. The cycle is extended by 1 order of magnitude, depending on the distance and the number of qubits; b) Increase in total runtime when the optimal demultiplexing ratio is applied. The increase in running time is advantageous up to some point (the optimal point) because the reduction in instrumentation is much larger than the increase in running time for the same demultiplexing ratio. Furthermore, although the cycle is extended by 1 order of magnitude, the running time is extended by a much smaller factor, because the decoding interruption time changes from no demultiplexing to with demultiplexing; c \& d) Decoder interruption time with optimal demultiplexing and without demultiplexing, respectively.}
\end{figure*}

On the other hand, the change in the decoding energy arises from the fact that the decoding interruption time $M$ depends indirectly on the cycle time. As a first approximation, increasing the effective cycle time gives the decoder proportionally more time to complete the decoding of each window. The required decoding power is then reduced by the same factor, leaving the resulting decoding energy approximately unchanged. However, the decoding interruption time $M$, measured in number of cycles, does not remain constant with the effective cycle time. As the cycle becomes longer, syndrome data is generated at a lower rate, while, at the same time, giving the decoder more time within each cycle and thus reducing the number of cycles required for decoding. This implies that the overall decoding energy is not constant, needing to be calculated for each demultiplexing ratio.

When these three components are considered together, the demultiplexing ratio establishes a trade-off between the direct reduction in instrumentation energy and indirect dependence of the cooling and decoding energies on the increased cycle time. A lower demultiplexing ratio might limit the increase in running time, but also provides a smaller reduction in instrumentation energy. Conversely, a higher demultiplexing ratio allows for a greater reduction in instrumentation energy, at the cost of a larger increase in effective cycle time. This interplay leads to the existence of an optimal demultiplexing ratio, as shown in Figs.~\ref{mltplx_d13q100} and \ref{mltplx_d25q100}. Notably, this optimal ratio is higher in Fig.~\ref{mltplx_d25q100} than in Fig.~\ref{mltplx_d13q100}, indicating that it shifts towards higher values as the code distance increases. This is exactly because a larger distance requires more instrumentation, making energetic benefit of reducing its power through the demultiplexing ratio increasingly significant. At the optimal ratio, the three main contributions become balanced and the energy cost is minimized.

\begin{figure}[t]
  \centering
  \subfloat[]{
\includegraphics[width=0.45\textwidth]{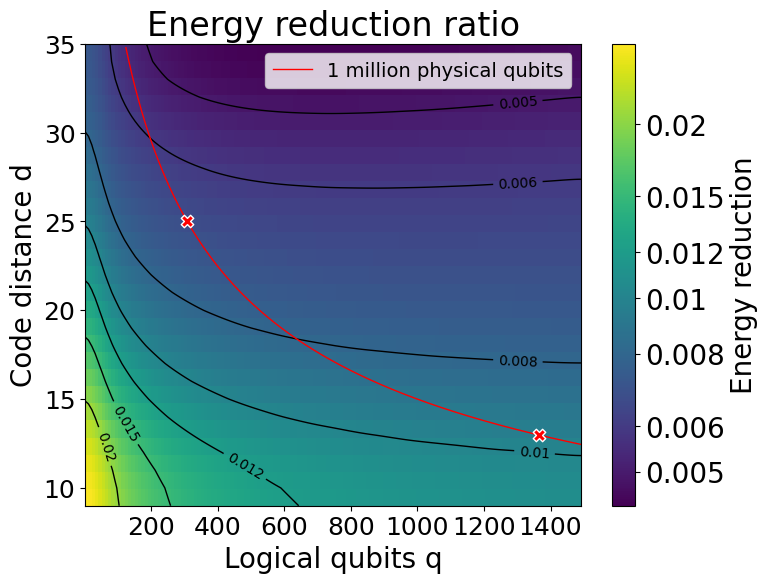}
    \label{mltplx_optimal_reduction}    
    }
    \hfill
    \subfloat[]{
\includegraphics[width=0.45\textwidth]{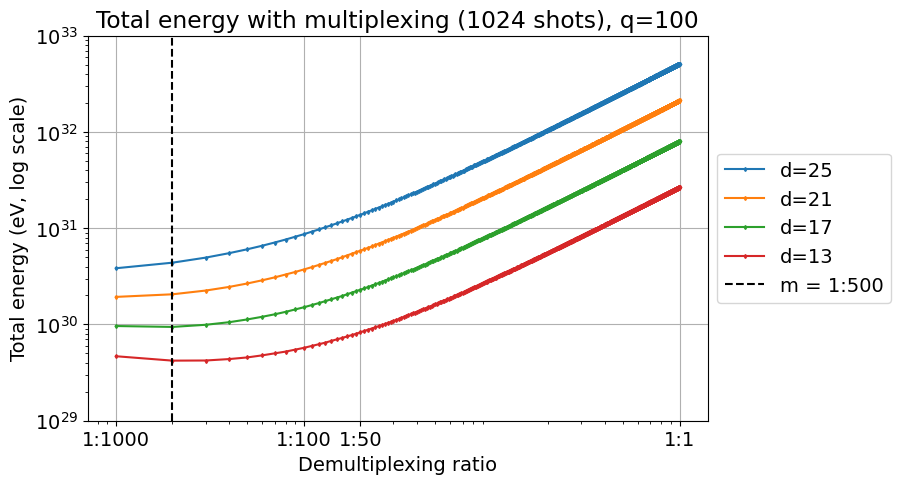}
    \label{mltplx_optimal_100} 
    }
\caption{a) Ratio between the energy with optimal demultiplexing and no demultiplexing: if demultiplexing is applied optimally, the total energy can be reduced by a factor of 100, depending on the distance and number of logical qubits; b) Total energy (1024 shots) with for 100 qubits and 4 different values of distance with varying demultiplexing ratio. As the demultiplexing ratio is reduced from 1:1 to the optimal point, the total energy is minimized. The vertical line represents the demultiplexing ratio $m=1:500$.}  
\end{figure}

By finding this balance, we determine the optimal demultiplexing ratio (Fig.~\ref{mltplx_optimal_ratio}), and the corresponding total energy (Fig.~\ref{mltplx_optimal_energy}). This optimization results in a reduction of energy that depends on the distance and number of logical qubits (as seen in Fig.~\ref{mltplx_optimal_reduction}), which can be by a factor of up to two orders of magnitude, as shown in Fig.~\ref{mltplx_optimal_100}. Moreover, as seen in Fig.~\ref{mltplx_optimal_power}, optimal demultiplexing allows even large systems of more than 1 million physical qubits to operate with a total power consumption well below 1 MW, substantially lower than the power requirements of the highest performing modern supercomputers listed in the TOP500 ranking \cite{top500}.

Fig.~\ref{mltplx_optimal_power} also shows that the optimized power reaches a maximum at $d\approx 25$ and decreases for larger code distances. This does not imply that the system becomes more efficient at larger code distances because, as shown in Fig.~\ref{mltplx_optimal_energy}, the total energy consumption increases with the distance and the number of qubits, even when using the optimal demultiplexing ratio. The decrease in optimized power results from using the optimal demultiplexing ratio for each configuration $(d,q)$. If, instead, the demultiplexing ratio was fixed, e.g. at $1:500$, the extension in time would be constant and the power requirements would increase with the distance and the number of qubits.

Interestingly, the fact that optimizing the demultiplexing ratio for energy directly affects the running time implies that the phenomenon that was verified before will be accentuated: the crossing point between the QFT and the full FFT benchmark in terms of time happens after the crossing point in energy. Therefore, even with error correction, there is a regime, for a certain interval of the number of qubits, in which the QFT falls below the classical FFT benchmark despite not outperforming it in terms of running time.

Furthermore, as displayed in Fig.~\ref{mltplx_optimal_cycle}, when optimal demultiplexing is applied, the effective cycle time is extended by up to an order of magnitude, depending on the distance and number of qubits. However, the total running time (Fig.~\ref{mltplx_optimal_time}) is extended by a considerably smaller factor. This is because the total runtime depends not only on the effective cycle time, but also on the decoding interruption time $M$, which, as explained previously, ends up being significantly different, with and without demultiplexing, as seen in Figs.~\ref{mltplx_optimal_M} and \ref{original_M} respectively. Just like it is observed in the optimal power, the decoding interruption time also reaches a maximum at $d\approx 25$, but this is a result of using the optimal demultiplexing ratio for each configuration $(d,q)$. If, instead, the demultiplexing ratio was fixed, e.g. at $1:500$, the interruption time would exhibit a similar dependence on distance and the number of qubits as in the case without demultiplexing, in Fig.~\ref{original_M}.

\subsection{Practical implementation and cooling budget}

The optimal demultiplexing ratio was obtained solely from an energetics perspective. However, there are hardware constraints which determine the demultiplexing ratio that can be practically implemented. An additional limitation arises from the qubit dephasing time, as extending the cycle increases the idle time of the physical qubits, increasing the probability of dephasing and leading to a larger value of $p$. To minimize the effect of dephasing noise, the extension of the QEC cycle should be limited to a duration shorter than the dephasing time $T_2^*\approx 100\,\mu$s \cite{gravier2025}.


For this reason, and considering a clock signal at 1 GHz, which can be generated on-chip or provided from room temperature electronics, and the shortest gate time of 500 ns, we fix our control demultiplexing ratio to 1:500, which is highlighted by a vertical line in Fig.~\ref{mltplx_optimal_100}. For a system with 1 million physical qubits, corresponding to 3 million electrode gates, this corresponds to around 6000 demultiplexers for control. Furthermore, noting that $m=500$ implies $t_{\text{cycle}}^{\text{demux}}=25\,\mu$s, a system with 1 million qubits with, for example, $d=25$ (for which we can take $n=800$ and $q=308$) would also require $(n(d^2-1)+qd^2)\frac{1\,\mu\text{s}}{t_{\text{cycle}}^{\text{demux}}}=27644$ readout time-division multiplexers.

\begin{figure}[h]
  \centering
  \subfloat[]{
\includegraphics[width=0.4\textwidth]{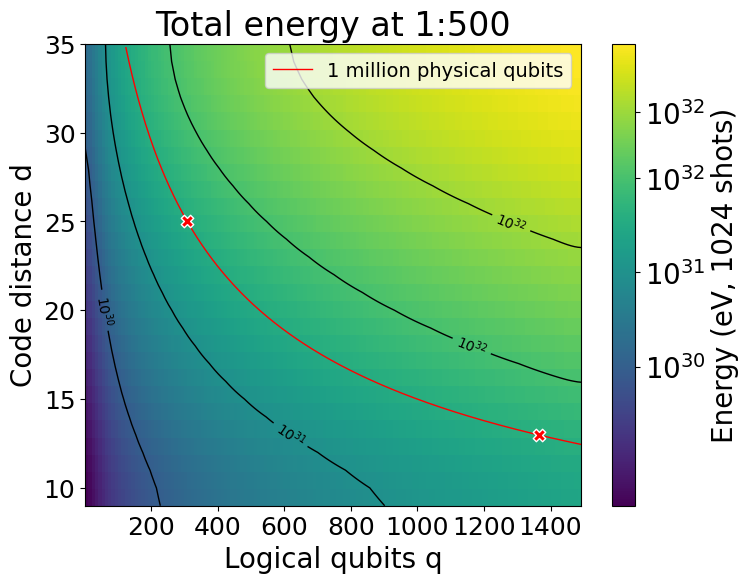}
    \label{optimal_energy_500}    
    }
    \hfill
    \subfloat[]{
\includegraphics[width=0.45\textwidth]{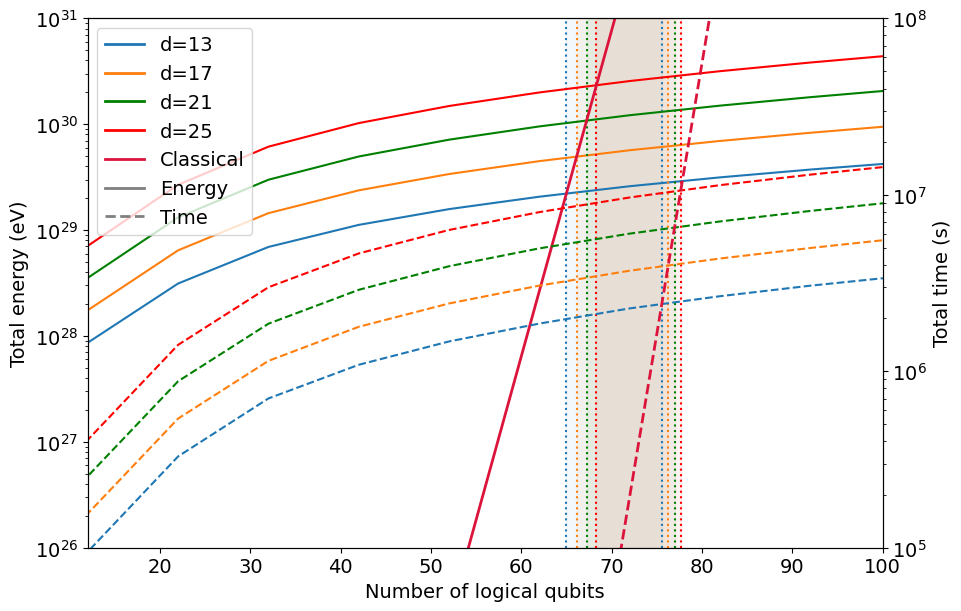}
    \label{optimal_comparison_500} 
    }
\caption{a) Total energy using a demultiplexing ratio $m=1:500$. The red line corresponds to the configuration with approximately $10^6$ physical qubits, with the highlighted points indicating the $d=13,q=1365$ and $d=25,q=308$ configurations; b) Total energy and runtime using a demultiplexing ratio $m=1:500$ as a function of the number of qubits for 4 different distances and comparison with the classical FFT. Solid lines represent the total energy (left axis), while dashed lines represent the runtime (right axis). The classical time is calculated using the El Capitan data, while the classical energy is calculated the KAIROS data. The vertical lines indicate the number of qubits at which the QFT is below the reference line associated with the full FFT benchmark. For every distance, the crossing point in terms of energy is verified before the crossing point in terms of time, as verified in the case without QEC. The shaded region represents the interval between these two points.}  
\end{figure}

If a configuration with $d=13$ ($n=2967$ and $q=1365$) is used instead, this number becomes $(n(d^2-1)+qd^2)\frac{1\,\mu\text{s}}{t_{\text{cycle}}^{\text{demux}}}=29173$. The multiplexers can then be chained to a frequency multiplexer (1:100) in order to reduce the cabling further by a factor of 100 \cite{Jadot2023}.

\begin{figure}[h]
  \centering
  \subfloat[]{
\includegraphics[width=0.45\textwidth]{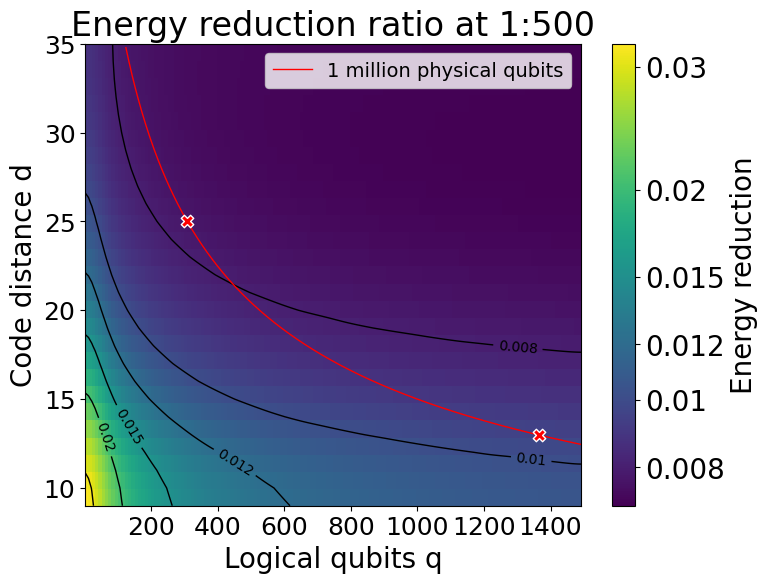}
    \label{optimal_reduction_500}    
    }
    \hfill
    \subfloat[]{
\includegraphics[width=0.45\textwidth]{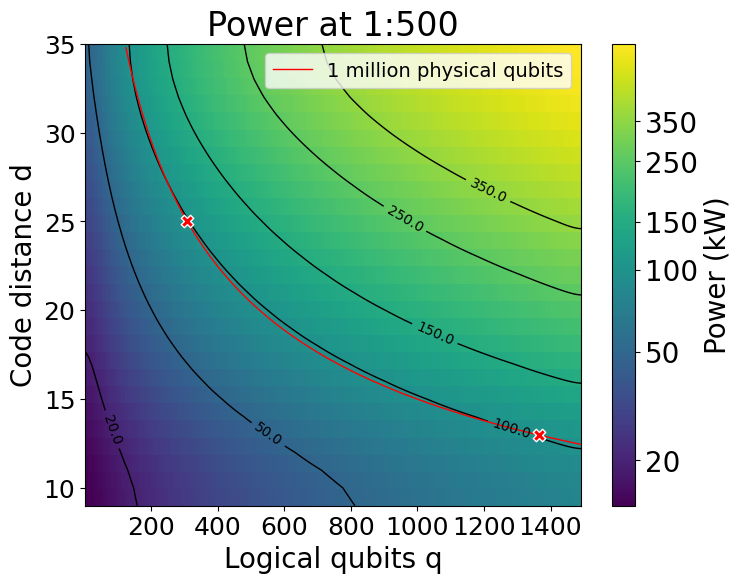}
    \label{optimal_power_500} 
    }
\caption{a) Ratio between the energy using a demultiplexing ratio $m=1:500$ and no demultiplexing: the total energy is still reduced by up to 2 orders of magnitude. The red line corresponds to the configuration with approximately $10^6$ physical qubits, with the highlighted points indicating the $d=13,q=1365$ and $d=25,q=308$ configurations; b) Total power using a demultiplexing ratio $m=1:500$: a system with 1 million physical qubits, corresponding to the points ($q,d$) on the red line, can be operated using 100 kW, substantially lower than the power requirements of modern supercomputers.}  
\end{figure}

The total energy of this system, for different values of distance and number of qubits is displayed in Fig.~\ref{optimal_energy_500}, with a comparison with the classical result being shown in Fig.~\ref{optimal_comparison_500}. Although not optimal, this configuration still provides a reduction in energy by 2 orders of magnitude, as shown in Fig.~\ref{optimal_reduction_500} and allows a system of 1 million physical qubits to be operated at a power of approximately 100 kW, as shown in Fig.~\ref{optimal_power_500}, substantially lower than the power requirements of modern supercomputers.

We must also estimate the power dissipated by the cryoelectronics to verify that the proposed architecture can operate within the available cooling budget, as this constraint was the primary motivation for the inclusion of cryoelectronics, rather than simply a means of optimizing the energetic requirements of the system. The complete derivation is presented in Appendix~\ref{sec:app_cooling}. The total dissipation is estimated by accounting for the control and readout circuitry as well as the cabling. The former consists of logic and pulse driver blocks, based on CMOS transistor-level logic whose dissipation is estimated based on the switching activity and circuit capacitances. For a configuration with 1 million physical qubits, the total cryoelectronics dissipation is estimated to be $\approx700$ mW. Thus the proposed system remains compatible with the cooling budget of commercial high-powered cryostats.

\section{Conclusions}

In this work, we estimated the energy and power requirements of implementing the QFT on a semiconductor spin qubit quantum computer. The calculation accounted for the four operational stages: the initialization of the qubits, the computation costs associated with the execution of the gates, the readout of the qubits and the baseline costs associated with the cryostat and room temperature instrumentation. Among these contributions, the baseline, particularly the cryogenic energy cost, was found to be the biggest fraction of energy consumption without QEC. 

The energetic evaluation reveals that, under the assumptions considered in this work, the QFT can fall below the energy requirements of the full classical FFT, which was used as a benchmark, at practically attainable system sizes. Moreover, the analysis shows that the energy crossover can be verified before the running-time crossover, highlighting that energetic benefits relative to a classical reference can emerge before any time advantage has been realized. The main reason for this difference is the fact that, although a quantum gate is both more energetically expensive and more time-consuming than one classical FLOP, the relative difference in energy is much smaller than the relative difference in time. As long as this relation remains true, this phenomenon would be observed for any example in which the quantum algorithm provides an algorithmic advantage, that is, in any situation in which crossing points are expected to occur. As such, this also remains valid when including factors not considered in this work, such as the initial state preparation and variable number of readout shots, provided that these contributions do not alter the computational complexity of the algorithm and that the total workload remains polynomially bounded.

We also considered the inclusion of the necessary SWAP gates to establish connectivity between different qubits, revealing that the native SWAP gate increases the energetic requirements of the QFT by 10\%, while a decomposed SWAP gates increases the requirements by a factor of $\approx5/2$.

Dynamical decoupling and surface-code were also considered, revealing that error suppression and correction increase the total cost by orders of magnitude, with this increase being expected to become larger as the code distance increases. The increase comes especially from the additional contributions that are necessary once error correction is included: the error correction cycles for every logical gate; the synthesis algorithm for rotations; the additional number of physical qubits; and the classical computation necessary for decoding using a MWPM model, which were found to be a significant component of energy expenditure, surpassing the cooling cost. It was verified that the synthesis algorithm increases the number of gates by a significant factor, depending on the synthesis error; furthermore, it also introduces a cutoff for small angles, thereby reducing the complexity of the QFT from quadratic to linear. The increased number of qubits due to the QEC overhead contributes to a significant increase in the instrumentation requirements, elevating them to be the dominant contribution in terms of energetic cost. Furthermore, this imposes limitations on the brute-force scalability of the QPU, as the resulting instrumentation overhead exceeds the cooling capacity of the cryostat, making the use of cryoelectronics a necessity for a large-scale system. 

To mitigate the increase resulting from error correction, one can increase the gate speeds, thereby decreasing the total execution time and, consequently, the baseline cost, in the case of unconstrained cooling power of a cryostat. However, the system is limited to a certain thermal budget, so, to facilitate the scalability, one can take advantage of cryogenic control electronics, specifically demultiplexers, which decrease the required instrumentation with a slight increase in running time. The utilization of demultiplexers allows to scale the system to a larger number of qubits, while staying within the thermal budget determined by the cooling capacity of the cryostat, and is fundamental for energetically optimal large-scale systems. Under optimized conditions, we found that the use of cryoelectronics can reduce the energetic requirements by 2 orders of magnitude, and enables even a large quantum processor with 1 million physical qubits to operate with a total power consumption well below 1 MW, up to an order of magnitude lower than the power requirements of the highest-performing supercomputers in the world.

It is important to note that the techniques presented in this work correspond to a specific modality of semiconductor spin-qubits, the single-spin Loss-DiVincenzo spin qubits, and there are numerous alternative approaches to semiconductor qubits \cite{Burkard2023} and control mechanisms. Platforms based on shuttling \cite{Knne2024, Li2026} and Exchange-Only (EO) qubits \cite{Abraham2026, Mdzik2025, Li2026} are two other architectures, which have shown potential for scaling up, in addition to the dense array considered in this work. EO qubits work at low magnetic fields and with baseband control signals, lowering the power requirements on these aspects. However, considering that each qubit consists of three physical qubits, and the requirement for a leakage removal step in the QEC cycle, the required overhead is increased. Similarly, shuttling architectures can increase the instrumentation requirement to enable the movement of spins in the QPU. However, these can be reduced using common sinusoidal signaling techniques using clavier gates \cite{Knne2024}. And although shuttling can add to the QEC cycle time, by exploiting the long-range connectivity it may be possible to use non-local quantum Low-Density Parity-Check (qLDPC) codes \cite{chadwick2026cablecarefficientlyschedulingqldpc}, which encode a higher number of logical qubits per physical qubit compared to standard 2D surface code. Thus, a power inefficiency in one area is compensated by an improvement in another in both these platforms, and this warrants a detailed study. However, we do not expect the differences to significantly affect the results presented in this work, as the main considerations remain applicable. 

Spin qubits are currently being fabricated on the well-established semiconductor industrial lines \cite{Paz2024,Fuentes2025,Edlbauer2025} and many components of these systems are not yet optimized for energy, which leaves many potential areas of improvement, including the electronics for control and readout, cryogenics, scalability and density of qubits, connectivity and the frequency of operations. Building on this work, and considering that the energetic performance always depends on the specific details about a platform, an ongoing effort is being made in order to evaluate how the results presented evolve over time and compare to other implementations.

Moreover, while the execution of the quantum gates that comprise the algorithm, the baseline costs and the control costs were all extensively studied with consideration for a range of variables, there are many factors that were not considered in our study. For example, the number of repetitions was always fixed at 1024, but, in reality, the minimum number of repetitions required to extract the output within a certain accuracy depends on the number of qubits, and how it scales could become an important consideration for large scale systems. The same can be said about preparing the input state, as the creation of non-trivial initial states involves a sizable investment in terms of operations, which was not considered in this study. Finally we did not consider a multi-core architecture with QPUs placed in different cryostats. For algorithms with long execution times, we also did not consider any online calibration that might be needed during the computation.


A clear area of potential improvement, in terms of evolution of the energetics of quantum systems, is the cost of decoding for QEC, which, for large code distances, clearly becomes a significant fraction of the total cost, comparable to the baseline energy. 


With these ideas in mind, we intend to explore whether quantum computers could be used for tasks where the possibility of energy advantage becomes a practical benefit, once the energetic feasibility is established. Furthermore, as quantum computing systems mature and become capable of solving problems of larger sizes with longer execution times, optimizing the energetics at a system level becomes critical for the pervasive adoption of this technology. For this purpose, future progress in the field requires new benchmarking methodologies that extend the analysis to a broader range of algorithms beyond the QFT that enable quantum computers to be compared both between themselves and with classical machines. In light of the fact that, within the constraints of our benchmarking, the energy crossing point was found to appear before the running time crossing point, we believe that there might exist other examples where quantum computers might not necessarily be faster for a certain problem size, but rather more efficient. 


\section{Acknowledgments}
JS, MP, YO thank the support from FCT -- Funda\c{c}\~{a}o para a Ci\^{e}ncia e a Tecnologia (Portugal), namely through project  UID/PRR/04540/2025 and contract LA/P/0095/2020. JN, VS, and TM would like to thank the support from European Innovation Council through the project MCSquare grant number 101136414. The authors thank European project SPINS, as well as the software team at Quobly for discussions related to energy consumption of instruments.

\bibliography{main}

\newpage

\appendix

\section{Gate decompositions}
\label{sec:app_gates}

\begin{figure}[H]
  \centering
  \subfloat[H]{
    \includegraphics[width=0.35\textwidth]{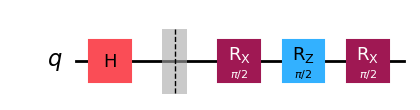}
    \label{fig:image1}    
    }
  \hfill
    \subfloat[CZ]{
    \includegraphics[width=0.35\textwidth]{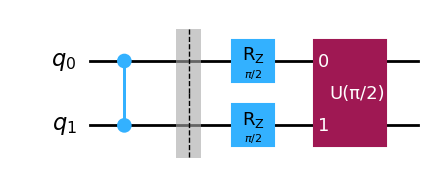}
    \label{fig:image2}
    }
    \hfill
    \subfloat[CNOT]{
    \includegraphics[width=0.35\textwidth]{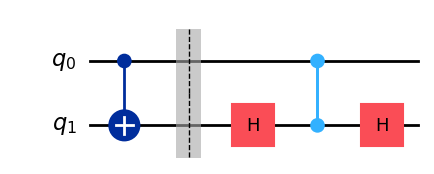}
    \label{fig:image2}
    }
    \hfill
    \subfloat[SWAP]{
    \includegraphics[width=0.35\textwidth]{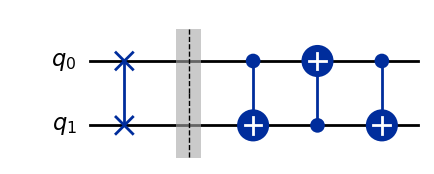}
    \label{fig:image2}
    }
    \hfill
    \subfloat[CP($\theta$)]{
    \includegraphics[width=0.35\textwidth]{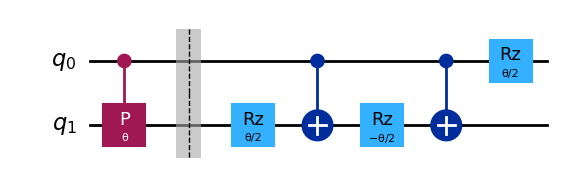}
    \label{fig:image2}
    }
\caption{Decompositions of $H$, $CZ$, $CNOT$, $SWAP$ and $CP(\theta)$ in Qiskit~\cite{Qiskit}}
\end{figure}

\section{Gate synthesis error}
\label{sec:app_synthesis}

The synthesis algorithm is used to approximate the $R_z(\theta)$ gate as a sequence of gates in $\{Z,X,H,T,S\}$. In this work, the approximation sequence is composed of 200-250 gates: $Rz\approx S\cdot H\cdot T\cdot H\cdot T\cdot H\cdot T ...$.

The synthesis error is determined from the distance between the approximated gate $\tilde R_z(\theta)$ and the target rotation $R_z(\theta)$: $\epsilon=||\tilde R_z(\theta)-R_z(\theta)||$. The distance is defined as the maximum deviation over all normalized states: $||\tilde R_z(\theta)-R_z(\theta)||=\sup_{\ket\psi}\frac{||(\tilde R_z(\theta)-R_z(\theta))\ket\psi||}{||\ket\psi||}$ which, for $z$ rotations, is obtained for any state on the equator of the Bloch sphere, $\ket\psi=\frac{1}{\sqrt2}(\ket0+e^{i\phi}\ket{1})$. Choosing $\ket\psi=\ket+$ for simplicity, $\epsilon=||\tilde R_z(\theta)-R_z(\theta)||=||(\tilde R_z(\theta)-R_z(\theta))\ket+||$.

Using the algorithm described in Ref.~\cite{ross2016}, the error $\epsilon$ can be achieved with a sequence whose gate count scales as $O(\log(1/\epsilon))$. However, if the angle of the target rotation is small enough, the best approximation is simply the identity: $||I-R_z(\theta)||<\epsilon$. In particular, for $\theta<\epsilon$, we have:

$||I-R_z(\theta)||=|(I-R_z(\theta))\ket+||
=\frac{1}{\sqrt2}||((1-e^{-i\theta/2})\ket0+(1-e^{i\theta/2})\ket1)||
=\sqrt{2-e^{-i\theta/2}-e^{i\theta/2}}=\sqrt{2-2\cos(\theta/2)}=\sqrt{4\sin^2(\theta/4)}=2\sin(\theta/4)\approx \theta/2<\epsilon/2<\epsilon$.

\section{Error probabilities}
\label{sec:app_errors}

Table \ref{error_probs} shows the algorithm-level error due to faulty $T$ gates, $P_T$, and the error due to logical errors, $P_L$, for different numbers of qubits $q$, values of $p$ and values of distance $d$ with a synthesis error of $\epsilon=10^{-8}$. Beyond the errors displayed in the table, one should consider the algorithm-level error associated with the approximation algorithm, quantified as $\approx \epsilon q^2$, and the error arising from the decoding, which depends on the decoding interruption time.

\begin{table}[h]
\begin{tabular}{c|c|c|c|c|c|c|c}
\multicolumn{4}{c|}{$10^6$ physical qubits} & \multicolumn{2}{c|}{$p=10^{-3}$} & \multicolumn{2}{c}{$p=10^{-4}$} \\
\cline{1-8}
d  & n    & q    & T count    & $P_T$             & $P_L$             & $P_T$             & $P_L$ \\
35 & 408  & 122  & 798659     & $2.8\cdot10^{-2}$ & $3.5\cdot10^{-7}$ & $2.8\cdot10^{-5}$ & $3.5\cdot10^{-25}$\\
33 & 459  & 146  & 955976     & $3.3\cdot10^{-2}$ & $5.3\cdot10^{-6}$ & $3.3\cdot10^{-5}$ & $5.4\cdot10^{-23}$\\
31 & 520  & 175  & 1145677    & $4.0\cdot10^{-2}$ & $8.2\cdot10^{-5}$ & $4.0\cdot10^{-5}$ & $8.2\cdot10^{-21}$\\
29 & 594  & 210  & 1377066    & $4.8\cdot10^{-2}$ & $1.3\cdot10^{-3}$ & $4.8\cdot10^{-5}$ & $1.3\cdot10^{-18}$\\
27 & 686  & 254  & 1663040    & $5.8\cdot10^{-2}$ & $2.0\cdot10^{-2}$ & $5.8\cdot10^{-5}$ & $2.0\cdot10^{-16}$\\
25 & 800  & 308  & 2021893    & $7.1\cdot10^{-2}$ & ------------    & $7.1\cdot10^{-5}$ & $3.2\cdot10^{-12}$\\
23 & 946  & 379  & 2480263    & $8.6\cdot10^{-2}$ & ------------    & $8.6\cdot10^{-5}$ & $5.2\cdot10^{-12}$\\
21 & 1135 & 470  & 3078140    & $1.1\cdot10^{-1}$ & ------------    & $1.1\cdot10^{-5}$ & $8.7\cdot10^{-10}$ \\
19 & 1386 & 592  & 3877771    & $1.3\cdot10^{-1}$ & ------------    & $1.3\cdot10^{-4}$ & $1.5\cdot10^{-7}$ \\
17 & 1733 & 761  & 4980339    & $1.7\cdot10^{-1}$ & ------------    & $1.7\cdot10^{-4}$ & $2.8\cdot10^{-5}$ \\
15 & 2227 & 1002 & 6559280    & $2.2\cdot10^{-1}$ & ------------    & $2.2\cdot10^{-4}$ & $5.6\cdot10^{-3}$ \\
13 & 2967 & 1365 & 8915997    & $3.1\cdot10^{-1}$ & ------------    & $3.1\cdot10^{-4}$ & ------------ \\
\end{tabular}
\caption{Error due to faulty $T$ gates ($P_T=T_{\text{count}}\cdot 35p^3$) and error due to faulty gates ($P_L\approx 0.1\cdot(\frac{p}{0.01})^{\frac{d+1}{2}}\cdot G\cdot n\cdot d$) for different number of qubits $q$, values of $p$ and values of distance $d$. The gate count $G_{\text{count}}\approx 16500\cdot q$ and $T_{\text{count}}\approx 6500\cdot q$ are approximately linear with the number of qubits. The total number of patches is given by $n=2q+\sqrt{8q}+133$ and is fixed such that the total number of physical qubits is 1000000: $n\cdot (2d^2-1)=10^6$.}
    \label{error_probs}
\end{table}

\section{Construction of the cycle}
\label{sec:app_cycle}

\begin{figure*}[ht]
    \centering
    \includegraphics[width=\linewidth]{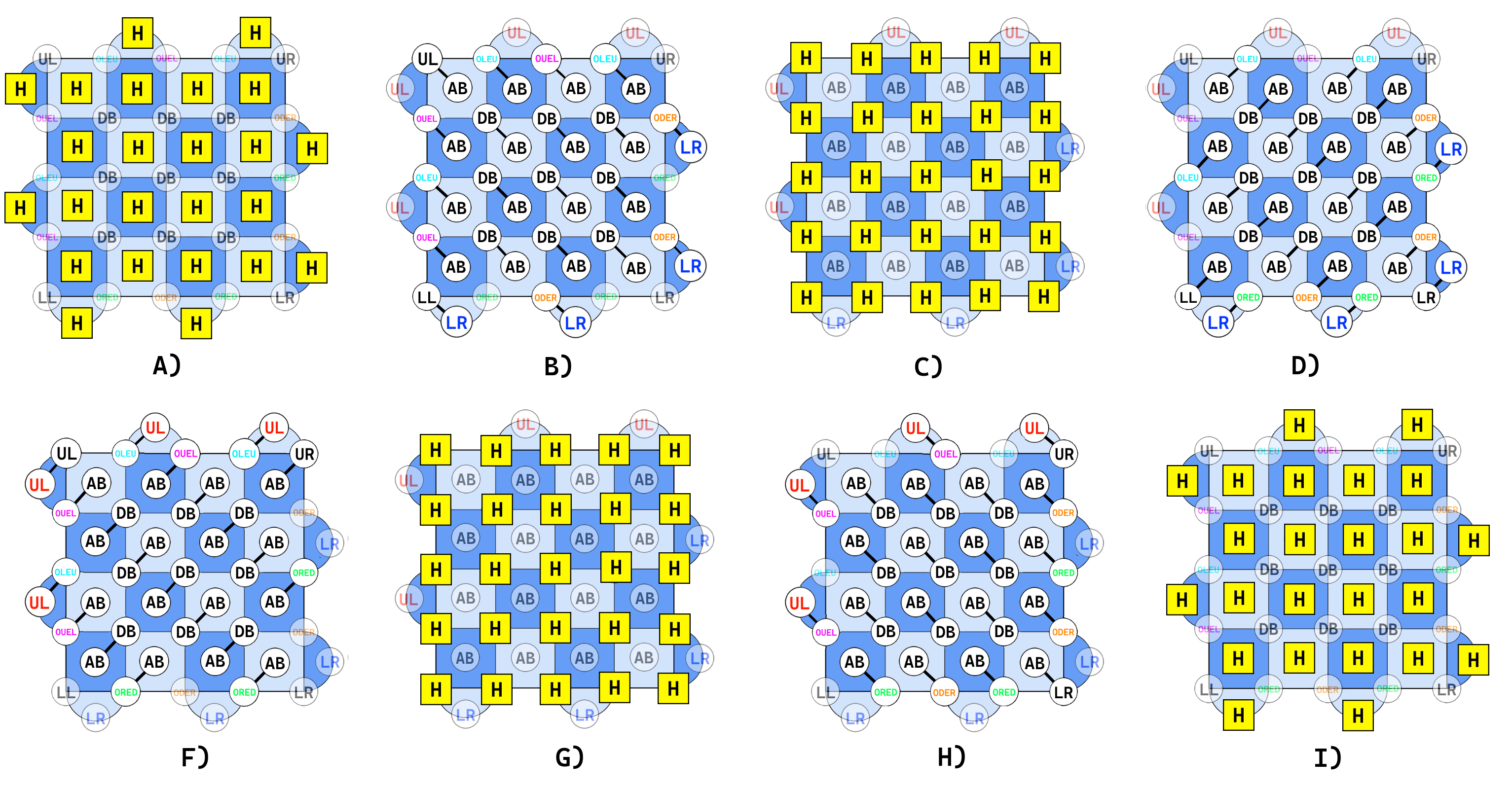}
    \caption{QEC cycle, based on Ref.~\cite{google2024}, for a patch of $d=5$. The cycle is composed of 8 layers of H and CZ gates. A line connecting an ancilla qubit (center of squares) and a data qubit (vertices) represents a CZ gate. A transparent qubit denotes a qubit which is inactive in a particular layer. We can separate the ancilla qubits in 3 groups: the ancillas in the bulk of the patch, denoted \textbf{AB} in black, which are active in every CZ layer; the ancillas in the upper and left boundaries, denoted as \textbf{UL} in red, which are inactive in layers b and d; and the ancillas in the lower and right boundaries, denoted as \textbf{LR} in blue, which are inactive in layers f and h. The data qubits can be separated in 9 groups: the data qubits in the bulk of the patch, denoted \textbf{DB} in black, which are active in every CZ layer; the data qubit in the upper-left corner, denoted \textbf{UL} in black, which is inactive in layers d and h; the data qubit in the upper-right corner, denoted \textbf{UR} in black, which is inactive in layers b and d; the data qubit in the lower-left corner, denoted \textbf{LL} in black, which is inactive in layers f and h; and the data qubit on the lower-right corner, denoted \textbf{LR} in black, which is inactive in layers b and f. Excluding the data qubits on the bulk and in the corners, we have: the data qubits in the odd rows in the left boundary and even columns in the upper boundary, denoted \textbf{OLEU} (odd-left even-up) in light blue, which are inactive in layer h; the data qubits in the odd columns in the upper boundary and even rows in the left boundary, denoted \textbf{OUEL} (odd-up even-left) in purple, which are inactive in layer d; the data qubits in the odd rows in the right boundary and even columns in the lower boundary, denoted \textbf{ORED} (odd-rigth even-down) in green, which are inactive in layer b; and the data qubits in the odd columns in the lower boundary and even rows in the right boundary, denoted \textbf{ODER} (odd-down even-right) in orange, which are inactive in layer f.}
    \label{cycle_draw}
\end{figure*}

We can estimate the energy and time of 1 QEC cycle for 1 logical qubit from Fig. S12 of Ref.~\cite{google2024}, which displays a ZXXZ cycle obtained from alternating layers of parallel H and CZ gates. It must be noted that not all ancilla qubits participate in every layer of the cycle. Each layer of the cycle is represented in Fig.~\ref{cycle_draw}. Looking at every data and ancilla qubit, we can separate the ancillas in 3 groups:
\begin{itemize}
    \item the ancillas in the bulk of the patch, denoted \textbf{AB} in black, which are active in every CZ layer; 
    \item the ancillas in the upper and left boundaries, denoted as \textbf{UL} in red, which are inactive in layers b and d; 
    \item the ancillas in the lower and right boundaries, denoted as \textbf{LR} in blue, which are inactive in layers f and h
\end{itemize}

The data qubits can be separated in 9 groups:
\begin{itemize}
    \item the data qubits in the bulk of the patch, denoted \textbf{DB} in black, which are active in every CZ layer;
    \item the data qubit in the upper-left corner, denoted \textbf{UL} in black, which is inactive in layers d and h;
    \item the data qubit in the upper-right corner, denoted \textbf{UR} in black, which is inactive in layers b and d;
    \item the data qubit in the lower-left corner, denoted \textbf{LL} in black, which is inactive in layers f and h;
    \item the data qubit on the lower-right corner, denoted \textbf{LR} in black, which is inactive in layers b and f
\end{itemize}

Excluding the data qubits on the bulk and in the corners, we also have: 
\begin{itemize}
    \item the data qubits in the odd rows in the left boundary and even columns in the upper boundary, denoted \textbf{OLEU} (odd-left even-up) in light blue, which are inactive in layer h;
    \item the data qubits in the odd columns in the upper boundary and even rows in the left boundary, denoted \textbf{OUEL} (odd-up even-left) in purple, which are inactive in layer d;
    \item the data qubits in the odd rows in the right boundary and even columns in the lower boundary, denoted \textbf{ORED} (odd-rigth even-down) in green, which are inactive in layer b;
    \item and the data qubits in the odd columns in the lower boundary and even rows in the right boundary, denoted \textbf{ODER} (odd-down even-right) in orange, which are inactive in layer f.
\end{itemize}

The operations involved in each layer are also presented in table \ref{table_cycle_1}.

\begin{table}
\begin{tabular}{c|c|c|c|c|c|c|c|c|c}
Qubit      & a & b  & c & d  & f  & g & h  & i & Number of qubits\\
DB         &   & CZ & H & CZ & CZ & H & CZ &   & $(d-2)^2$ \\
OLEU       &   & CZ & H & CZ & CZ & H &    &   & $d-2$ \\
OUEL       &   & CZ & H &    & CZ & H & CZ &   & $d-2$ \\
ORED       &   &    & H & CZ & CZ & H & CZ &   & $d-2$ \\
ODER       &   & CZ & H & CZ &    & H & CZ &   & $d-2$ \\
UL         &   & CZ & H &    & CZ & H &    &   & $1$ \\
UR         &   &    & H &    & CZ & H & CZ &   & $1$ \\
LL         &   & CZ & H & CZ &    & H &    &   & $1$ \\
LR         &   &    & H & CZ &    & H & CZ &   & $1$ \\
Ancilla AB & H & CZ &   & CZ & CZ &   & CZ & H & $(d-1)^2$ \\
Ancilla UL & H &    &   &    & CZ &   & CZ & H & $d-1$ \\
Ancilla LR & H & CZ &   & CZ &    &   &    & H & $d-1$ \\
\end{tabular}
\caption{Operations involved in the ZXXZ QEC cycle from Ref.~\cite{google2024}, presented in Fig.~\ref{cycle_draw}.}
    \label{table_cycle_1}
\end{table}

The cycle can then be simplified \cite{gravier2025}. First, we can decompose the Hadamard gate as (read from right to left in matrix form):
\begin{equation}
    H=R_y(\pi/2)Z=ZR_y(-\pi/2)
\end{equation}
Second, we note that the CZ gate can be decomposed in one $U(-\pi/2)$ gate and two $R_z(-\pi/2)$ gates (one in the target qubit and one in the control qubit) \cite{gravier2025}. Because the $U(-\pi/2)$ gate commutes with the $R_z$ gates, these 3 gates can be applied in any order: 
\begin{equation*}
    CZ=(R_z(-\pi/2)\otimes(R_z(-\pi/2))U(-\pi/2)
\end{equation*}
\begin{equation*}
    CZ=U(-\pi/2)(R_z(-\pi/2)\otimes(R_z(-\pi/2))
\end{equation*}
\begin{equation*}
    CZ=(R_z(-\pi/2)\otimes I)U(-\pi/2)(I\otimes R_z(-\pi/2))
\end{equation*}
Thus, every CZ entry of table \ref{table_cycle_1} can be interpreted as a block (read from left to right as gates) $U(-\pi/2)R_Z(-\pi/2)$ or $R_Z(-\pi/2)U(-\pi/2)$. From this point forward, $R_z(\pi/2)\equiv S$, $R_z(-\pi/2)\equiv S^*$, $R_y(\pi/2)\equiv R_y$, $R_y(-\pi/2)\equiv R_y^*$ and $U(-\pi/2)\equiv U$. This leads to the decomposition in table \ref{table_cycle_2}.

\begin{table}
\begin{tabular}{c|c|c|c|c|c|c|c|c|}
                    & a & b  & c & d  & f  & g & h  & i \\ 
DB         &          & $US^*$ & $R_y^*Z$ & $S^*U$ & $US^*$ & $ZR_y$   & $S^*U$ &        \\
OLEU       &          & $US^*$ & $R_y^*Z$ & $S^*U$ & $US^*$ & $ZR_y$   &        &        \\
OUEL       &          & $US^*$ & $ZR_y$   &        & $US^*$ & $ZR_y$   & $S^*U$ &        \\
ORED       &          &        & $R_y^*Z$ & $S^*U$ & $US^*$ & $ZR_y$   & $S^*U$ &        \\
ODER       &          & $US^*$ & $R_y^*Z$ & $S^*U$ &        & $R_y^*Z$ & $S^*U$ &        \\
UL         &          & $US^*$ & $ZR_y$   &        & $US^*$ & $ZR_y$   &        &        \\
UR         &          &        & $R_y^*Z$ &        & $S^*U$ & $R_y^*Z$ & $S^*U$ &        \\
LL         &          & $US^*$ & $ZR_y$   & $US^*$ &        & $ZR_y$   &        &        \\
LR         &          &        & $R_y^*Z$ & $S^*U$ &        & $R_y^*Z$ & $S^*U$ &        \\
Ancilla AB & $R_y^*Z$ & $S^*U$ &          & $US^*$ & $S^*U$ &          & $US^*$ & $ZR_y$ \\
Ancilla UL & $R_y^*Z$ &        &          &        & $S^*U$ &          & $US^*$ & $ZR_y$ \\
Ancilla LR & $R_y^*Z$ & $S^*U$ &          & $US^*$ &        &          &        & $ZR_y$ \\
\end{tabular}
\caption{Operations involved in the ZXXZ QEC cycle from Ref.~\cite{google2024} displayed in Fig.~\ref{cycle_draw}, using the decompositions of $CZ$ as one $U(-\pi/4)\equiv U$ and 2 $(R_z(-\pi/2)\equiv S^*$, each in one qubit, in any order, and H as $H=ZR_y=R_y^*Z$ (read from left to right as gates) where $R_y\equiv R_y(\pi/2)$ and $R_y^*\equiv R_y(-\pi/2)$.}
    \label{table_cycle_2}
\end{table}

\begin{table}
\begin{tabular}{c|c|c|c|c|c|c|c|c|c}
                     & a & b  & c & d  & f  & g & h  & i & \\ 
DB         &         & $U$ & $R_x$ & $U$ & $U$ & $R_x$ & $U$ &       & $(d-2)^2$ \\
OLEU       &         & $U$ & $R_x$ & $U$ & $U$ & $R_x$ & $S$ &       & $d-2$ \\
OUEL       &         & $U$ & $R_x$ & $S$ & $U$ & $R_x$ & $U$ &       & $d-2$ \\
ORED       &         & $S$ & $R_x$ & $U$ & $U$ & $R_x$ & $U$ &       & $d-2$ \\
ODER       &         & $U$ & $R_x$ & $U$ & $S$ & $R_x$ & $U$ &       & $d-2$ \\
UL         &         & $U$ & $R_x$ & $Z$ & $U$ &       &     & $R_y$ & $1$ \\
UR         & $R_y^*$ &     &       & $Z$ & $U$ & $R_x$ & $U$ &       & $1$ \\
LL         &         & $U$ & $R_x$ & $U$ & $Z$ &       &     & $R_y$ & $1$ \\
LR         & $R_y^*$ &     &       & $U$ & $Z$ & $R_x$ & $U$ &       & $1$ \\
Ancilla AB & $R_y^*$ & $U$ &       & $U$ & $U$ &       & $U$ & $R_y$ & $(d-1)^2$ \\
Ancilla UL & $R_y^*$ &     &       & $Z$ & $U$ &       & $U$ & $R_y$ & $d-1$ \\
Ancilla LR & $R_y^*$ & $U$ &       & $U$ & $Z$ &       &     & $R_y$ & $d-1$ \\
\end{tabular}
\caption{Operations involved in the ZXXZ QEC cycle from Ref.~\cite{google2024}, using the decompositions $CZ=(R_z(-\pi/2)\otimes(R_z(-\pi/2))U(-\pi/2)=U(-\pi/2)(R_z(-\pi/2)\otimes(R_z(-\pi/2))=(R_z(-\pi/2)\otimes I)U(-\pi/2)(I\otimes R_z(-\pi/2))$ and $H=ZR_y(\pi/2)=R_y(-\pi/2)Z$.}
    \label{table_cycle_3}
\end{table}

\begin{figure*}[ht]
    \centering
    \includegraphics[width=\linewidth]{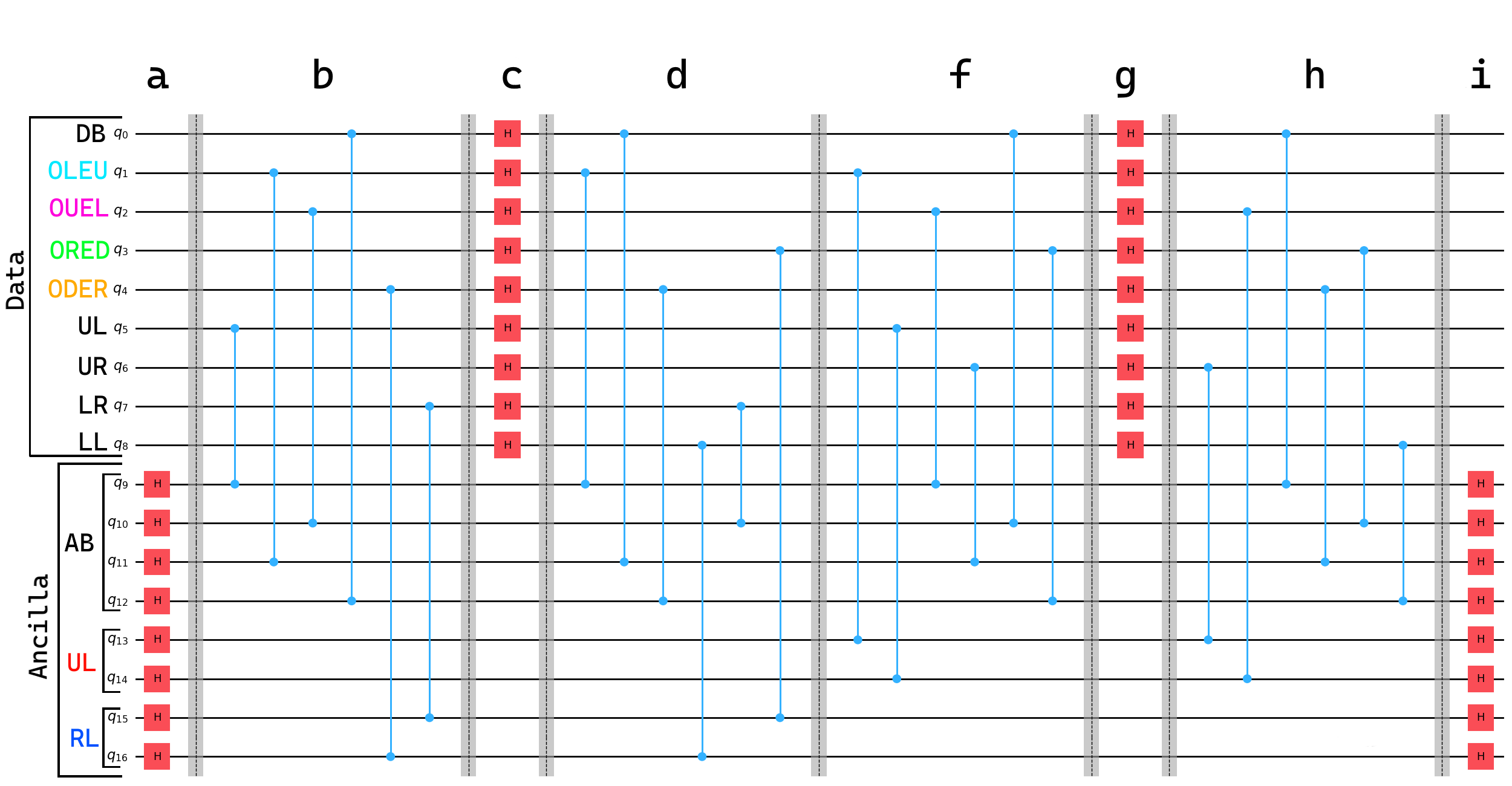}
    \caption{Operations involved in the ZXXZ QEC cycle from Ref.~\cite{google2024}, presented in Fig.~\ref{cycle_draw} before the decomposition and simplification, for a patch of $d=3$. Here, a line connecting 2 qubits represents a CZ gate.}
    \label{cycle_circuit_before}
\end{figure*}

\begin{figure*}[ht]
    \centering
    \includegraphics[width=\linewidth]{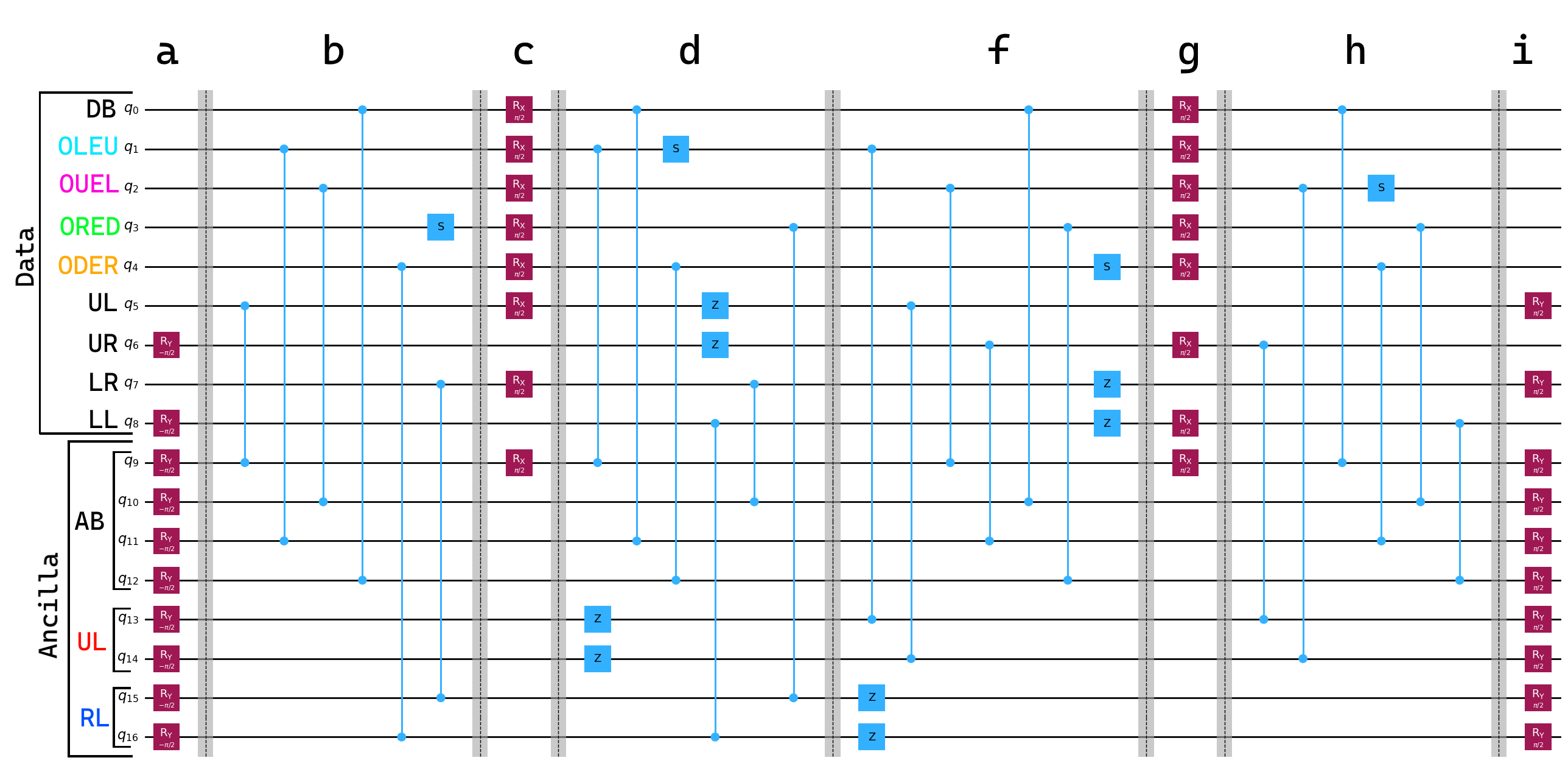}
    \caption{Operations involved in the ZXXZ QEC cycle from Ref.~\cite{google2024}, presented in Fig.~\ref{cycle_draw} after the decomposition and simplification, for a patch of $d=3$. Here, a line connecting 2 qubits represents a U gate.}
    \label{cycle_circuit_after}
\end{figure*}

At this point, all the rotations in the $z$ axis that follow one after the other, in particular $ZS^*$ or $S^*Z$, can be compressed into $S$. Furthermore, $S^*R_y^*S=SR_yS^*=R_x(-\pi/2)\equiv R_x^*$. Finally, the $U$ gate commutes with $S$ and $S^*$, so some gates can be added into one unique $Z$ gate. From these 3 simplifications, we obtain the decomposition in table \ref{table_cycle_3}. We notice that in 4 of the 8 layers, $R_{x/y}$ gates are applied, which can be done in the same axis $x$ or $y$, provided that additional virtual $S,S^*$ gates are handled via software, so that they can be executed using a single global ESR line. The remaining 4 layers are applications of the two-qubit gate between data and ancilla qubits. The circuits before and after the simplification, for patches of $d=3$, are presented in Figs.~\ref{cycle_circuit_before} and \ref{cycle_circuit_after}, respectively.

The refocused $U$ gate \cite{gravier2025}  that is considered in this work is
\begin{equation}
    U_{\text{refocused}}=(X_1\otimes X_2)\sqrt{U}(X_1\otimes X_2)\sqrt{U}
\end{equation}
in which the $\sqrt{U}$ gate can be implemented by halving the application time of the $U$ gate. This enables the cycle to not only be used to correct logical errors, but also to extend the dephasing time of the qubits, effectively serving the role of a DD sequence. This extends the time of the $U$ gate from 1 $\mu$s to 3 $\mu$s, and its energy from $3.125$ MeV to $18.75$ MeV, excluding the energy necessary to sustain the ESR field. 

Furthermore, we can assume that the time of the $R_{x/y}(\pi/2)$ rotations are half of the time of the $R_{x/y}(\pi)$ rotations. Thus, adding the time of each layer of parallel operations and the time of performing 1 measurement of the ancilla qubits at the end, we obtain $t_{\text{cycle}}=15\,\mu$s.

In terms of energy, the cycle is composed by $4(d-2)^2+12(d-2)+8+4(d-1)^2+4(d-1)=8d^2-8d$ $U$ gates, adding up to $8(d^2-d)\cdot18.75$ MeV, and $2(d-2)^2+12(d-2)+12+2(d-1)^2+6(d-1)=4d^2+6d-8$ rotations, each requiring $3.125$ MeV, which adds up to $(4d^2+6d-8)\cdot 3.125$ MeV. The $d^2-1$ measurements at the end of the sequence account for $(d^2-1)\cdot 49.39$ MeV. The total cost of the cycle is displayed in Fig.~\ref{cycle_energy}. If the gates in each layer are performed in parallel, the global ESR field is used 12 times in the cycle (4 times for the $R_{x/y}$ gates and 8 times for the refocusing sequences during the U gates) for a total time of $10\,\mu$s. However, if the cycles across logical qubits have a small delay with respect to each other, which is the case if readout and control multiplexing is used, the ESR field is, in practice, active for the entire duration of the cycles, with the different qubits being brought in and out of resonance at each time-step, representing an additional energy of $1.6\,\mu\text{W}\cdot t_{cycle}$.

\section{Estimating the decoding costs}
\label{sec:app_decoding}

We consider the parallel window MWPM decoder from Ref.~\cite{Camps2024}. To estimate the resources required for decoding, an example sequence of gates $H$, $S$, $CX$, $T$ is considered, which requires a total of $18d+2M$ cycles. Here, $M$ is the maximum number of cycles that the decoder is allowed to take before being interrupted. As such, $18d+2M$ encapsulates both the cycles necessary for the execution of the gates, as well as the cycles dedicated to decoding.

The data generated during the cycles is divided in two rounds of windows of size $3d$, as seen in Fig. 7 of Ref.~\cite{Camps2024}. The total number number of windows is
\begin{equation}
    N_{\text{windows}}=\lceil\frac{18d+2M}{4d}\rceil
\end{equation}
In total, it is estimated that each window requires on average $34d^{4.3}$ FLOPs to be decoded. The time of decoding one window depends on the single-core performance of the CPU. An AMD EPYC 7763 has a peak performance of 3.584 TFLOPs/s distributed along 64 cores. Assuming that each window is decoded by 1 core at 10\% peak performance, the time of decoding 1 window is on average $34d^{4.3}/\frac{3.584\cdot10^{12}\cdot0.1}{64}=6.1\cdot10^{-9}\cdot d^{4.3}$ seconds.

Not all windows require this exact time to decode. The number of cycles to decode one window follows a binomial distribution, with a maximum of $N$ cycles and expected value $Np$, where the factor of $p$ reflects the increased difficulty of decoding, if non-zero syndrome events are more likely. The expected value (in cycles) should coincide with the average time (in seconds) obtained before. As such, the average number of cycles to decode 1 window is $Np=\frac{6.1\cdot10^{-9}d^{4.3}\text{ s}}{t_{\text{cycle}}}$. By dividing by $p$, we can determine the maximum number of cycles to decode the window $N=\frac{6.1\cdot10^{-9}d^{4.3}\text{ s}}{pt_{\text{cycle}}}$. Taking $p=10^{-4}$ and $t_{\text{cycle}}=15\,\mu$s, this gives approximately $N=4.1d^{4.3}$ cycles. 

Instead of allowing the decoder to take $N$ cycles for each window, the decoding is interrupted after $M\leq N$ cycles. A window that requires more than $M$ cycles to decode results in a failed decoding process. This will result in a decoding error per cycle $p_L^{\text{dec}}=p_L\cdot p_{\text{dec}}(k\leq M)+p\cdot p_{\text{dec}}(k>M)$ where $p_{\text{dec}}(k)=C_k^Np^k(1-p)^{N-k}$ is the binomial distribution giving the probability that the decoding takes $k$ cycles, $p_L=0.1(\frac{p}{0.01})^{\frac{d+1}{2}}$, and $p_{\text{dec}}(k>M)=1-p_{dec}(k\leq M)$. The total algorithm-level error due to this component can be estimated by multiplying the error probability per cycle by the number of patches and by total number of cycles, which in this case is approximately $(18d+2M)\cdot T_{\text{count}}$: $P_L^{\text{dec}}=n\cdot p_L^{\text{dec}}\cdot (18d+2M)\cdot T_{\text{count}}$. By setting this error to a target value of 0.01, if $P_L\leq0.01$, or to $P_L$, if $P_L>0.01$, we can estimate the appropriate interruption time $M$ and the resulting required decoding power in FLOPs/s \cite{Camps2024}
\begin{equation}
    C=20\frac{N_{\text{windows}}\cdot n\cdot34\cdot d^{4.3}}{M\cdot t_{cycle}}
\end{equation}
where the prefactor of 20 comes from assuming 10\% peak performance and two rounds of windows. By dividing the result by the FLOPs/s of the CPU and multiplying by its power (280 W for the AMD EPYC 7763), we can determine how many CPUs are needed and the resulting power in W.

\section{Cooling budget for a system with 1 million qubits}
\label{sec:app_cooling}

As we scale, we need to determine if the system can efficiently operate within the cooling budget set by the cryostat. As seen previously, brute force scaling with coaxial cables alone will be limiting, due to which cryogenic control and readout circuitry is a must. In order to calculate the dissipated power, we must take into account the dissipation from the RF cables and the cryogenic circuitry. For the cryogenic circuitry, we assume the operational power to be completely dissipated, giving us an upper bound on the dissipated power.

\begin{figure}[h]
  \centering
  \subfloat[]{
    \includegraphics[width=0.45\textwidth]{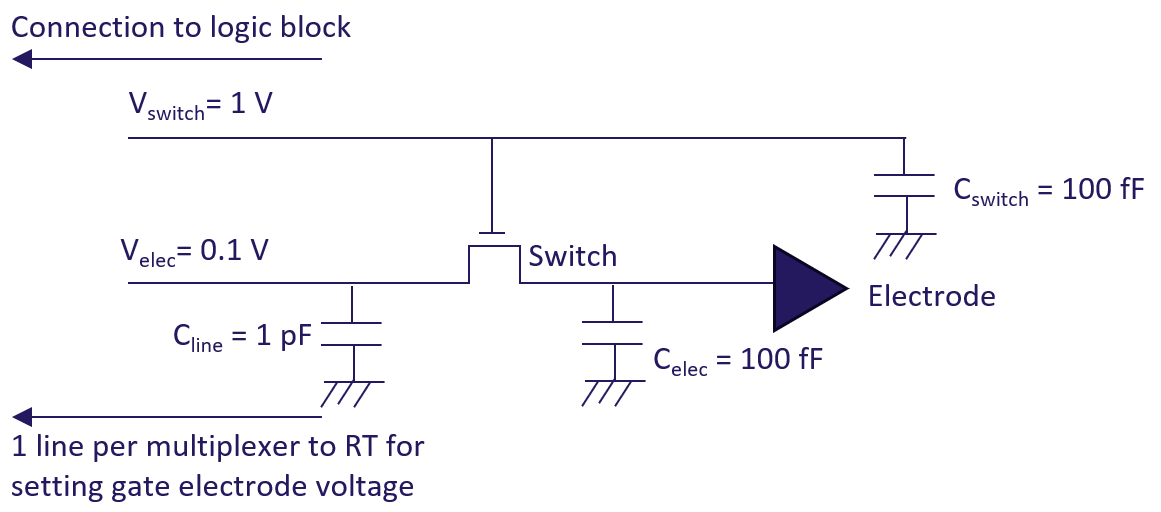}
    \label{cryoelec_circ}    
    }
    \hfill
    \subfloat[]{
    \includegraphics[width=0.45\textwidth]{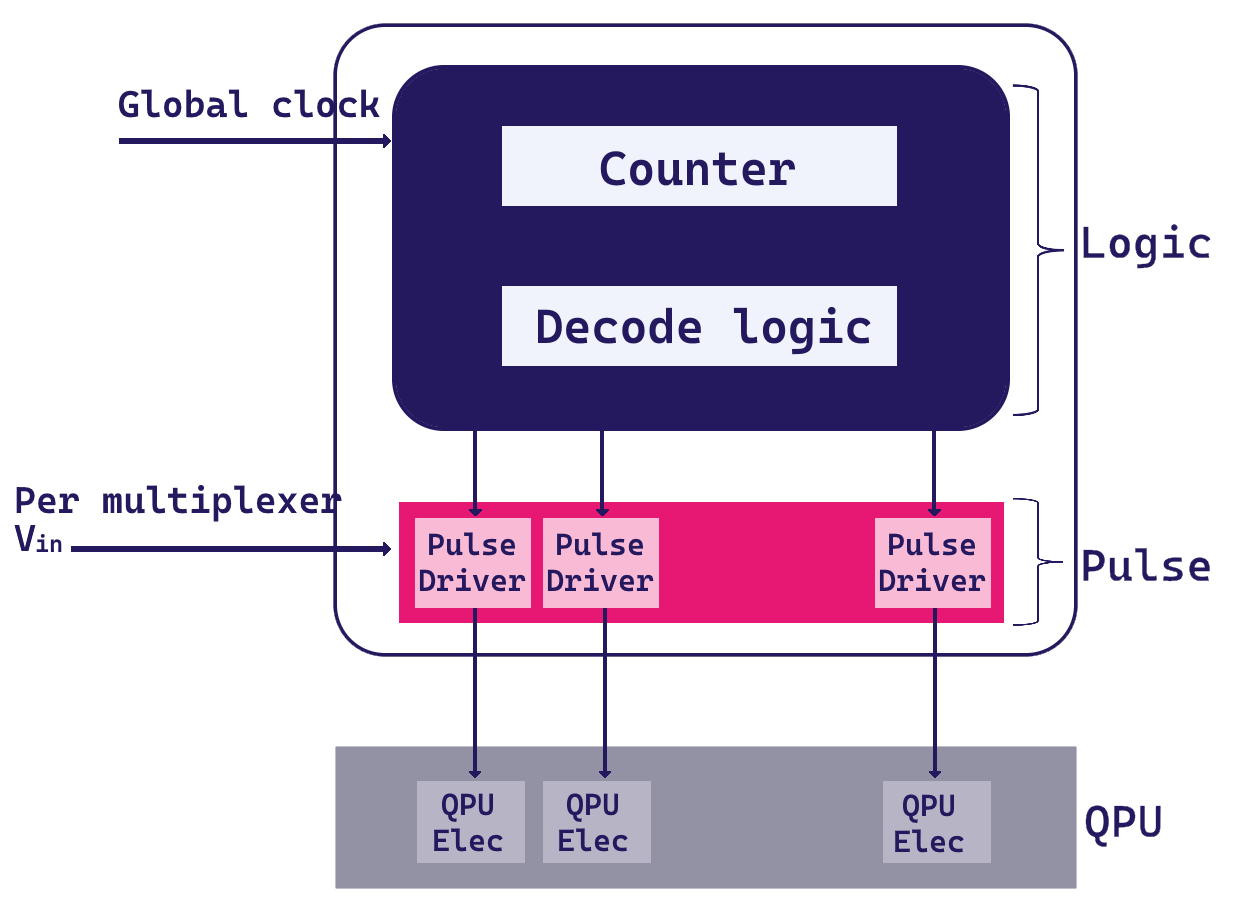}
    \label{cryoelec_blocks}
    }
\caption{a) Circuit diagram of the pulsing block of the multiplexer. There is 1 such block per gate electrode that needs to be controlled; b) The two main components of a cryogenic demultiplexer for control and multiplexer for readout, connected to the QPU.}  
\end{figure}

These cryogenic circuits, consisting of the control demultiplexer and the readout multiplexer, primarily consist of two components, displayed in Fig.~\ref{cryoelec_blocks} - a logic block consisting of a counter and a decoder, and a pulse driver block \cite{Jadot2023}. Such a circuit allows us to apply control pulses or readout individual gate electrodes sequentially, in a cyclical manner, thus giving us a time-division multiplexer/demultiplexer. The pulse driver consists of a switch/transistor that is controlled by the logic. This allows the circuit displayed in Fig.~\ref{cryoelec_circ} to select individual blocks with a voltage swing of $V_{\text{switch}}=1$ V. We assume a switching capacitance $C_{\text{switch}}$ of 100 fF, corresponding to CMOS technology. Once an electrode is selected by this switch, a gate voltage $V_{\text{elec}}=0.1$ V can be applied to the electrode. This voltage can be provided from an external connection (1 RF cable to each multiplexer), as considered here, or by on-chip generation. Here we assume a line capacitance $C_{\text{line}}\approx1$ pF. This voltage charges a hold capacitor, $C_{\text{elec}}$ considered to be around 100 fF. This results in a total power consumption of
\begin{equation}
	\begin{split}
		&P_{\text{pulse}}=P_{\text{switch}}+P_{\text{elec}} \\= &C_{\text{switch}}V_{\text{switch}}^2f_{\text{switch}} + C_{\text{line}}\parallel C_{\text{elec}} V_{elec}^2f_{\text{elec}}
	\end{split}
\end{equation}

The control and readout drivers are switched at frequencies of 2 MHz and 1 MHz respectively. This ensures that these signals stays ON for a duration of 500 ns and 1$\mu$ s respectively, prior to being refreshed. The QPU gate electrodes are controlled using signals up to a frequency of 1 GHz. This requires the driver switch to have fast switching time in ns. However, since these signal lines terminate into high-impedence, purely capacitive QPU gate electrodes, continuous RF conduction loss is zero. Thus, the total power dissipation of the switch is dominated by the dynamic gate control power given above. This results in a per individual driver block dissipation of 200 nW for control and 100 nW for readout. For control, this results in a dissipation of around 600 mW for addressing $10^6$ qubits with $3\cdot10^6$ gate electrodes in total. 

In order to readout $10^6$ qubits, assuming a multiplexing ratio $m=1:500$, we would need 29173 multiplexers for $d=13$ and 27644 multiplexers for $d=25$, resulting in a total of 2.92 mW and 2.76 mW respectively. We can further chain the readout to a frequency-division multiplexer \cite{Schmidt2024}, allowing us to further reduce the cabling by a factor of 100, at an expense of $292\cdot7.4\,\mu\text{W}=2.16$ mW (d=13) and $277\cdot7.4\,\mu\text{W}=2.05mW$ (d=25).

The logic block consists of a counter and a decoding block controlled by a clock running at at a frequency $F=1$ GHz. This allows us to switch between electrodes at a ns time scale. The readout circuits can run slower at 1 MHz. The power of such circuits are given by:
\begin{equation}
    P_{\text{logic}} = f\sum \alpha_iC_iV^2_{DD}
\end{equation}
Here, $f$ is the switching frequency, $\alpha$ is the activity factor, $C$ is the switching capacitor and $V_{DD}$ is the voltage swing. Considering 2 state changes of the counter per clock cycle ($\alpha=2$), and assuming a capacitance of 5 fF and a voltage swing of 1 V, we obtain $P_{\text{counter}}=10\,\mu$W for the control and $0.01\,\mu$W for the RO.

This counter then selects the correct gate electrode to apply the signal to, via the decode logic. The decode can consist of registers or a simple transistor-level logic, which is considered here. Considering CMOS technology, we assume an effective $C_{\text{decode}}$ of 5 fF, with a voltage swing of 1 V, resulting in $P_{\text{decode}}=5\,\mu$W and 5 nW for the control and readout respectively.

The final part to consider in the calculation of the power dissipation inside the cryostat is the thermal load brought in by the RF cables. Here, the static load is the dominant factor rather than the dynamic load \cite{Raicu2025}. Considering a dissipation of $2.4\,\mu$W at 1 K, we obtain 14.4 mW of disspated power by the cables.

\begin{table}[h]
\begin{tabular}{lll}
                                 & d=13, q=1365 & d=25, q=308 \\
number of control demultiplexers & 6002                 & 5988               \\
number of readout multiplexers   & 29173                & 27644              \\
$P_{\text{pulse}}$ control                  & 600.01 mW            & 598.77 mW          \\
$P_{\text{pulse}}$ readout                  & 2.92 mW              & 2.76 mW            \\
$P_{\text{logic}}$ control                  & 90.03 mW             & 89.82 mW           \\
$P_{\text{logic}}$ readout                  & 0.44 mW              & 0.41 mW            \\
RO cables                        & 2.16 mW              & 2.05 mW            \\
RF cables                        & 14.40 mW             & 14.37 mW           \\
Total                            & 710.08 mW            & 708.19 mW         
\end{tabular}
\caption{Cryoelectronics dissipation for two configurations ($d=13$ and $d=25$) with 1 million physical qubits.}
    \label{table_cryoelec}
\end{table}

Adding the three contributions $P_{\text{pulse total}}$, $P_{\text{logic total}}$ and $P_{\text{cable total}}$, for both the control demultiplexers and readout multiplexers, we can obtain the total dissipated power for the given configuratioon with 1 million physical qubits, which gives $\approx700$ mW. The dissipation for the two configurations considered ($d=13$ and $d=25$) is detailed in table.~\ref{table_cryoelec}. Thus accounting for the dissipation of the cryoelectronic circuitry, the proposed system remains compatible with the available cooling power.

\end{document}